\documentclass[11pt]{article}
\usepackage[margin=2.5cm,top=2cm,bottom=2.3cm,centering]{geometry}
\usepackage[utf8]{inputenc}
\usepackage[english]{babel}
\usepackage{graphicx}
\usepackage{amsmath}
\usepackage{amsfonts}
\usepackage{amssymb}
\usepackage{amsthm}
\usepackage{mathrsfs}
\usepackage{hyperref}
\usepackage{tikz}
\usepackage[hypcap]{caption}
\usepackage{bm}
\usepackage{cancel}
\usepackage{cite}
\usetikzlibrary{decorations.pathmorphing}
\usetikzlibrary{arrows,shapes,positioning}
\usetikzlibrary{decorations.markings}
\usetikzlibrary{patterns}

\hypersetup{
colorlinks=true,
citecolor=red,
filecolor=black,
linkcolor=blue,
urlcolor=blue,
pdfauthor={Adrien Poncelet and Philippe Ruelle},
pdftitle={Multipoint correlators in the Abelian sandpile model},
pdfstartview={XYZ null null 1.50},
bookmarksopen=true
}

\numberwithin{equation}{section}

\newcommand*\xbar[1]{%
  \hbox{%
    \vbox{%
      \hrule height 0.5pt % The actual bar
      \kern0.3ex%         % Distance between bar and symbol
      \hbox{%
        \kern-0.1em%      % Shortening on the left side
        \ensuremath{#1}%
        \kern-0.0em%      % Shortening on the right side
      }%
    }%
  }%
} 

\newcommand{\C}{\mathbb{C}}
\newcommand{\Z}{\mathbb{Z}}
\newcommand{\E}{\mathcal{E}}
\newcommand{\G}{\mathcal{G}}
\newcommand{\N}{\mathcal{N}}
\newcommand{\V}{\mathcal{V}}
\newcommand{\Gr}{\mathbf{G}}
\newcommand{\Zr}{\mathbf{Z}}
\newcommand{\bs}{\backslash}
\newcommand{\be}{\begin{equation}}
\newcommand{\ee}{\end{equation}}
\newcommand{\bea}{\begin{eqnarray}}
\newcommand{\eea}{\end{eqnarray}}
\newcommand\egal{&\!\!\!=\!\!\!&}
\newcommand\plus{&\!\!\!+\!\!\!&}

\newcommand\la{\langle}
\newcommand\ra{\rangle}

\renewcommand{\l}{\ell}
\renewcommand{\P}{\mathbb{P}}
\renewcommand{\O}{\mathcal{O}}

\renewcommand{\ge}{\geqslant}
\renewcommand{\le}{\leqslant}

\newtheorem*{grovethm}{Grove theorem}

\title{\bf Multipoint correlators in the\\ Abelian sandpile model}
\author{\normalsize \textsc{Adrien Poncelet}, \textsc{Philippe Ruelle}\medskip\\
{\normalsize
\begin{minipage}{0.95\textwidth}
\begin{center}
\textit{Universit\'e catholique de Louvain\\Institut de recherche en math\'ematique et physique\\Chemin du Cyclotron 2, 1348 Louvain-la-Neuve, Belgium}\\
\medskip
\href{mailto:adrien.poncelet@uclouvain.be}{\normalsize\texttt{adrien.poncelet@uclouvain.be}}, \href{mailto:philippe.ruelle@uclouvain.be}{\normalsize\texttt{philippe.ruelle@uclouvain.be}}
\end{center}
\end{minipage}}
}
\date{}

\begin{document}
\maketitle

\begin{abstract}
We revisit the calculation of height correlations in the two-dimensional Abelian sandpile model by taking advantage of a technique developed recently by Kenyon and Wilson. The formalism requires to equip the usual graph Laplacian, ubiquitous in the context of cycle-rooted spanning forests, with a complex connection. In the case at hand, the connection is constant and localized along a semi-infinite defect line (zipper). In the appropriate limit of a trivial connection, it allows one to count spanning forests whose components contain prescribed sites, which are of direct relevance for height correlations in the sandpile model. Using this technique, we first rederive known 1- and 2-site lattice correlators on the plane and upper half-plane, more efficiently than what has been done so far. We also compute explicitly the (new) next-to-leading order in the distances ($r^{-4}$ for 1-site on the upper half-plane, $r^{-6}$ for 2-site on the plane). We extend these results by computing new correlators involving one arbitrary height and a few heights 1 on the plane and upper half-plane, for the open and closed boundary conditions. We examine our lattice results from the conformal point of view, and confirm the full consistency with the specific features currently conjectured to be present in the associated logarithmic conformal field theory.

\end{abstract}

%%%%%%%%%%%%%%%%%%%%%%%%%%%%%%%%%%%%%%%%%%%%%%%%%%%%%%%%%%%%%%%%%%%%%%%%%%%%%%%%%%%%%%%%%%%%%%%%%%%%%%%%

\section{Introduction}

Much attention is being directed toward logarithmic conformal field theories (LCFTs), either for the specific mathematical challenges they raise, or as descriptions of statistical models. As for any two-dimensional conformal theory, the underlying symmetry is the Virasoro algebra. However, the distinctive feature of LCFTs is that the symmetry is realized through reducible indecomposable representations, which are notoriously hard to classify. In more physical terms, the Hamiltonian operator is not diagonalizable, with the consequence that the correlation functions of certain observables are no longer purely algebraic, but involve in addition products of logarithms. 

Steady progress has been made since the early days of LCFTs \cite{RS92,Gur93}. Mathematical understanding has gradually improved, and issues like the structure of indecomposable representations, fusion, vertex operators and modular transformations have, to some extent, been clarified. These advances have been paralleled by similar progress in statistical models realizing LCFTs in their scaling limit. Most of these models (among which the so-called minimal logarithmic models \cite{PRZ06}, which so far form the largest supply of such models) possess nonlocal degrees of freedom, believed to be the hallmark of LCFTs. These concrete models are of crucial importance, as they allow one to probe and to envision the general features of LCFTs. For further details and references, we refer the reader to the special issue published recently \cite{GRR13}, devoted to recent developments of LCFTs; in particular two complementary review articles, one on logarithmic conformal field theory itself \cite{CR13}, the other on the way it can be seen and understood on the lattice \cite{GJRSV13}.

As impressive as these results may be, the situation is admittedly much less satisfactory than for the rational, nonlogarithmic conformal theories. In particular, the efforts have been mainly concentrated, with a few exceptions, on chiral aspects of LCFTs. Indeed, most logarithmic statistical models lend themselves well to the analysis of their chiral features, but give little or no access to their nonchiral characteristics: bulk correlation functions are usually very difficult to compute, structural elements of their representations are very hard to obtain and (nonchiral) fusion therefore cannot be assessed. In addition, even if the seeds of the continuum conformal representations are believed to be present in finite systems, their spectrum are often understood in terms of auxiliary lattice algebras, which tend to have complicated representation theory in periodic geometries. It is therefore technically hard to infer from finite-size models what type of full conformal representations emerge in the scaling limit.

Unlike their rational cousins, fully fledged, nonchiral LCFTs are much more complicated than their chiral restrictions (themselves technically more complex than their rational analogues), and cannot be simply reduced to their two chiral halves. The triplet model at $c=-2$ is among the few---perhaps even the only one---local, nonchiral logarithmic field theories that is fully understood \cite{GK99,GR06}. It is also known as the free symplectic fermion model, since it has a Lagrangian realization in terms of free fermions. In addition, it possesses an extended symmetry, in the form of a W-algebra, a nonlinear extension of the Virasoro algebra. Although this specific LCFT is well under control, as a local field theory, it is not known yet for sure what lattice model it can describe in the scaling limit. 

Three well-studied, inequivalent lattice models have central charge $c=-2$ and would be natural candidates: the dense polymer model \cite{PR07}, the dimer model---at least for some of its features \cite{MRR15,MRR16,PV17}---and the Abelian sandpile model \cite{JPR06}. All three models have features that are very close to the symplectic fermion theory. However, for the sandpile model, it has been established that the underlying LCFT is not the symplectic fermion model, as will be recalled in Section 
\ref{sec7}. The question remains open for the other two models. 

It is in this context that the present paper draws its motivations. It has two main goals. The first one is to carry out explicit calculations of correlations in the lattice sandpile model, without referring to LCFTs. A number of computations have already been done using standard graph-theoretical methods, see for instance \cite{Pri94,JPR06,PGPR08,PGPR10}; however, in view of their technical complexity and clumsiness, these methods have somehow delivered all what they could. Recent ideas put forward by Kenyon and Wilson \cite{Ken11,KW15} to compute grove probabilities have much improved the situation, since they dramatically reduce the complexity of calculations\footnote{It is worth mentioning that the technique was not specifically designed for the sandpile model, and can be applied in many situations. It led in particular to impressive results on passage probabilities for the loop-erased random walk \cite{KW15}.}. These techniques are directly relevant to the evaluation of sandpile correlations and therefore allow one to obtain new explicit results, which can in turn be related to LCFT predictions.

This leads to our second main motivation. Unlike most of the other lattice models believed to be described by an LCFT, the sandpile model is particularly amenable to the exact and explicit computation of correlation functions. This implies the possibility to probe specific aspects of the underlying LCFT, which would otherwise be much harder to obtain from a transfer-matrix analysis. It should be all the more instructive that the underlying LCFT is a new one, albeit expected to share certain gross features with the triplet model. In particular, it is known that most indecomposable representations come with specific values of certain indecomposability parameters, which univoquely fix their equivalence class. These parameters affect the fusion rules and the operator product expansions, and therefore play an important role; they would however be hard to determine without the knowledge of correlation functions.  

The paper is organized as follows. Section~\ref{sec2} starts with a brief reminder of the Abelian sandpile model in terms of its natural, local variables, namely the height variables. We also recall how height configurations are in bijection with the set of spanning trees, and how multisite height probabilities can be traded for specific conditions on trees. We take this opportunity to generalize the known result to the case where several heights are strictly larger than the minimal value (1 in our conventions). The question is nontrivial since the most naive guess does not work, as has been already noticed before; we illustrate how the problem can be solved in the case of two-point joint probabilities.

In Section~\ref{sec3}, we review the main ideas of \cite{KW15} on the use of a line bundle Laplacian to compute the probabilities of so-called cycle-rooted groves. We recall the principal result, the grove  theorem, and illustrate how it can be used in concrete examples. Section~\ref{sec4} describes the way this theorem can be applied to the calculation of sandpile height correlations. Relying on this technique, we rederive one-point height probabilities $\P_a$ for $a=1,2,3,4$. Although the results are well known, the new method reduces the calculations to just a few lines\footnote{David Wilson has carried out similar calculations and has computed several multipoint height probabilities when the insertion points are neighboring sites (private communication). To our knowledge, these have remained largely unpublished; a typical result has however been reported in \cite{Rue13}.}, in stark contrast with earlier routes. In addition, it really forms the core of our subsequent calculations, which can all be viewed as one-point probabilities on modified lattices.

Sections \ref{sec5} and \ref{sec6} contain our new results on height correlations on the lattice. In Section~\ref{sec5}, we compute three-point probabilities $\P_{a,1,1}(\vec r_{12},\vec r_{13})$ on the full plane and the corresponding correlations $\sigma_{a,1,1}(\vec r_{12},\vec r_{13})$, for $a=1,2,3,4$. The latter require extending the calculation of two-point probabilities $\P_{a,1}(\vec r)$ to order $r^{-6}$, namely one order beyond what is presently known (the term in $r^{-4}$ is sufficient to compute the scaling limit of the 2-corrrelator $\sigma_{a,1}(\vec r)$, but due to subtractions, the next order $r^{-6}$ is needed to compute the scaling limit of the 3-correlator). Surprisingly, we find that the three-point correlators $\sigma_{a,1,1}(\vec r_{12},\vec r_{13})$ for $a=2,3,4$ are not logarithmic, unlike the two-point functions.

Section~\ref{sec6} concerns the computation of lattice correlations on the upper half-plane (UHP). The new correlations $\sigma_{a,1}$ we obtain are those of a height $a\ge 1$ in the bulk of the upper half-plane and a height 1, which is either also in the bulk or on the boundary of the UHP. They are all found to be logarithmic. The calculations are carried out for two boundary conditions, open and closed, and require extending the one-point bulk probabilities in the UHP to order $r^{-4}$. For these probabilities, we also consider a new form of UHP, the boundary of which is diagonal ($x=y$), and verify that, at dominant order ($r^{-2}$), the results are identical. 

The new lattice results obtained in Sections \ref{sec5} and \ref{sec6} are discussed in Section~\ref{sec7} in the framework of logarithmic conformal field theory. We recall the conjectural nature of the four height fields and of the conformal module they belong to, and discuss the peculiar way in which the lattice correlations are to be understood as conformal correlators. We then proceed to compute the conformal correlators and compare them with the corresponding lattice results, at dominant order. In all cases, we find a complete agreement.

Finally, two appendices collect technical material related to lattice Green functions and their derivatives with respect to the parameter defining the line bundle Laplacian, and to symmetries encountered in the calculation of certain fractions of spanning trees.

%%%%%%%%%%%%%%%%%%%%%%%%%%%%%%%%%%%%%%%%%%%%%%%%%%%%%%%%%%%%%%%%%%%%%%%%%%%%%%%%%%%%%%%%%%%%%%%%%%%%%%%%

\section{Review of the sandpile model}
\label{sec2}

We briefly recall the definition of the Abelian sandpile model introduced by Bak, Tang and Wiesenfeld \cite{BTW87}. We consider a two-dimensional unoriented grid $\G$ with a discrete random variable $h_i\in\mathbb{N}^*$ associated with each site $i\in\G$. A \emph{sandpile configuration} is a  set $\mathcal{C}=\left\{h_i\right\}_{i\in\G}$, where the \emph{height} $h_i \ge 1$ counts the number of grains of sand at site $i$.

The definition of the model on $\G$ requires to extend it to $\G^*=\G\cup\{s\}$ with an additional site $s$ called the \emph{sink} (or \emph{root}), and edges linking a nonempty subset $\mathcal{D}\subset\G$ to $s$ (multiple edges between a site and the sink are allowed, whereas all edges are simple in $\G$). Sites of $\mathcal{D}$ are called dissipative or \emph{open}, while those of $\G \backslash {\mathcal D}$ are \emph{closed}. We denote by $z_i^{}$ (resp. $z_i^*$) the degree of $i$ in $\G$ (resp. $\G^*$), so that $z_i^*=z_i^{}$ if $i$ is closed and $z_i^* > z_i^{}$ if $i$ is open. We shall use the notation $\{i,j\}$ (resp. $(i,j)$) for the undirected (resp. directed) edge from $i$ to $j$. We also define the symmetric \emph{toppling matrix} $\Delta$ on $\G$ as follows:
\begin{equation}
\Delta_{i,j}=\begin{cases}
z_i^*&\quad\textrm{if $i=j$,}\\
-1&\quad\textrm{if there exists an edge $\{i,j\}$ between $i$ and $j$,}\\
0&\quad\textrm{otherwise.}
\end{cases}
\end{equation}
Hence, $\Delta$ corresponds to the discrete Laplacian on $\G^*$ with Dirichlet boundary conditions at $s$. A configuration $\mathcal{C}=\{h_i\}$ is called \emph{stable} if $1\le h_i\le z_i^*=\Delta_{i,i}$ for each $i\in\G$.

The discrete, stochastic dynamics of the sandpile model is defined as follows. Let $\mathcal{C}_t$ be a stable configuration at time $t$. The configuration $\mathcal{C}_{t+1}$ is obtained in two steps:
\begin{enumerate}
\item [$(i)$] \textit{Seeding:} A grain of sand is dropped on a random site $i$ of $\G$, producing new height values, $h_j^{\text{new}}=h_j^{\text{old}}+\delta_{j,i}$. If $h_i^{\text{new}}\le\Delta_{i,i}$, the new configuration is stable, and defines $\mathcal{C}_{t+1}$.
\item [$(ii)$] \textit{Relaxation:} If $h_i^{\text{new}}>\Delta_{i,i}$, then the whole system is updated according to $h_j^{\text{new}}\rightarrow h_j^{\text{new}}-\Delta_{i,j}$ for each $j\in\G$, as well as $s \to s + z_i^* - z_i$. In other words, site $i$ \emph{topples}: it loses $z_i^*$ grains of sand and gives one grain to each of its neighbors on the graph $\G^*$. If one of the updated heights $h_j$ on $\G$ in turn exceeds $\Delta_{j,j}$, the toppling process is repeated until all sites are stable, thus defining $\mathcal{C}_{t+1}$ (see Fig.~\ref{topplings}). We note that each time a dissipative site is toppled, a certain number of grains leave $\G$ and are transferred to the sink.
\end{enumerate}

Since the sink $s$ is never toppled, its height is unboundedly increasing over time. This property implies that the dynamics is well defined, in the sense that the relaxation of an unstable configuration on $\G$ terminates after a finite number of topplings \cite{Dha90}. The Abelian property stems from the fact that the stable configuration obtained after all unstable sites have toppled does not depend on the order in which the topplings are carried out.

\begin{figure}[t]
\centering
\begin{tikzpicture}[scale=0.6]
\tikzstyle arrowstyle=[scale=1.3]
\tikzstyle directend=[postaction={decorate,decoration={markings,mark=at position 1 with {\arrow[arrowstyle]{stealth}}}}]

\begin{scope}[xshift=-0.5cm,font=\large]
\draw[dotted] (-0.75,-0.75) grid (3.75,3.75);
\draw (0,0) node {$2$};
\draw (1,0) node {$4$};
\draw (2,0) node {$3$};
\draw (3,0) node {$3$};
\draw (0,1) node {$3$};
\draw (1,1) node {$4$};
\draw (2,1) node {$1$};
\draw (3,1) node {$2$};
\draw (0,2) node {$1$};
\draw (1,2) node[red] {$\mathbf{4}$};
\draw (2,2) node {$2$};
\draw (3,2) node {$1$};
\draw (0,3) node {$4$};
\draw (1,3) node {$3$};
\draw (2,3) node {$2$};
\draw (3,3) node {$3$};
\draw (1.4,2.2) node[red] {\small $+1$};
\draw[thick] (-0.75,-0.75)--(3.75,-0.75)--(3.75,3.75)--(-0.75,3.75)--(-0.75,-0.75);
\draw (-1,1.5) node {$s$};
\draw[directend,thick] (4.25,1.5)--(5,1.5);
\end{scope}

\begin{scope}[xshift=6cm,font=\large]
\draw[dotted] (-0.75,-0.75) grid (3.75,3.75);
\draw (0,0) node {$2$};
\draw (1,0) node {$4$};
\draw (2,0) node {$3$};
\draw (3,0) node {$3$};
\draw (0,1) node {$3$};
\draw (1,1) node[blue] {$\mathbf{5}$};
\draw (2,1) node {$1$};
\draw (3,1) node {$2$};
\draw (0,2) node[blue] {$\mathbf{2}$};
\draw (1,2) node[red] {$\mathbf{1}$};
\draw (2,2) node[blue] {$\mathbf{3}$};
\draw (3,2) node {$1$};
\draw (0,3) node {$4$};
\draw (1,3) node[blue] {$\mathbf{4}$};
\draw (2,3) node {$2$};
\draw (3,3) node {$3$};
\draw[directend,blue] (0.7,2)--(0.3,2);
\draw[directend,blue] (1,2.3)--(1,2.7);
\draw[directend,blue] (1.3,2)--(1.7,2);
\draw[directend,blue] (1,1.7)--(1,1.3);
\draw[directend,thick] (4.25,1.5)--(5,1.5);
\end{scope}

\begin{scope}[xshift=12.5cm,font=\large]
\draw[dotted] (-0.75,-0.75) grid (3.75,3.75);
\draw (0,0) node {$2$};
\draw (1,0) node[blue] {$\mathbf{5}$};
\draw (2,0) node {$3$};
\draw (3,0) node {$3$};
\draw (0,1) node[blue] {$\mathbf{4}$};
\draw (1,1) node[red] {$\mathbf{1}$};
\draw (2,1) node[blue] {$\mathbf{2}$};
\draw (3,1) node {$2$};
\draw (0,2) node {$2$};
\draw (1,2) node[blue] {$\mathbf{2}$};
\draw (2,2) node {$3$};
\draw (3,2) node {$1$};
\draw (0,3) node {$4$};
\draw (1,3) node {$4$};
\draw (2,3) node {$2$};
\draw (3,3) node {$3$};
\draw[directend,blue] (0.7,1)--(0.3,1);
\draw[directend,blue] (1,1.3)--(1,1.7);
\draw[directend,blue] (1.3,1)--(1.7,1);
\draw[directend,blue] (1,0.7)--(1,0.3);
\draw[directend,thick] (4.25,1.5)--(5,1.5);
\end{scope}

\begin{scope}[xshift=19cm,font=\large]
\draw[dotted] (-0.75,-0.75) grid (3.75,3.75);
\draw (0,0) node[blue] {$\mathbf{3}$};
\draw (1,0) node[red] {$\mathbf{1}$};
\draw (2,0) node[blue] {$\mathbf{4}$};
\draw (3,0) node {$3$};
\draw (0,1) node {$4$};
\draw (1,1) node[blue] {$\mathbf{2}$};
\draw (2,1) node {$2$};
\draw (3,1) node {$2$};
\draw (0,2) node {$2$};
\draw (1,2) node {$2$};
\draw (2,2) node {$3$};
\draw (3,2) node {$1$};
\draw (0,3) node {$4$};
\draw (1,3) node {$4$};
\draw (2,3) node {$2$};
\draw (3,3) node {$3$};
\draw[directend,blue] (0.7,0)--(0.3,0);
\draw[directend,blue] (1,0.3)--(1,0.7);
\draw[directend,blue] (1.3,0)--(1.7,0);
\draw[directend,blue] (1,-0.3)--(1,-0.7);
\end{scope}

\end{tikzpicture}
\caption{Relaxation of a configuration on a $4\times 4$ square grid with dissipative sites on its (geometrical) boundary after a grain is dropped on a site $i$ with height $h_i=4$. Its toppling creates another height $h_j=5$, which in turn topples. On the last step, the grain of sand ``falling off the grid'' arrives into the sink $s$, marked as the contour line surrounding the grid. The dotted lines indicate the edges in $\G^*$.}
\label{topplings}
\end{figure}

The dynamics described above defines a discrete Markov chain on a finite state space, namely the space of stable configurations. It can be shown \cite{Dha90} that it has a unique invariant measure $\P^*_{\G}$, which is moreover uniform on the subset $\mathcal{R}$ of recurrent configurations:
\begin{equation}
\P^*_{\G}(\mathcal{C})=\begin{cases}
\frac{1}{|\mathcal{R}|}&\quad\textrm{if $\mathcal{C}\in\mathcal{R}$,}\\
0&\quad\textrm{if $\mathcal{C}\notin\mathcal{R}$.}
\end{cases}
\end{equation}
Hence, recurrent configurations are the only ones to keep reoccurring in the repeated image of the dynamics; they all do so infinitely often, with equal frequency. In the continuum limit, the measures $\P^*_{\G}$ are expected to converge to the field-theoretical measure $\P$ of a conformal field theory of central charge $c=-2$ \cite{MD92}, which is believed to be logarithmic, see the recent review \cite{Rue13}.

In practice, recurrent versus nonrecurrent (transient) configurations can be characterized in terms of certain subconfigurations. For example, the subconfiguration consisting in two neighboring 1s cannot be part of a recurrent configuration, since it cannot be produced by topplings (because a height 5 that topples to become a 1 gives one grain of sand to each of its neighbors, making their height at least equal to 2). Similarly, the block 121 is also forbidden because its production by topplings means that it contained a block 11 before. More generally, a \emph{forbidden subconfiguration} (FSC) $F$ is such that its heights satisfy
\be
h_i \le z_i^{(F)} = \#\{\text{neighbors of $i$ in $F$}\}, \quad \text{for all $i \in F$}.
\ee
Then a stable configuration is recurrent if and only if it contains no FSCs \cite{Dha90,MD92}. The condition for a configuration to be recurrent is nonlocal, because one has to scan the whole grid to make sure that no such subconfigurations appear. For example, the stable configuration $h_i=z_i$ for each $i\in\G$ is not recurrent and contains $F=\G$ as the smallest FSC.

The recurrence criterion in terms of FSCs shows that the configuration obtained by increasing the height  of any site in a recurrent configuration (without making it unstable) is recurrent. Decreasing the heights in a recurrent configuration may however result in nonrecurrent configurations. 

%%%%%%%%%%%%%%%%%%%%%%%%%%%%%%%%%%%%%%%%%%%%%%%%%%%%%%%%%%%%%%%%%%%%%%%%%%%%%%%%%%%%%%%%%%%%%%%%%%%%%%%%

\subsection{The burning algorithm}
\label{burning}

Height variables are the most natural degrees of freedom in the Abelian sandpile model, and evolve according to local toppling rules. However, as observed above, the characterization of recurrent configurations in terms of FSCs is highly nonlocal.  

\begin{figure}[t]
\centering
\begin{tikzpicture}[scale=0.7,very thick,blue]
\draw[dotted,semithick] (0,0) grid (8,5);
\foreach \y in {0,1,...,5}{\foreach \x in {0,1,...,8}{\filldraw (\x,\y) circle (0.075cm);};}
\draw (0,0)--(2,0)--(2,1)--(5,1)--(5,0);
\draw (4,1)--(4,0)--(3,0);
\draw (1,0)--(1,3)--(2,3);
\draw (0,1)--(0,2)--(3,2)--(3,4)--(0,4)--(0,3);
\draw (2,4)--(2,5);
\draw (0,5)--(1,5);
\draw (2,5)--(4,5);
\draw (3,3)--(6,3)--(6,2);
\draw (7,2)--(7,4)--(4,4);
\draw (6,4)--(6,5)--(5,5);
\draw (7,3)--(8,3)--(8,5)--(7,5);
\draw (4,3)--(4,2);
\draw (5,3)--(5,2);
\draw (1,5)--( 1,5.7);
\draw (8,2)--(8.7,2);
\draw (3,-0.7)--(3,0);
\draw (8,3)--(8,0)--(6,0)--(6,1)--(7,1);
\foreach \y in {0,1,...,5} {\draw[dotted,semithick] (-0.7,\y)--(0,\y);}
\foreach \y in {0,1,...,5} {\draw[dotted,semithick] (8,\y)--(8.7,\y);}
\foreach \x in {0,1,...,8} {\draw[dotted,semithick] (\x,-0.7)--(\x,0);}
\foreach \x in {0,1,...,8} {\draw[dotted,semithick] (\x,5)--(\x,5.7);}
\draw[thick,black] (-0.7,-0.7)--(8.7,-0.7)--(8.7,5.7)--(-0.7,5.7)--(-0.7,-0.7);
\draw (-1,3) node[black] {$s$};
\end{tikzpicture}
\caption{Spanning tree on a rectangular $9\times 6$ grid, extended by a unique sink $s$.}
\label{USTex}
\end{figure}
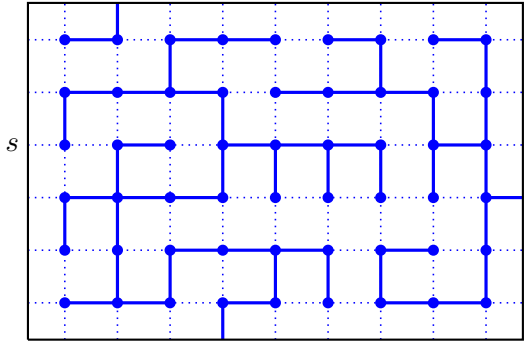

The \emph{burning algorithm} \cite{MD92} gives a bijection between recurrent sandpile configurations on $\G$ and \emph{spanning trees} on $\G^*$, which turn out to be more convenient for concrete calculations. A spanning tree on a connected graph is a connected subgraph containing all the vertices such that there are no loops (see Fig.~\ref{USTex}). This property implies that there is a unique path between any two vertices of the graph, this path possibly passing through $s$. We shall view the sink $s$ as the root of the tree, since the burning algorithm will produce trees as growing from $s$. We call the resulting subgraph a \emph{rooted spanning tree}.

%\begin{figure}[t]
%\centering
%\begin{tikzpicture}[scale=0.7,very thick]
%\draw[dotted,thin] (0,0) grid (8,5);
%\foreach \y in {0,1,...,5}{\foreach \x in {0,1,...,8}{\filldraw (\x,\y) circle (0.075cm);};}
%\draw (0,0)--(2,0)--(2,1)--(5,1)--(5,0);
%\draw (4,1)--(4,0)--(3,0);
%\draw (1,0)--(1,3)--(2,3);
%\draw (0,1)--(0,2)--(3,2)--(3,4)--(0,4)--(0,3);
%\draw (2,4)--(2,5);
%\draw (0,5)--(1,5);
%\draw (2,5)--(4,5);
%\draw (3,3)--(6,3)--(6,2);
%\draw (7,2)--(7,4)--(4,4);
%\draw (6,4)--(6,5)--(5,5);
%\draw (7,3)--(8,3)--(8,5)--(7,5);
%\draw (4,3)--(4,2);
%\draw (5,3)--(5,2);
%\draw (1,5)--( 1,5.7);
%\draw (8,2)--(8.7,2);
%\draw (3,-0.7)--(3,0);
%\draw (8,3)--(8,0)--(6,0)--(6,1)--(7,1);
%\foreach \y in {0,1,...,5} {\draw[dotted,thin] (-0.7,\y)--(0,\y);}
%\foreach \y in {0,1,...,5} {\draw[dotted,thin] (8,\y)--(8.7,\y);}
%\foreach \x in {0,1,...,8} {\draw[dotted,thin] (\x,-0.7)--(\x,0);}
%\foreach \x in {0,1,...,8} {\draw[dotted,thin] (\x,5)--(\x,5.7);}
%\draw[thick] (-0.7,-0.7)--(8.7,-0.7)--(8.7,5.7)--(-0.7,5.7)--(-0.7,-0.7);
%\draw (-1,3) node {$s$};
%\end{tikzpicture}
%\caption{Spanning tree on a rectangular $9\times 6$ grid, extended by a unique sink $s$.}
%\label{USTex}
%\end{figure}

Given a stable configuration on $\G$, the burning algorithm describes the propagation of a fire front throughout the grid $\G^*$, starting from the root; when the algorithm stops, the various fire lines form a rooted spanning tree on $\G^*$ if the configuration we started from is recurrent. The fire propagation depends on the height values of the  configuration one considers, and so does the resulting spanning tree. It is defined as follows. 

At any time\footnote{This time variable is used to describe the propagation of the fire front, and should not be confused with the time used in the previous subsection to describe the dynamics of the model.} $t$, the sites of $\G^*$ belong to one of the following three disjoint sets, ${\mathcal U}_t$, ${\mathcal B}_t$ and ${\mathcal E}_t$ (for more clarity, we omit the explicit dependence of these sets on the configuration one is looking at). ${\mathcal U}_t$ contains the sites that have not burnt yet at time $t$, ${\mathcal B}_t$ contains those that are burning at time $t$ and the rest goes in ${\mathcal E}_t$, which contains the extinct sites. At the initial time $t=0$, only the sink/root is burning while all other sites are unburnt, so ${\mathcal U}_0 = \G$, ${\mathcal B}_0 = \{s\}$ and ${\mathcal E}_0 = \varnothing$. From then on, the three sets evolve by the following rules.

The sites in ${\mathcal B}_t$ keep burning for just one time unit, so that ${\mathcal E}_{t+1} = {\mathcal E}_t \cup {\mathcal B}_t$. The unburnt sites in ${\mathcal U}_t$ whose height is strictly larger than their number of neighbors in ${\mathcal U}_t$ move to ${\mathcal B}_{t+1}$, whereas the others form ${\mathcal U}_{t+1}$:
\be
{\mathcal B}_{t+1} = \{i \in {\mathcal U}_t \;:\; h_i > z_i^{({\mathcal U}_t)} \}, \qquad 
{\mathcal U}_{t+1} = {\mathcal U}_{t} \,\bs\, {\mathcal B}_{t+1}.
\ee
Therefore, the size of the set ${\mathcal U}_t$ decreases in time, while that of ${\mathcal E}_t$ increases. The algorithm stops when the three sets no longer change, that is, at the first time $\tau$ for which ${\mathcal B}_\tau = \varnothing$ (and ${\mathcal E}_{\tau+1} = {\mathcal E}_{\tau}$, ${\mathcal U}_{\tau} = {\mathcal U}_{\tau-1}$). From the definitions, we see that the set ${\mathcal U}_\tau$, if nonempty, is an FSC. A configuration is therefore recurrent if and only if the algorithm stops after all sites have burnt, namely ${\mathcal E}_\tau = \G^*$ or ${\mathcal U}_\tau = \varnothing$.

Let us observe that for all times $0 < t < \tau$, every site of ${\mathcal B}_t$ has at least one nearest neighbor that was in ${\mathcal B}_{t-1}$, since otherwise that site would have been burning at an earlier time. For instance, at time $t=1$, the sites of ${\mathcal B}_1$ must be dissipative sites, i.e. neighbors of the sink. In this way, we can imagine the fire as propagating along edges connecting nearest neighbors; these edges eventually form the spanning tree. 

The spanning tree starts from the root, and connects it to the dissipative sites in ${\mathcal B}_1$. Edges are subsequently added as the algorithm is running. If a site $j$ in ${\mathcal B}_t$ has only one nearest neighbor $i$ in ${\mathcal B}_{t-1}$, we say that $j$ catches fire from $i$, and we add the edge connecting $i$ to $j$ to the tree under construction. More generally, if $j$ has $k \ge 1$ nearest neighbors in ${\mathcal B}_{t-1}$, the height at $j$ satisfies 
\be
z_j^{(\mathcal{U}_{t-1})} + 1 \le h_j \le z_j^{(\mathcal{U}_{t-2})} = z_j^{(\mathcal{U}_{t-1})} + k,
\ee 
where the lower bound is because $j$ is in ${\mathcal B}_{t}$ and the upper bound is because $j$ was not in ${\mathcal B}_{t-1}$. Therefore, the height $h_j$ can possibly take $k$ different values. If the actual value is $h_j = z_j^{(\mathcal{U}_{t-1})} + m$, we use the additional prescription that $j$ catches fire from its $m$th neighbor in ${\mathcal B}_{t-1}$, once these neighbors are ordered clockwise starting from the northern one. We also add the corresponding edge to the tree.

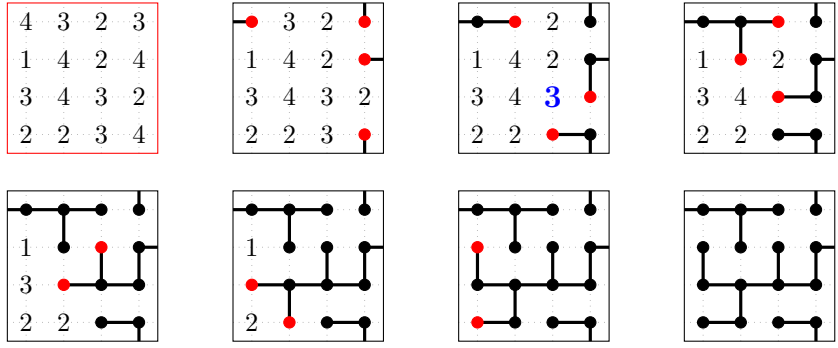
\begin{figure}[t]
\centering
\begin{tikzpicture}[scale=0.5]

\begin{scope}[xshift=0cm,font=\large]
\draw[dotted,help lines] (-0.5,-0.5) grid (3.5,3.5);
\draw[red] (-0.5,-0.5) rectangle (3.5,3.5);
\draw (0,0) node {\small $2$};
\draw (1,0) node {\small $2$};
\draw (2,0) node {\small $3$};
\draw (3,0) node {\small $4$};
\draw (0,1) node {\small $3$};
\draw (1,1) node {\small $4$};
\draw (2,1) node {\small $3$};
\draw (3,1) node {\small $2$};
\draw (0,2) node {\small $1$};
\draw (1,2) node {\small $4$};
\draw (2,2) node {\small $2$};
\draw (3,2) node {\small $4$};
\draw (0,3) node {\small $4$};
\draw (1,3) node {\small $3$};
\draw (2,3) node {\small $2$};
\draw (3,3) node {\small $3$};
\end{scope}

\begin{scope}[xshift=6cm,font=\large]
\draw[dotted,help lines] (-0.5,-0.5) grid (3.5,3.5);
\draw (-0.5,-0.5) rectangle (3.5,3.5);
\draw (0,0) node {\small $2$};
\draw (1,0) node {\small $2$};
\draw (2,0) node {\small $3$};
\draw (0,1) node {\small $3$};
\draw (1,1) node {\small $4$};
\draw (2,1) node {\small $3$};
\draw (3,1) node {\small $2$};
\draw (0,2) node {\small $1$};
\draw (1,2) node {\small $4$};
\draw (2,2) node {\small $2$};
\draw (1,3) node {\small $3$};
\draw (2,3) node {\small $2$};
\draw[very thick] (3,-0.5)--(3,0);
\draw[very thick] (3.5,2)--(3,2);
\draw[very thick] (-0.5,3)--(0,3);
\draw[very thick] (3,3.5)--(3,3);
\draw[fill=red,red] (3,0) circle (0.15cm);
\draw[fill=red,red] (3,2) circle (0.15cm);
\draw[fill=red,red] (0,3) circle (0.15cm);
\draw[fill=red,red] (3,3) circle (0.15cm);
\end{scope}

\begin{scope}[xshift=12cm,font=\large]
\draw[dotted,help lines] (-0.5,-0.5) grid (3.5,3.5);
\draw (-0.5,-0.5) rectangle (3.5,3.5);
\draw (0,0) node {\small $2$};
\draw (1,0) node {\small $2$};
\draw (0,1) node {\small $3$};
\draw (1,1) node {\small $4$};
\draw[thick,blue] (2,1) node {$\mathbf{3}$};
\draw (0,2) node {\small $1$};
\draw (1,2) node {\small $4$};
\draw (2,2) node {\small $2$};
\draw (2,3) node {\small $2$};
\filldraw (3,0) circle (0.15cm);
\filldraw (3,2) circle (0.15cm);
\filldraw (0,3) circle (0.15cm);
\filldraw (3,3) circle (0.15cm);
\draw[very thick] (3,-0.5)--(3,0);
\draw[very thick] (3.5,2)--(3,2);
\draw[very thick] (-0.5,3)--(0,3);
\draw[very thick] (3,3.5)--(3,3);
\draw[very thick] (0,3)--(1,3);
\draw[very thick] (3,2)--(3,1);
\draw[very thick] (3,0)--(2,0);
\draw[fill=red,red] (2,0) circle (0.15cm);
\draw[fill=red,red] (3,1) circle (0.15cm);
\draw[fill=red,red] (1,3) circle (0.15cm);
\end{scope}

\begin{scope}[xshift=18cm,font=\large]
\draw[dotted,help lines] (-0.5,-0.5) grid (3.5,3.5);
\draw (-0.5,-0.5) rectangle (3.5,3.5);
\draw (0,0) node {\small $2$};
\draw (1,0) node {\small $2$};
\draw (0,1) node {\small $3$};
\draw (1,1) node {\small $4$};
\draw (0,2) node {\small $1$};
\draw (2,2) node {\small $2$};
\filldraw (3,0) circle (0.15cm);
\filldraw (3,2) circle (0.15cm);
\filldraw (0,3) circle (0.15cm);
\filldraw (3,3) circle (0.15cm);
\filldraw (2,0) circle (0.15cm);
\filldraw (3,1) circle (0.15cm);
\filldraw (1,3) circle (0.15cm);
\draw[very thick] (3,-0.5)--(3,0);
\draw[very thick] (3.5,2)--(3,2);
\draw[very thick] (-0.5,3)--(0,3);
\draw[very thick] (3,3.5)--(3,3);
\draw[very thick] (0,3)--(1,3);
\draw[very thick] (3,2)--(3,1);
\draw[very thick] (3,0)--(2,0);
\draw[very thick] (3,1)--(2,1);
\draw[very thick] (1,3)--(1,2);
\draw[very thick] (1,3)--(2,3);
\draw[fill=red,red] (2,1) circle (0.15cm);
\draw[fill=red,red] (1,2) circle (0.15cm);
\draw[fill=red,red] (2,3) circle (0.15cm);
\end{scope}

\begin{scope}[xshift=0cm,yshift=-5cm,font=\large]
\draw[dotted,help lines] (-0.5,-0.5) grid (3.5,3.5);
\draw (-0.5,-0.5) rectangle (3.5,3.5);
\draw (0,0) node {\small $2$};
\draw (1,0) node {\small $2$};
\draw (0,1) node {\small $3$};
\draw (0,2) node {\small $1$};
\filldraw (3,0) circle (0.15cm);
\filldraw (3,2) circle (0.15cm);
\filldraw (0,3) circle (0.15cm);
\filldraw (3,3) circle (0.15cm);
\filldraw (2,0) circle (0.15cm);
\filldraw (3,1) circle (0.15cm);
\filldraw (1,3) circle (0.15cm);
\filldraw (2,1) circle (0.15cm);
\filldraw (1,2) circle (0.15cm);
\filldraw (2,3) circle (0.15cm);
\draw[very thick] (3,-0.5)--(3,0);
\draw[very thick] (3.5,2)--(3,2);
\draw[very thick] (-0.5,3)--(0,3);
\draw[very thick] (3,3.5)--(3,3);
\draw[very thick] (0,3)--(1,3);
\draw[very thick] (3,2)--(3,1);
\draw[very thick] (3,0)--(2,0);
\draw[very thick] (3,1)--(2,1);
\draw[very thick] (1,3)--(1,2);
\draw[very thick] (1,3)--(2,3);
\draw[very thick] (2,1)--(1,1);
\draw[very thick] (2,1)--(2,2);
\draw[fill=red,red] (1,1) circle (0.15cm);
\draw[fill=red,red] (2,2) circle (0.15cm);
\end{scope}

\begin{scope}[xshift=6cm,yshift=-5cm,font=\large]
\draw[dotted,help lines] (-0.5,-0.5) grid (3.5,3.5);
\draw (-0.5,-0.5) rectangle (3.5,3.5);
\draw (0,0) node {\small $2$};
\draw (0,2) node {\small $1$};
\filldraw (3,0) circle (0.15cm);
\filldraw (3,2) circle (0.15cm);
\filldraw (0,3) circle (0.15cm);
\filldraw (3,3) circle (0.15cm);
\filldraw (2,0) circle (0.15cm);
\filldraw (3,1) circle (0.15cm);
\filldraw (1,3) circle (0.15cm);
\filldraw (2,1) circle (0.15cm);
\filldraw (1,2) circle (0.15cm);
\filldraw (2,3) circle (0.15cm);
\filldraw (1,1) circle (0.15cm);
\filldraw (2,2) circle (0.15cm);
\draw[very thick] (3,-0.5)--(3,0);
\draw[very thick] (3.5,2)--(3,2);
\draw[very thick] (-0.5,3)--(0,3);
\draw[very thick] (3,3.5)--(3,3);
\draw[very thick] (0,3)--(1,3);
\draw[very thick] (3,2)--(3,1);
\draw[very thick] (3,0)--(2,0);
\draw[very thick] (3,1)--(2,1);
\draw[very thick] (1,3)--(1,2);
\draw[very thick] (1,3)--(2,3);
\draw[very thick] (2,1)--(1,1);
\draw[very thick] (2,1)--(2,2);
\draw[very thick] (1,1)--(1,0);
\draw[very thick] (1,1)--(0,1);
\draw[fill=red,red] (1,0) circle (0.15cm);
\draw[fill=red,red] (0,1) circle (0.15cm);
\end{scope}

\begin{scope}[xshift=12cm,yshift=-5cm,font=\large]
\draw[dotted,help lines] (-0.5,-0.5) grid (3.5,3.5);
\draw (-0.5,-0.5) rectangle (3.5,3.5);
\filldraw (3,0) circle (0.15cm);
\filldraw (3,2) circle (0.15cm);
\filldraw (0,3) circle (0.15cm);
\filldraw (3,3) circle (0.15cm);
\filldraw (2,0) circle (0.15cm);
\filldraw (3,1) circle (0.15cm);
\filldraw (1,3) circle (0.15cm);
\filldraw (2,1) circle (0.15cm);
\filldraw (1,2) circle (0.15cm);
\filldraw (2,3) circle (0.15cm);
\filldraw (1,1) circle (0.15cm);
\filldraw (2,2) circle (0.15cm);
\filldraw (1,0) circle (0.15cm);
\filldraw (0,1) circle (0.15cm);
\draw[very thick] (3,-0.5)--(3,0);
\draw[very thick] (3.5,2)--(3,2);
\draw[very thick] (-0.5,3)--(0,3);
\draw[very thick] (3,3.5)--(3,3);
\draw[very thick] (0,3)--(1,3);
\draw[very thick] (3,2)--(3,1);
\draw[very thick] (3,0)--(2,0);
\draw[very thick] (3,1)--(2,1);
\draw[very thick] (1,3)--(1,2);
\draw[very thick] (1,3)--(2,3);
\draw[very thick] (2,1)--(1,1);
\draw[very thick] (2,1)--(2,2);
\draw[very thick] (1,1)--(1,0);
\draw[very thick] (1,1)--(0,1);
\draw[very thick] (1,0)--(0,0);
\draw[very thick] (0,1)--(0,2);
\draw[fill=red,red] (0,0) circle (0.15cm);
\draw[fill=red,red] (0,2) circle (0.15cm);
\end{scope}

\begin{scope}[xshift=18cm,yshift=-5cm,font=\large]
\draw[dotted,help lines] (-0.5,-0.5) grid (3.5,3.5);
\draw (-0.5,-0.5) rectangle (3.5,3.5);
\filldraw (3,0) circle (0.15cm);
\filldraw (3,2) circle (0.15cm);
\filldraw (0,3) circle (0.15cm);
\filldraw (3,3) circle (0.15cm);
\filldraw (2,0) circle (0.15cm);
\filldraw (3,1) circle (0.15cm);
\filldraw (1,3) circle (0.15cm);
\filldraw (2,1) circle (0.15cm);
\filldraw (1,2) circle (0.15cm);
\filldraw (2,3) circle (0.15cm);
\filldraw (1,1) circle (0.15cm);
\filldraw (2,2) circle (0.15cm);
\filldraw (1,0) circle (0.15cm);
\filldraw (0,1) circle (0.15cm);
\filldraw (0,0) circle (0.15cm);
\filldraw (0,2) circle (0.15cm);
\draw[very thick] (3,-0.5)--(3,0);
\draw[very thick] (3.5,2)--(3,2);
\draw[very thick] (-0.5,3)--(0,3);
\draw[very thick] (3,3.5)--(3,3);
\draw[very thick] (0,3)--(1,3);
\draw[very thick] (3,2)--(3,1);
\draw[very thick] (3,0)--(2,0);
\draw[very thick] (3,1)--(2,1);
\draw[very thick] (1,3)--(1,2);
\draw[very thick] (1,3)--(2,3);
\draw[very thick] (2,1)--(1,1);
\draw[very thick] (2,1)--(2,2);
\draw[very thick] (1,1)--(1,0);
\draw[very thick] (1,1)--(0,1);
\draw[very thick] (1,0)--(0,0);
\draw[very thick] (0,1)--(0,2);
\end{scope}

\end{tikzpicture}
\caption{Construction of the spanning tree associated with a recurrent configuration on a $4\times 4$ grid (the box represents the sink). The burning algorithm stops at time $\tau=7$. Each panel shows a snapshot for times $t$ between 0 and 7: sites in sets ${\mathcal U}_t$ are shown by their height, those in ${\mathcal B}_t$ and ${\mathcal E}_t$ are marked by red and black dots, respectively.}
\label{burnalg}
\end{figure}

The algorithm is illustrated in Fig.~\ref{burnalg}. The site in bold is in ${\mathcal U}_2$ where it has two neighbors, then belongs to ${\mathcal B}_3$ with two of its neighbors, E and S, in ${\mathcal B}_2$. Its height is $h_j = 3 = z_j^{(\mathcal{U}_{2})} + 1$ and so $m=1$. Therefore, it catches fire from its E neighbor, the first of his neighbors in ${\mathcal B}_2$ with respect to the ordering N-E-S-W. The same prescription has been used at $t=1$ to decide which edges propagate the fire to the three burning corner sites.

The burning algorithm clearly defines an injective map from the set $\mathcal{R}$ of recurrent configurations to the set $\mathcal{T}$ of spanning trees on $\G^*$ rooted at the sink. Since the two sets have equal size \cite{MD92}, the map is bijective. It follows that the probability measure on spanning trees induced by that on recurrent configurations is uniform. Moreover the size of each set defines the partition function,
\be
Z = |\mathcal R| = |\mathcal T| = \det \Delta.
\ee

It is important to bear in mind that the above algorithm, which we shall refer to as the standard algorithm, is only one among the many possible burning algorithms. One may decide to choose an ordering different from N-E-S-W, or even use a distinct but fixed  ordering at each site. One can also delay the burning of a portion of the grid, something we shall do in the next section. All these different algorithms are perfectly admissible but would typically assign trees with many different shapes to a given height configuration. Each one however provides a bijective map if consistently applied to all height configurations.

%%%%%%%%%%%%%%%%%%%%%%%%%%%%%%%%%%%%%%%%%%%%%%%%%%%%%%%%%%%%%%%%%%%%%%%%%%%%%%%%%%%%%%%%%%%%%%%%%%%%%%%%

\subsection{Height probabilities}

\begin{figure}[t]
\centering
\begin{tikzpicture}[scale=0.7,font=\small]
\tikzstyle arrowstyle=[scale=0.9]
\tikzstyle directed=[postaction={decorate,decoration={markings,mark=at position 0.7 with {\arrow[arrowstyle]{stealth}}}}]
\draw[densely dotted] (-0.5,-0.5) grid (3.5,3.5);
\draw[thick] (-0.5,-0.5) rectangle (3.5,3.5);
\foreach \x in {0,...,3}{\foreach \y in {0,...,3}{\filldraw (\x,\y) circle (0.1cm);}}
\draw[very thick,directed] (3,0)--(3,-0.5);
\draw[very thick,directed] (3,2)--(3.5,2);
\draw[very thick,directed] (0,3)--(-0.5,3);
\draw[very thick,directed] (3,3)--(3,3.5);
\draw[very thick,directed] (1,3)--(0,3);
\draw[very thick,directed] (3,1)--(3,2);
\draw[very thick,directed] (2,0)--(3,0);
\draw[very thick,directed] (2,1)--(3,1);
\draw[very thick,directed] (1,2)--(1,3);
\draw[very thick,directed] (2,3)--(1,3);
\draw[very thick,directed] (1,1)--(2,1);
\draw[very thick,directed] (2,2)--(2,1);
\draw[very thick,directed] (1,0)--(1,1);
\draw[very thick,directed] (0,1)--(1,1);
\draw[very thick,directed] (0,0)--(1,0);
\draw[very thick,directed] (0,2)--(0,1);
\filldraw[red] (2,1) circle (0.15cm) node[below right,font=\large] {$i$};
\filldraw[red] (0,1) circle (0.15cm) node[below right,font=\large] {$j$};
\end{tikzpicture}
\caption{Oriented spanning tree rooted at the sink, where $j$ is a predecessor of $i$.}
\label{USTar}
\end{figure}
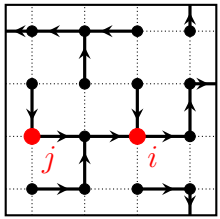

The burning algorithm associates a rooted spanning tree with each recurrent configuration. Since there is a unique path connecting every site to the root, the edges of the spanning tree can be oriented toward the root, as pictured in Fig.~\ref{USTar} (this orientation is opposite to the fire propagation). We define the notion of \emph{predecessor} by saying that $j$ is a predecessor of $i$ on a rooted spanning tree if the path from $j$ to the root goes through $i$.

Since the stationary sandpile measure $\P=\P_{\G}^*$ is uniform over the set $\mathcal{R}$ of recurrent configurations, the probability $\P_a(i) \equiv \P(h_i=a)$ that a given site $i$ has height $h_i=a$ is given by the ratio of the number of recurrent configurations with $h_i=a$ to the total number of recurrent configurations. To compute these probabilities relative to the site $i$, we partition $\mathcal{R}$ into four disjoint subsets $\mathcal{R}_k(i)$, $1\le k\le 4$, defined as 
\begin{equation}
\mathcal{R}_k(i) = \{\text{configurations that are recurrent for $k\le h_i\le 4$, transient for $1\le h_i\le k-1$}\}.
\end{equation}
Since each $\mathcal{R}_k(i)$ contains an equal number of configurations where $h_i=k,k+1,\ldots,4$, one readily finds that the probabilities are given by
\begin{equation}
\begin{split}
\P_1(i)&=\frac{|\mathcal{R}_1(i)|}{4|\mathcal{R}|}, \quad 
\P_2(i)=\P_1(i)+\frac{|\mathcal{R}_2(i)|}{3|\mathcal{R}|},\\
\P_3(i)&=\P_2(i)+\frac{|\mathcal{R}_3(i)|}{2|\mathcal{R}|},\quad
\P_4(i)=\P_3(i)+\frac{|\mathcal{R}_4(i)|}{|\mathcal{R}|}.
\label{onesite}
\end{split}
\end{equation}

The sets $\mathcal{R}_k(i)$ can be characterized in terms of spanning trees \cite{Pri94}. To see this, we use a slight modification of the standard algorithm. Namely, we keep the reference site $i$ in the set of unburnt sites as long as possible, that is, until no other site is ready to burn. At that point, $i$ and possibly other sites form a cluster $F_i$ of unburnt sites. The whole of $F_i$ will then subsequently burn following the standard procedure. With respect to this modified algorithm, and since $i$ is necessarily the first site in $F_i$ to burn, all the other sites of $F_i$ are predecessors of $i$. Now let $\mathcal C$ be a configuration in $\mathcal{R}_k(i)$, to which we apply the modified burning algorithm just described. Since the cluster $F_i$ will eventually burn for whatever value of $h_i \ge k$, $F_i$ must contain exactly $k-1$ nearest neighbors of $i$.

Conversely, let us consider a two-component spanning forest on the extended graph $\G^*$, with a tree $T_s$ rooted at the sink $s$ and a tree $T_i$ rooted at the reference site $i$, such that $T_i$ contains $k-1$ nearest neighbors of $i$. Each such forest can be made into a spanning tree on $\G^*$ by adding an extra edge between $i$ and any of its $4-(k-1)$ nearest neighbors in $T_s$. In the resulting spanning tree, $i$ has therefore $k-1$ predecessors among its neighbors. On the other hand, the forest $T_s\cup T_i$ is associated with a unique sandpile configuration $\mathcal{C}_i$ on $\G\backslash\{i\}$ through the modified burning algorithm. This configuration $\mathcal{C}_i$ can be extended to a configuration $\mathcal{C}$ on the full grid $\G$ by specifying the height $h_i$.

Let us show that $\mathcal{C}\equiv\mathcal{C}_i\cup\left\{h_i\right\}$ is recurrent if and only if $h_i\ge k$. First, observe that there are no FSCs in the subset $F_s$ of sites that are burnt before $i$ and in the subset $F_i\backslash\{i\}$, since they give rise to spanning trees through the standard burning algorithm, applied to both separately. However, the subconfiguration $\mathcal{C}\left(F_i\right)$ on $F_i$ (including $i$) is an FSC if and only if $h_i$ is less than or equal to the number of its neighbors in $F_i$, which is precisely $k-1$.

Therefore, if $X_k(i)$ denotes the fraction of spanning trees in which $i$ has exactly $k$ predecessors among its nearest neighbors, for $0 \le k \le 3$, we have shown that $X_k(i) = |\mathcal{R}_{k+1}(i)|/|{\mathcal R}|$. The one-site height probabilities can then be written as
\begin{equation}
\P_a(i)=\sum_{k=0}^{a-1}\frac{X_k(i)}{4-k}.
\label{hpred}
\end{equation}

%\begin{figure}[t]
%\centering
%\begin{tikzpicture}[scale=0.7,font=\small]
%\tikzstyle arrowstyle=[scale=0.9]
%\tikzstyle directed=[postaction={decorate,decoration={markings,mark=at position 0.7 with {\arrow[arrowstyle]{stealth}}}}]
%\draw[densely dotted] (-0.5,-0.5) grid (3.5,3.5);
%\draw[thick] (-0.5,-0.5) rectangle (3.5,3.5);
%\foreach \x in {0,...,3}{\foreach \y in {0,...,3}{\filldraw (\x,\y) circle (0.1cm);}}
%\draw[very thick,directed] (3,0)--(3,-0.5);
%\draw[very thick,directed] (3,2)--(3.5,2);
%\draw[very thick,directed] (0,3)--(-0.5,3);
%\draw[very thick,directed] (3,3)--(3,3.5);
%\draw[very thick,directed] (1,3)--(0,3);
%\draw[very thick,directed] (3,1)--(3,2);
%\draw[very thick,directed] (2,0)--(3,0);
%\draw[very thick,directed] (2,1)--(3,1);
%\draw[very thick,directed] (1,2)--(1,3);
%\draw[very thick,directed] (2,3)--(1,3);
%\draw[very thick,directed] (1,1)--(2,1);
%\draw[very thick,directed] (2,2)--(2,1);
%\draw[very thick,directed] (1,0)--(1,1);
%\draw[very thick,directed] (0,1)--(1,1);
%\draw[very thick,directed] (0,0)--(1,0);
%\draw[very thick,directed] (0,2)--(0,1);
%\draw (2,1) circle (0.2cm) node[below right,font=\large] {$i$};
%\draw (0,1) circle (0.2cm) node[below right,font=\large] {$j$};
%\end{tikzpicture}
%\caption{Oriented spanning tree rooted at the sink, where $j$ is a predecessor of $i$.}
%\label{USTar}
%\end{figure}

%%%%%%%%%%%%%%%%%%%%%%%%%%%%%%%%%%%%%%%%%%%%%%%%%%%%%%%%%%%%%%%%%%%%%%%%%%%%%%%%%%%%%%%%%%%%%%%%%%%%%%%%

\subsection{Multisite height probabilities}
\label{sec2.3}

We provide here a detailed discussion for two-site probabilities $\P_{a,b}(i,j)\equiv\P(h_i=a,h_j=b)$, assuming the reference sites $i$ and $j$ are not neighbors and do not share common neighbors. By analogy with one-site probabilities, it seems natural to define the following subsets \cite{Iva94},
\begin{equation}
\begin{split}
\mathcal{R}_{k,\l}(i,j)&=\{\text{configurations that are recurrent if $k\le h_i\le 4$}\\
&\qquad\text{and $\l \le h_j\le 4$, and transient otherwise}\},
\end{split}
\end{equation}
and try to characterize the trees contributing to $\P_{a,b}(i,j)$ in terms of these sets.

However, it was observed in \cite{Jen05b} that some recurrent configurations do not belong to any $\mathcal{R}_{k,\l}$. Consider for example a rectangular grid $\G$ of arbitrary size, with reference sites $i$ and $j$ respectively located at the top left and bottom right corners of the grid, and the configuration $\mathcal{C}_0$ such that $h_k=z_k$ for any site $k\neq i,j$, and $h_i=z_i+1=3$, $h_j=z_j+1=3$ (see Fig.~\ref{Rab}). The configuration $\mathcal{C}_0$ is recurrent, and remains recurrent if \emph{either} $h_i$ or $h_j$ is decreased by one or two. However, $\mathcal{C}_0$ becomes transient if \emph{both} $h_i,h_j$ are set to $2$. In particular $\mathcal{C}_0$ does not belong to $\mathcal{R}_{1,3}(i,j)$, since $\left\{h_i\ge 1,h_j\ge 3\right\}$ is a sufficient but not \emph{necessary} condition for $\mathcal{C}_0$ to be recurrent; similarly it does not belong to $\mathcal{R}_{3,1}(i,j)$ either. More generally, we see that $\mathcal{C}_0$ does not belong to any of the subsets $\mathcal{R}_{k,\l}(i,j)$, which therefore do not form a partition of the set of recurrent configurations.

To solve this issue, Jeng proposed an alternative division of $\mathcal{R}$ \cite{Jen05b}, which we do not state here explicitly. Instead we propose the following definition, equivalent to that of Jeng:
\begin{equation}
\begin{split}
\widetilde{\mathcal{R}}_{k,\l}(i,j)&=\{\text{configurations that are recurrent for $(h_i,h_j)=(k,\l)$,}\\
&\qquad\qquad\text{transient for $(h_i,h_j)=(k-1,\l)$ or $(k,\l-1)$}\}.
\end{split}
\end{equation}
It follows that $\mathcal{R}_{k,l}(i,j)\subset\widetilde{\mathcal{R}}_{k,\l}(i,j)$. However, the two subsets are not equal: $\widetilde{\mathcal{R}}_{k,\l}(i,j)\backslash\mathcal{R}_{k,\l}(i,j)$ contains configurations that remain recurrent for certain heights $h_i,h_j$ such that $h_i>k$ and $h_j<\l$, or $h_i<k$ and $h_j>\l$.

While any recurrent configuration belongs to a subset $\widetilde{\mathcal{R}}_{k,\l}(i,j)$, it might not be unique. For instance, the configuration $\mathcal{C}_0$ described above belongs to both $\widetilde{\mathcal{R}}_{1,3}(i,j)$ and $\widetilde{\mathcal{R}}_{3,1}(i,j)$. The subsets $\widetilde{\mathcal{R}}_{k,\l}$ can therefore be used for sandpile computations, but prove to be impractical (see for example the discussion of $\P_{2,2}(i,j)$ along a closed boundary on the upper half-plane in \cite{Jen05b}).

%\begin{figure}[t]
%\centering
%\begin{tikzpicture}[scale=0.5]
%\filldraw[black!10!white] (0,8) rectangle (1,9);
%\filldraw[black!10!white] (15,0) rectangle (16,1);
%\draw (0,0) rectangle (16,9);
%\draw (1,0)--(1,9);
%\draw (15,0)--(15,9);
%\draw (0,1)--(16,1);
%\draw (0,8)--(16,8);
%\node at (0.5,0.5) {$2$};
%\node at (15.5,8.5) {$2$};
%\node at (0.5,8.5) {$\mathbf{3}$};
%\node at (15.5,0.5) {$\mathbf{3}$};
%\node at (0.5,1.5) {$3$};
%\node at (0.5,7.5) {$3$};
%\node at (15.5,1.5) {$3$};
%\node at (15.5,7.5) {$3$};
%\node at (1.5,0.5) {$3$};
%\node at (14.5,0.5) {$3$};
%\node at (1.5,8.5) {$3$};
%\node at (14.5,8.5) {$3$};
%\node at (1.5,1.5) {$4$};
%\node at (1.5,7.5) {$4$};
%\node at (14.5,1.5) {$4$};
%\node at (14.5,7.5) {$4$};
%\draw[loosely dashed,thick] (0.5,3.75)--(0.5,5.25);
%\draw[loosely dashed,thick] (15.5,3.75)--(15.5,5.25);
%\draw[loosely dashed,thick] (7.25,0.5)--(8.75,0.5);
%\draw[loosely dashed,thick] (7.25,8.5)--(8.75,8.5);
%\draw[loosely dashed,thick] (1.5,3.75)--(1.5,5.25);
%\draw[loosely dashed,thick] (14.5,3.75)--(14.5,5.25);
%\draw[loosely dashed,thick] (7.25,1.5)--(8.75,1.5);
%\draw[loosely dashed,thick] (7.25,7.5)--(8.75,7.5);
%\draw[loosely dashed,thick] (7,5)--(9,4);
%\end{tikzpicture}
%\caption{Recurrent sandpile configuration $\mathcal{C}_0$ on a rectangular grid, with reference sites $i,j$ located at the top left and bottom right corners, that does not belong to any $\mathcal{R}_{k,\l}(i,j)$, but to $\widetilde{\mathcal{R}}_{1,3}(i,j)$ and $\widetilde{\mathcal{R}}_{3,1}(i,j)$.}
%\label{Rab}
%\end{figure}

\begin{figure}[t]
\centering
\begin{tikzpicture}[scale=0.5]
\filldraw[red!15!white] (0,8) rectangle (1,9);
\filldraw[red!15!white] (15,0) rectangle (16,1);
\draw (0,0) rectangle (16,9);
\draw (1,0)--(1,9);
\draw (15,0)--(15,9);
\draw (0,1)--(16,1);
\draw (0,8)--(16,8);
\node at (0.5,0.5) {$2$};
\node at (15.5,8.5) {$2$};
\node[red] at (0.5,8.5) {$\mathbf{3}$};
\node[red] at (15.5,0.5) {$\mathbf{3}$};
\node at (0.5,1.5) {$3$};
\node at (0.5,7.5) {$3$};
\node at (15.5,1.5) {$3$};
\node at (15.5,7.5) {$3$};
\node at (1.5,0.5) {$3$};
\node at (14.5,0.5) {$3$};
\node at (1.5,8.5) {$3$};
\node at (14.5,8.5) {$3$};
\node at (1.5,1.5) {$4$};
\node at (1.5,7.5) {$4$};
\node at (14.5,1.5) {$4$};
\node at (14.5,7.5) {$4$};
\draw[loosely dashed,thick] (0.5,3.75)--(0.5,5.25);
\draw[loosely dashed,thick] (15.5,3.75)--(15.5,5.25);
\draw[loosely dashed,thick] (7.25,0.5)--(8.75,0.5);
\draw[loosely dashed,thick] (7.25,8.5)--(8.75,8.5);
\draw[loosely dashed,thick] (1.5,3.75)--(1.5,5.25);
\draw[loosely dashed,thick] (14.5,3.75)--(14.5,5.25);
\draw[loosely dashed,thick] (7.25,1.5)--(8.75,1.5);
\draw[loosely dashed,thick] (7.25,7.5)--(8.75,7.5);
\draw[loosely dashed,thick] (7,5)--(9,4);
\end{tikzpicture}
\caption{Recurrent sandpile configuration $\mathcal{C}_0$ on a rectangular grid, with reference sites $i,j$ located at the top left and bottom right corners, that does not belong to any $\mathcal{R}_{k,\l}(i,j)$, but to $\widetilde{\mathcal{R}}_{1,3}(i,j)$ and $\widetilde{\mathcal{R}}_{3,1}(i,j)$.}
\label{Rab}
\end{figure}

Instead we set up a specific one-to-one correspondence between recurrent configurations and spanning trees based on a slight modification of the burning algorithm. We begin by making the following observation: the classification of rooted spanning trees according to the number of predecessors of $i$ and $j$ is not sufficient to compute two-site probabilities if the heights at the two reference sites are both strictly larger than 1. Indeed, let us define $X_{k,\l}(i,j)$ as the fraction of spanning trees in which $i$ and $j$ have respectively $k$ and $\ell$ predecessors among their own nearest neighbors. Then trees in a given class $X_{k,\ell}(i,j)$ can contribute differently to two-site probabilities.

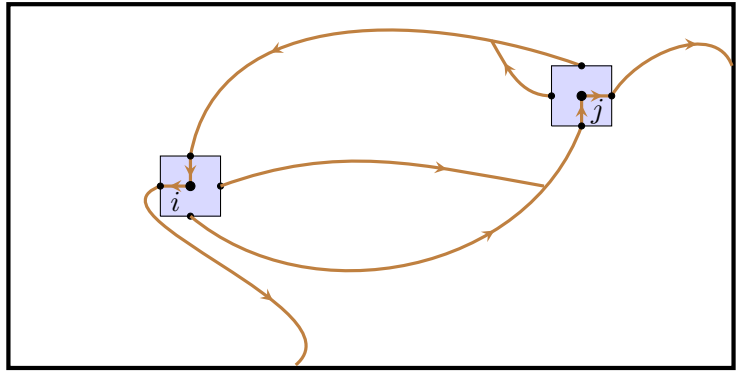
\begin{figure}[t]
\centering
\begin{tikzpicture}[scale=0.8]
\tikzstyle arrowstyle=[scale=0.9]
\tikzstyle directed=[postaction={decorate,decoration={markings,mark=at position 0.7 with {\arrow[arrowstyle]{stealth}}}}]
\draw[ultra thick] (-3.025,-3.025) rectangle (9.025,3.025);
\draw[fill=blue!15!white] (-0.5,-0.5)--(-0.5,0.5)--(0.5,0.5)--(0.5,-0.5)--(-0.5,-0.5);
\draw[very thick,directed,brown] (6.5,2)to[out=160,in=80](0,0.5);
\draw[very thick,directed,brown] (6,1.5)to[out=180,in=300](5,2.425);
\draw[very thick,directed,brown] (0,0.5)--(0,0);
\draw[very thick,directed,brown] (0,0)--(-0.5,0);
\draw[very thick,directed,brown] (-0.5,0)to[out=200,in=40](1.75,-2.975);
\filldraw (0.5,0) circle (0.05cm);
\filldraw (0,0.5) circle (0.05cm);
\filldraw (-0.5,0) circle (0.05cm);
\filldraw (0,-0.5) circle (0.05cm);
\filldraw (0,0) circle (0.075cm);
\draw (-0.25,-0.25) node {$i$};
\draw[fill=blue!15!white] (6,1)--(7,1)--(7,2)--(6,2)--(6,1);
\draw[very thick,directed,brown] (0,-0.5)to[out=320,in=250](6.5,1);
\draw[very thick,directed,brown] (0.5,0)to[out=20,in=170](5.875,0);
\draw[very thick,directed,brown] (6.5,1)--(6.5,1.5);
\draw[very thick,directed,brown] (6.5,1.5)--(7,1.5);
\draw[very thick,directed,brown] (7,1.5)to[out=60,in=110](9,2);
\filldraw (6.5,1.5) circle (0.075cm);
\draw (6.77,1.25) node {$j$};
\filldraw (7,1.5) circle (0.05cm);
\filldraw (6,1.5) circle (0.05cm);
\filldraw (6.5,2) circle (0.05cm);
\filldraw (6.5,1) circle (0.05cm);
\end{tikzpicture}
\caption{Schematic representation of a rooted spanning tree contributing to $X_{1,1}(i,j)$ with $2$ neighbors of $i$ (resp. $j$) that are predecessors of $j$ (resp. $i$). Here, only the paths connecting $i,j$, their neighbors and the root are shown.}
\label{X11_an}
\end{figure}

Consider for example the trees making the fraction $X_{1,1}(i,j)$. A tree in that set such that $i$ and its neighbors are not predecessors of $j$ or any of its neighbors, and vice versa, will contribute equally to $\P_{a,b}(i,j)$ for all values $a,b\ge 2$. The situation is different for those trees such that $i$ and $j$ are not predecessors of each other, but where two neighbors of $i$ (resp. $j$) are predecessors of $j$ (resp. $i$), as illustrated in Fig.~\ref{X11_an}. Using the standard burning algorithm (or a modified version of it), we see that $i$ or $j$ (or both) must be burnable at a time when only one of its neighbors is burnt. It follows that $h_i$ and/or $h_j$ must necessarily be equal to $4$, and therefore these spanning trees in $X_{1,1}(i,j)$ do not contribute to $\P_{2,2}$, $\P_{2,3},\P_{3,2}$ or $\P_{3,3}$.

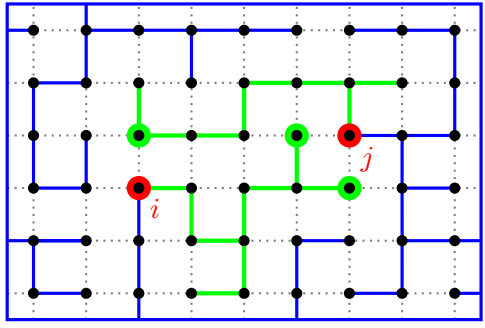
\begin{figure}[t]
\centering
\begin{tikzpicture}[scale=0.7,very thick]
\draw[help lines,thick,dotted] (-0.5,-0.5) grid (8.5,5.5);
\draw[blue] (-0.5,-0.5) rectangle (8.5,5.5);
\node[below right,red] at (2,2) {$i$};
\node[below right,red] at (6,3) {$j$};
\draw[ultra thick,green] (2,3)--(3,3)--(4,3)--(4,4)--(6,4)--(6,3);
\draw[blue] (6,3)--(7,3)--(7,2)--(7,1)--(8.5,1);
\draw[ultra thick,green] (5,3)--(5,2)--(4,2)--(4,1)--(3,1)--(3,2)--(2,2);
\draw[blue] (2,2)--(2,-0.5);
\draw[blue] (6,0)--(7,0)--(7,1);
\draw[blue] (6,5)--(8,5)--(8,3)--(7,3);
\draw[ultra thick,green] (6,4)--(7,4);
\draw[ultra thick,green] (6,2)--(5,2);
\draw[blue] (8,0)--(8.5,0);
\draw[ultra thick,green] (3,0)--(4,0)--(4,1);
\draw[blue] (1,0)--(0,0)--(0,1)--(1,1)--(-0.5,1);
\draw[blue] (1,3)--(1,2)--(0,2)--(0,4)--(1,4)--(1,5.5);
\draw[blue] (3,4)--(3,5)--(1,5);
\draw[blue] (5,5)--(3,5);
\draw[blue] (-0.5,5)--(0,5);
\draw[blue] (6,1)--(5,1)--(5,-0.5);
\draw[ultra thick,green] (2,4)--(2,3);
\draw[blue] (8,2)--(7,2);
\filldraw[red] (2,2) circle (0.2cm);
\filldraw[red] (6,3) circle (0.2cm);
\filldraw[green] (5,3) circle (0.2cm);
\filldraw[green] (2,3) circle (0.2cm);
\filldraw[green] (6,2) circle (0.2cm);
\foreach \y in {0,1,...,5}{\foreach \x in {0,1,...,8} {\filldraw (\x,\y) circle (0.075cm);};}
\end{tikzpicture}
\caption{Spanning tree accounted for in $X_{1,1}^{1,2}(i|j)$. The northern neighbor of $i$ is a predecessor of $j$, while the southern and western neighbors of $j$ are predecessors of $i$.}
\label{Xkl_ex}
\end{figure}

We see that a naive generalization of the characterization of recurrent configurations in terms of spanning trees (established for one-site probabilities) fails. A more detailed characterization including the predecessorship between $i,j$ and their neighbors is therefore required. We partition the set of spanning trees on $\G^*$ according to whether one of the reference sites is a predecessor of the other. More precisely, we define the following three quantities, seen as refined versions of the $X_{k,\ell}(i,j)$'s. In the three cases, $k$ and $\ell$ denote the number of nearest neighbors that are predecessors of $i$ and $j$ respectively.\\[-5mm]
\begin{itemize}
\item $X_{k,\ell}^{m,0}(i \to j)$ is the fraction of spanning trees on $\G^*$ in which $i$ is a predecessor of $j$, and $m$ nearest neighbors of $i$ are predecessors of $j$ but not of $i$ itself; so $1 \le \ell \le3$, $1 \le m \le 4$ and $1 \le k+m \le 4$.
\item $X_{k,\ell}^{0,n}(i \leftarrow j)$ is similarly the fraction of spanning trees on $\G^*$ in which $j$ is a predecessor of $i$, and $n$ nearest neighbors of $j$ are predecessors of $i$ but not of $j$; so $1 \le k \le 3$, $1 \le n \le 4$ and $1 \le \ell+n \le 4$.
\item $X_{k,\ell}^{m,n}(i|j)$ is the fraction of spanning trees in which neither $i$ nor $j$ is a predecessor of the other, and in which $m$ neighbors of $i$ are predecessors of $j$, and $n$ neighbors of $j$ are predecessors of $i$; here $0\le k+m \le 3$ and $0\le \l + n\le 3$, with the additional conditions that $m=0$ if $\l=0$ and $n=0$ if $k=0$. A tree of this type, contributing to $X_{1,1}^{1,2}(i|j)$, is pictured in Fig.~\ref{Xkl_ex}.
\end{itemize}

%\begin{figure}[t]
%\centering
%\begin{tikzpicture}[scale=0.7,very thick]
%\draw[thin,help lines,dotted] (-0.5,-0.5) grid (8.5,5.5);
%\draw (-0.5,-0.5) rectangle (8.5,5.5);
%\node[below right] at (2,2) {$i$};
%\node[below right] at (6,3) {$j$};
%\draw[ultra thick] (2,3)--(3,3)--(4,3)--(4,4)--(6,4)--(6,3);
%\draw (6,3)--(7,3)--(7,2)--(7,1)--(8.5,1);
%\draw[ultra thick] (5,3)--(5,2)--(4,2)--(4,1)--(3,1)--(3,2)--(2,2);
%\draw (2,2)--(2,-0.5);
%\draw (6,0)--(7,0)--(7,1);
%\draw (6,5)--(8,5)--(8,3)--(7,3);
%\draw[ultra thick] (6,4)--(7,4);
%\draw[ultra thick] (6,2)--(5,2);
%\draw (8,0)--(8.5,0);
%\draw (3,0)--(4,0)--(4,1);
%\draw (1,0)--(0,0)--(0,1)--(1,1)--(-0.5,1);
%\draw (1,3)--(1,2)--(0,2)--(0,4)--(1,4)--(1,5.5);
%\draw (3,4)--(3,5)--(1,5);
%\draw (5,5)--(3,5);
%\draw (-0.5,5)--(0,5);
%\draw (6,1)--(5,1)--(5,-0.5);
%\draw[ultra thick] (2,4)--(2,3);
%\draw (8,2)--(7,2);
%\filldraw (2,2) circle (0.2cm);
%\filldraw (6,3) circle (0.2cm);
%\draw[fill=white] (5,3) circle (0.2cm);
%\draw[fill=white] (2,3) circle (0.2cm);
%\draw[fill=white] (6,2) circle (0.2cm);
%\foreach \y in {0,1,...,5}{\foreach \x in {0,1,...,8} {\filldraw (\x,\y) circle (0.075cm);};}
%\end{tikzpicture}
%\caption{Spanning tree accounted for in $X_{1,1}^{1,2}(i|j)$. The northern neighbor of $i$ is a predecessor of $j$, while the southern and western neighbors of $j$ are predecessors of $i$.}
%\label{Xkl_ex}
%\end{figure}

In order to relate these fractions to two-site probabilities, we modify the standard burning algorithm in a way similar to what we did in the previous subsection, so that any recurrent configuration on $\G$ is now associated with a three-component spanning forest on $\G^*$. We proceed in three steps.
\begin{enumerate}
\item First we let the fire propagate on the grid except for $i$ and $j$, which we prevent from burning. We denote by $\tau$ the time at which $i$ and/or $j$ are the only burnable sites left.
\item If both $i$ and $j$ are burnable at that time, they burn simultaneously and propagate the fire to the remaining unburnt sites.
\item If $i$ is burnable at time $\tau$ but $j$ is not, we burn every burnable site except for $j$. Then we allow $j$ to burn until all sites of the grid are burnt. Otherwise $j$ is burnable at time $\tau$ and $i$ is not, then we burn $j$ first and $i$ second.
\end{enumerate}

Let us consider the three-component spanning forest $F$ on $\G^*$ whose components are the trees rooted at $s,i,j$ respectively: $F=T_s\cup T_i\cup T_j$. We assume that $T_i$ (resp. $T_j$) contains $k$ (resp. $m$) nearest neighbors of $i$ and $n$ (resp. $\ell$) nearest neighbors of $j$; then $T_s$ contains $4{-}k{-}m$ neighbors of $i$ and $4{-}\l{-}n$ neighbors of $j$.

The spanning forest $F$ can be extended into a spanning tree on $\G^*$ by adding two extra edges: one between $i$ and one of its neighbors, and one between $j$ and one of its neighbors (so that no loop is formed). If $i$ is linked to one of its $4{-}k{-}m$ neighbors in $T_s$, and $j$ is linked to one of its $4{-}\l{-}n$ neighbors in $T_s$, the resulting spanning tree on $\G^*$ contributes to the fraction $X_{k,\l}^{m,n}(i|j)$. If rather $i$ is linked to $T_s$ and $j$ to $T_i$, the spanning tree is included in $X_{k+m,\l}^{0,n}(i \leftarrow j)$. Likewise if $i$ is grafted to $T_j$ and $j$ to $T_s$, the tree belongs to $X_{k,\l+n}^{m,0}(i\to j)$. It follows that
\begin{equation}
\frac{1}{(4{-}k{-}m)(4{-}\l{-}n)}X_{k,\l}^{m,n}(i|j) = \frac{1}{(4{-}k{-}m)n}X_{k+m,\l}^{0,n}(i \leftarrow j)=\frac{1}{m(4{-}\l{-}n)}X_{k,\l+n}^{m,0}(i\to j),
\label{tree_subclass}
\end{equation}
since these three quantities (if the denominators do not vanish) equal the number of spanning forests of the type specified above.

On the other hand, a three-component spanning forest $F$, together with the information that $i|j$, $i{\to}j$ or $i{\leftarrow}j$, is in one-to-one correspondence with a sandpile configuration $\mathcal{C}_{i,j}$ on $\G\backslash\{i,j\}$ through the standard burning algorithm (starting from the roots $s$, $i$ and $j$). Let us now discuss the possible values of the pair $(h_i,h_j)$ such that the configuration $\mathcal{C}\equiv\mathcal{C}_{i,j}\cup\left\{h_i,h_j\right\}$ on the whole grid $\G$ is recurrent:
\begin{itemize}
\item If $F$ is extended to a tree contributing to $X_{k,\l}^{m,n}(i|j)$, both $i$ and $j$ must be burnable after the first step of the modified burning algorithm described above. Since at that time, $i$ (resp. $j$) has $k+m$ (resp. $\l+n$) unburnt neighbors, we obtain the inequality $k+m+1 \le h_i\le 4$ (resp. $\l+n+1 \le h_j \le 4$).
\item If $F$ is extended to a tree contributing to $X_{k+m,\l}^{0,n}(i \leftarrow j)$, $i$ must be burnable after the first step of the burning algorithm while $j$ is not burnable at that time. However $j$ must be burnable after the second step of the algorithm. Therefore, $k+m+1 \le h_i\le 4$ and $\l+1\le h_j\le \l+n$.
\item Similarly if $F$ is extended to a tree contributing to $X_{k,\l+n}^{m,0}(i\to j)$, then $k+1 \le h_i\le k+m$ and $\l+n+1 \le h_j \le 4$.
\end{itemize}
In all three cases, the number of admissible heights for $i$ and $j$ is equal to the number of ways the trees $T_i$ and $T_j$ can be grafted to one another and/or to $T_s$ to form a spanning tree on $\G^*$ of a given subclass $i|j$, $i{\leftarrow}j$ or $i{\to}j$. Therefore, a given fraction $X$ contributes equally to all pairs of admissible heights.

It is then straightforward to express two-site probabilities $\P_{a,b}(i,j)$ in terms of these fractions:
\be
\begin{split}
\P_{a,b}(i,j)&=\sum^*_{\substack{0\le k+m \le a-1 \\ 0\le\l+n \le b-1}} \frac{X_{k,\l}^{m,n}(i|j)}{(4-k-m)(4-\ell-n)} + 
\sum_{\substack{(k,m) \in U(a) \\ (\l,n) \in V(b)}} \frac{X_{k+m,\l}^{0,n}(i \leftarrow j)}{(4-k-m)n}\\
&\qquad+\sum_{\substack{(k,m) \in V(a) \\ (\l,n) \in U(b)}} \frac{X_{k,\l+n}^{m,0}(i \to j)}{m(4-\l-n)},
\label{Pab_Xkl}
\end{split}
\ee
where the symbol $^*$ over the sum is a reminder for the conditions that $k=0$ implies $n=0$, and $\l=0$ implies $m=0$. $U(h)$ and $V(h)$ are subsets of $\{0,1,2,3,4\}^2$ defined by 
\be
U(h) = \{(x,y) \,:\, 1 \le x+y \le h-1\}\,, \qquad V(h) = \{(x,y) \,:\, x+1 \le h \le x+y \le 4\}.
\ee
If $b=1$, then Eq.~\eqref{Pab_Xkl} simplifies to
\begin{align}
&\P_{a,1}(i,j)=\frac{X_{0,0}^{0,0}(i|j)}{16}&\quad\textrm{for $a=1$},\\
&\P_{a,1}(i,j)=\sum_{k=0}^{a-1}\frac{X_{k,0}^{0,0}(i|j)}{4(4-k)}+\sum_{k=1}^{a-1}\sum_{n=1}^{4}\frac{X_{k,0}^{0,n}(i\leftarrow j)}{(4-k)n}&\quad\textrm{for $a>1$.}
\end{align}
Alternatively, these probabilities can be written in terms of the fractions
\begin{equation}
X_{0,0}(i,j)=X_{0,0}^{0,0}(i|j),\quad X_{k,0}(i,j)=\sum_{n=0}^{3}X_{k,0}^{0,n}(i|j)+\sum_{n=1}^{4}X_{k,0}^{0,n}(i\leftarrow j)\textrm{ for $k>0$}.
\end{equation}
They are easier to compute, since they only take into account the number of predecessors of $i$ among its neighbors (with $j$ having none among its own neighbors). Using Eq.~\eqref{tree_subclass} with $\l=m=0$, we find that
\begin{align}
\P_{a,1}(i,j)&=\sum_{k=0}^{a-1}\frac{X_{k,0}^{0,0}(i|j)}{4(4-k)}+\sum_{k=1}^{a-1}\sum_{n=1}^{4}\frac{X_{k,0}^{0,n}(i\leftarrow j)}{4(4-k)}+\sum_{k=1}^{a-1}\sum_{n=1}^{3}\frac{(4-n)X_{k,0}^{0,n}(i\leftarrow j)}{4(4-k)n}\nonumber \\
&=\sum_{k=0}^{a-1}\frac{X_{k,0}^{0,0}(i|j)}{4(4-k)}+\sum_{k=1}^{a-1}\sum_{n=1}^{4}\frac{X_{k,0}^{0,n}(i\leftarrow j)}{4(4-k)}+\sum_{k=1}^{a-1}\sum_{n=1}^{3}\frac{X_{k,0}^{0,n}(i|j)}{4(4-k)}=\sum_{k=0}^{a-1}\frac{X_{k,0}(i,j)}{4(4-k)}.
\label{Pab_Xk0}
\end{align}

In principle, it is possible to write similar relations between $n$-site probabilities $\P_{a_1,\ldots,a_n}$ and fractions of spanning trees with various types of connectivities. Clearly, the discussion is already quite involved for $n=2$, and will certainly get more complicated for a general value of $n>2$ (indeed the number of classes of trees grows exponentially with $n$). It is however possible to deal rather simply with the particular case $a_j=1$ for $2\le j\le n$ (all the probabilities computed in this article are of this form). Following \cite{MD91}, we define a modified graph $\widetilde{\G}$ by removing three edges around each site $i_2,\ldots,i_n$, so that these sites have only one neighbor left on $\widetilde{\G}$. In doing so, the degree of each site belonging to a removed edge is decreased by 1; in particular the degree of the sites $i_2,\ldots,i_n$ is 1 on $\widetilde{\G}$. It is not difficult to see that there is a one-to-one correspondence between recurrent configurations on $\G$ with $h_{i_2}=\ldots=h_{i_n}=1$, and recurrent configurations on $\widetilde{\G}$. Therefore, the probability that a recurrent configuration on $\G$ has heights 1 at $i_2,\ldots,i_n$ is given by the ratio $\det\Delta_{\scriptsize\widetilde{\G}}/\det\Delta_{\G}$. Furthermore, it does not depend on which specific edges have been removed around the reference sites. The $n$-point probability on the original graph can therefore be written as
\begin{equation}
\P^{\G}_{a,1,\ldots,1}(i_1,i_2,\ldots,i_n)=\P^{\G}_a \big(i_1 \big| h_{i_j}{=}1,\,2{\le}j{\le}n\big)\times \P^{\G}_{1,\ldots,1}(i_2,\ldots,i_n)=\P^{\scriptsize\widetilde{\G}}_a(i_1)\times\frac{\det\Delta_{\scriptsize\widetilde{\G}}}{\det\Delta_{\G}}.
\label{Pa111}
\end{equation}
We see that evaluating an $n$-site probability with $n{-}1$ heights 1 amounts to computing a one-site probability on a modified graph, which is how we proceed for the calculations of Sections \ref{sec5} and \ref{sec6}.

Equivalently, we can write the probability \eqref{Pa111} in terms of spanning tree fractions, as computed in \eqref{Pab_Xk0} for $n=2$. Consider a spanning tree contributing to $X^{\scriptsize\widetilde{\G}}_k(i_1)$: sites $i_2,\ldots,i_n$ have only one neighbor on $\widetilde{\G}$, so they are necessarily leaves in that tree (that is, they have no predecessors). On the other hand, spanning trees in $X^{\G}_{k,0,\ldots,0}(i_1,i_2,\ldots,i_n)$ are such that $i_{j \ge 2}$ are leaves, which can be connected to any of their four neighbors. It follows that
\begin{equation}
\det\Delta_{\G}\times X^{\G}_{k,0,\ldots,0}(i_1,i_2,\ldots,i_n) = 4^{n-1} \times \det\Delta_{\scriptsize\widetilde{\G}}\times X^{\scriptsize\widetilde{\G}}_k(i_1).
\end{equation}
Using \eqref{hpred} applied to $\widetilde{\G}$, one obtains the relation
\begin{equation}
\P^{\G}_{a,1,\ldots,1}(i_1,i_2,\ldots,i_n) = \frac{\det\Delta_{\scriptsize\widetilde{\G}}}{\det\Delta_{\G}}\times\sum_{k=0}^{a-1}\frac{X^{\scriptsize\widetilde{\G}}_k(i_1)}{4-k} = \sum_{k=0}^{a-1}\frac{X^{\G}_{k,0,\ldots,0}(i_1,i_2,\dots,i_n)}{4^{n-1}(4-k)},
\end{equation}
which indeed coincides with \eqref{Pab_Xk0} for $n=2$.

%%%%%%%%%%%%%%%%%%%%%%%%%%%%%%%%%%%%%%%%%%%%%%%%%%%%%%%%%%%%%%%%%%%%%%%%%%%%%%%%%%%%%%%%%%%%%%%%%%%%%%%%

\section{Groves and the line bundle Laplacian}
\label{sec3}

In this section, we review the formalism and the results obtained by Kenyon and Wilson in \cite{KW15} that are relevant to the computations we want to present in later sections. Indeed, this formalism, which relies on the use of a line bundle Laplacian, constitutes an efficient framework to make explicit calculations of sandpile probabilities and correlations. We essentially follow the presentation of their article.

%%%%%%%%%%%%%%%%%%%%%%%%%%%%%%%%%%%%%%%%%%%%%%%%%%%%%%%%%%%%%%%%%%%%%%%%%%%%%%%%%%%%%%%%%%%%%%%%%%%%%%%%

\subsection{Parallel transport}

Let $\G^*=(\V,\E)$ be an undirected graph with vertices $\V$ and edges $\E$, then the usual graph Laplacian $\Delta_0$ is given by
\begin{equation}
\Delta_0=D-A,
\label{usual}
\end{equation}
where the \emph{degree matrix} $D$ and the \emph{adjacency matrix} $A$ are defined by
\begin{align}
D_{v,w}&=\delta_{v,w}\deg_{\G^*}(v),\\
A_{v,w}&=\begin{cases} 1\quad\textrm{if $v$ and $w$ are connected by an edge in $\E$},\\
0\quad\textrm{otherwise},
\end{cases}
\end{align}
where $\deg_{\G^*}(v)$ is the number of edges of $\G^*$ that contain $v$. The matrix $\Delta_0$ is symmetric and has vanishing row and column sums.

The graph Laplacian, acting on complex-valued functions on the graph, can be generalized in the following way. For a fixed vector space $V$, a \emph{vector bundle} $B$ over the graph $\G^*$ is the assignment to each vertex $v\in\V$ of a vector space $V_v$ isomorphic to $V$. A \emph{section} $f \in B$ is an element of $V_{\G^*}=\bigoplus_v V_v$. We focus here on the one-dimensional case $V_v\simeq V=\C$ for each $v\in\V$, which was first envisaged in \cite{For93} (the higher-dimensional analogue has been considered in \cite{Ken11}). Since $\dim V=1$, the vector bundle $B$ is called a \emph{line bundle}.

\begin{figure}[t]
%\centering
\begin{minipage}{0.5\textwidth}
\centering
\begin{tikzpicture}[scale=1.25]
\draw[very thick] (0,0)--(2,0)--(2,2)--(0,2)--(0,0);
\draw[very thick] (0,0)--(2,2);
\filldraw (0,0) circle (0.1cm) node [below left] {\large $1$};
\filldraw (2,0) circle (0.1cm) node [below right] {\large $2$};
\filldraw (2,2) circle (0.1cm) node [above right] {\large $3$};
\filldraw (0,2) circle (0.1cm) node [above left] {\large $4$};
\draw[thick] (0.75,-0.25)--node[below]{$\phi_{1,2}$}(1.25,-0.25)--(1.125,-0.2)--(1.25,-0.25)--(1.125,-0.3);
\draw[thick] (2.25,0.8)--node[right]{$\phi_{2,3}$}(2.25,1.2)--(2.2,1.075)--(2.25,1.2)--(2.3,1.075);
\draw[thick] (1.25,2.25)--node[above]{$\phi_{3,4}$}(0.75,2.25)--(0.875,2.3)--(0.75,2.25)--(0.875,2.2);
\draw[thick] (-0.25,0.8)--node[left]{$\phi_{1,4}$}(-0.25,1.2)--(-0.3,1.075)--(-0.25,1.2)--(-0.2,1.075);
\draw[thick] (0.55,0.85)--(0.9,1.2)--(0.75,1.15)--(0.9,1.2)--(0.86,1.0575);
\node at (0.55,1.4){$\phi_{1,3}$};
\end{tikzpicture}
\end{minipage}
%\hfill
\begin{minipage}{0.4\textwidth}
\centering
\begin{equation*}
\mathbf{\Delta}_0=\begin{pmatrix}
3 & -\phi^{-1}_{1,2} & -\phi^{-1}_{1,3} & -\phi^{-1}_{1,4}\\
-\phi_{1,2} & 2 & -\phi^{-1}_{2,3} & 0\\
-\phi_{1,3} & -\phi_{2,3} & 3 & -\phi^{-1}_{3,4}\\
-\phi_{1,4} & 0 & -\phi_{3,4} & 2
\end{pmatrix}
\end{equation*}
\end{minipage}
\caption{On the left: four-vertex graph equipped with the most general complex-valued connection. On the right: the corresponding line bundle Laplacian.}
\label{grcon}
\end{figure}
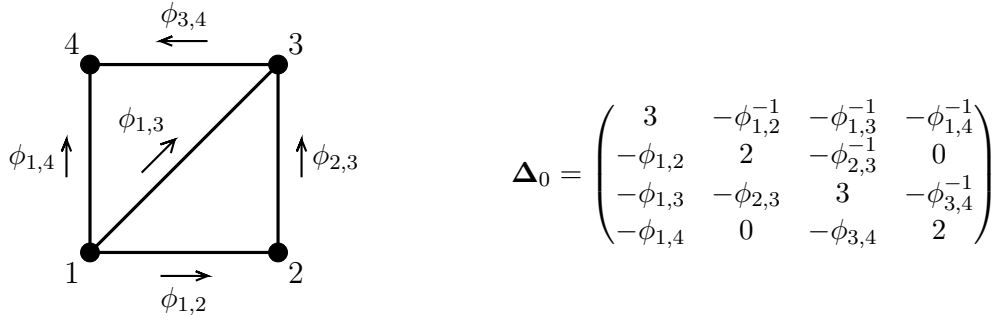

A \emph{connection} $\Phi$ is the choice, for each directed edge $e=(v,w)\in\mathcal{E}$, of an isomorphism $\phi_{v,w}:V_v\rightarrow V_w$ called a \emph{parallel transport}, such that $\phi_{w,v}=\phi^{-1}_{v,w}$. For $V=\C$, the parallel transport is just the multiplication by a nonzero complex number, $\phi_{v,w}\in\C^*$. The line bundle Laplacian with a $\C^*$-connection reads:
\begin{equation}
\mathbf{\Delta}_0 f(v)=\sum_{w\sim v}\big[f(v)-\phi_{w,v}f(w)\big] = \deg_{\G^*}(v)\,f(v)-\sum_{w\sim v}\phi_{w,v}f(w),
\label{LBLdef}
\end{equation}
where the sum is over the vertices $w$ such that the directed edge $e=(w,v)\in\mathcal{E}$. An example of a line bundle Laplacian on a graph with four vertices is given in Fig.~\ref{grcon}. Clearly the usual graph Laplacian \eqref{usual} is recovered when $\phi_{v,w}=1$ for each $e=(v,w)\in\E$.

To make contact with the toppling matrix encountered in Section~\ref{sec2}, we also need to discuss Dirichlet boundary conditions. Let $\mathcal B \subset\V$ be a subset of vertices of the graph $\G^*$, which we call the \emph{roots}. The Laplacian with Dirichlet boundary conditions at $\mathcal B$ is defined as above in \eqref{LBLdef} and restricted to those sections that vanish on $\mathcal B$:
\begin{equation}
\mathbf{\Delta}_{\V \bs \mathcal B}f(v) = \sum_{\substack{w\sim v\\w\in\V}}f(v)-\sum_{\substack{w\sim v\\w\in\V}}\phi_{w,v}f(w) = \deg_{\G^*}(v) f(v)-\sum_{\substack{w\sim v\\w\in\V\bs\mathcal B}}\phi_{w,v}f(w),
\quad v \in \V \bs \mathcal B.
\end{equation}
As a matrix, $\mathbf{\Delta}_{\V\bs\mathcal B}$ is simply the submatrix of $\mathbf{\Delta}_0\equiv\mathbf{\Delta}_{\V}$ obtained by removing the rows and columns labeled by the vertices of $\mathcal B$. In the context of the sandpile model, the elements of $\mathcal B$ are usually collectively identified as being one single root, denoted by $s$. The set $\cal D$ of dissipative sites are then the vertices connected to the root, and $\V\bs\mathcal B$ forms the set of vertices of the graph we have called $\G$.

%%%%%%%%%%%%%%%%%%%%%%%%%%%%%%%%%%%%%%%%%%%%%%%%%%%%%%%%%%%%%%%%%%%%%%%%%%%%%%%%%%%%%%%%%%%%%%%%%%%%%%%%

\subsection{Cycle-rooted groves}

\begin{figure}[t]
\centering
\begin{tikzpicture}[scale=0.60,font=\Large]
\draw[help lines,thick,dotted] (0,0) grid (10,7);
\draw[line width=0.1cm,blue] (3,5)--(2,5)--(2,6)--(3,6)--(4,6)--(4,5)--(3,5);
\draw[very thick,blue] (2,5)--(0,5)--(0,6);
\draw[very thick,blue] (0,5)--(0,4)--(1,4)--(1,3)--(0,3)--(0,2);
\draw[very thick,blue] (0,7)--(1,7)--(1,5);
\draw[very thick,blue] (1,7)--(4,7);
\draw[very thick,blue] (3,5)--(3,4)--(4,4)--(5,4)--(5,3)--(6,3);
\draw[very thick,blue] (3,5)--(4,5)--(4,6)--(5,6)--(5,7)--(6,7);
\draw[very thick,orange] (1,1)--(1,2)--(2,2)--(2,3)--(4,3)--(4,0)--(2,0)--(2,1);
\draw[very thick,orange] (3,1)--(3,2)--(6,2)--(6,1)--(8,1)--(8,2)--(7,2)--(7,3);
\draw[very thick,orange] (2,3)--(2,4);
\draw[very thick,orange] (1,1)--(0,1)--(0,0)--(1,0);
\draw[very thick,orange] (5,2)--(5,1);
\draw[very thick,orange] (7,1)--(7,0)--(5,0);
\draw[very thick,orange] (8,1)--(9,1)--(9,0);
\draw[very thick,orange] (8,0)--(10,0);
\draw[very thick,red] (6,4)--(6,6)--(8,6)--(8,5)--(9,5)--(9,4)--(8,4)--(7,4);
\draw[very thick,red] (7,5)--(7,7)--(9,7)--(9,6)--(10,6)--(10,7);
\draw[very thick,red] (6,5)--(5,5);
\draw[very thick,red] (9,5)--(10,5)--(10,4);
\draw[very thick,red] (8,4)--(8,3)--(10,3)--(10,1);
\draw[very thick,red] (10,2)--(9,2);
\filldraw[orange] (3,2) circle (0.15cm) node[above right,orange] {$1$};
\filldraw[orange] (7,1) circle (0.15cm) node[below right,orange] {$3$};
\filldraw[orange] (1,0) circle (0.15cm) node[above right,orange] {$4$};
\filldraw[red] (10,5) circle (0.15cm) node[above right,red] {$2$};
\filldraw[red] (8,4) circle (0.15cm) node[above right,red] {$5$};
\end{tikzpicture}
\caption{Cycle-rooted grove of type $\sigma=134|25$ with $n=5$ nodes.}
\label{CRGex}
\end{figure}
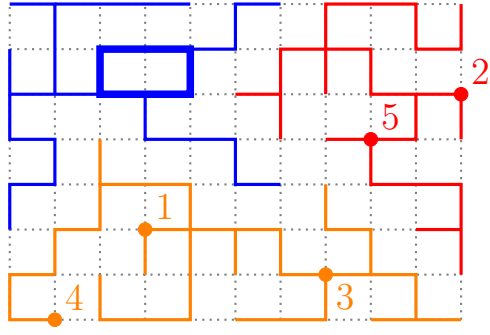

Let us select a subset $\N$ of distinguished vertices of $\G^*$ and call them \emph{nodes}, which we number from $1$ to $n=|\N|$. The vertices of $\V\bs\N$ are called \emph{interior vertices}. A \emph{cycle-rooted grove} (CRG), in the sense of \cite{KW15}, is a subgraph of $\G^*$ containing all vertices such that each component is either a cycle-rooted tree (also called a unicycle, i.e. a subgraph containing exactly one loop) without any node or a tree containing at least one node\footnote{Similar geometrical objects, comprising trees and cycle-rooted trees, were considered in the context of the monomer-dimer model \cite{BBGJ07,PPR08}, where they arose from the Temperley correspondence, and were called spanning webs. In particular, for the dimer model on a cylinder, the distribution of the number of loops was obtained by using a technique closely related to a connection \cite{BPPR14}.}. CRGs on a graph can be classified according to the way nodes are distributed in the trees. A specific way to distribute the nodes will be called a partition; the nodes belonging to distinct components will be separated by bars. For instance $\sigma=134|25$ specifies CRGs with two tree components, containing respectively nodes $\{1,3,4\}$ and $\{2,5\}$ (see the example in Fig.~\ref{CRGex}). Let us note that, depending on the positions of the nodes, certain partitions cannot be realized by CRGs, for topological reasons. If $\N = \varnothing$ (no node), all components of the CRGs are unicycles.

With respect to a given $\C^*$-connection, the \emph{monodromy} $\omega_{\gamma}$ of the unique cycle $\gamma = (v_1,v_2,\ldots,v_{\l},v_1)$  in a cycle-rooted tree is the product of all parallel transports along the cycle,
\begin{equation}
\omega_{\gamma}=\prod_{i=1}^{\ell}\phi_{v_i,v_{i+1}},
\end{equation}
with $v_{\ell+1}\equiv v_1$. Since $\phi_{v_i,v_{i+1}}$ are complex numbers, the starting point $v_1$ of the cycle does not matter, and the property $\phi_{w,v}=\phi^{-1}_{v,w}$ implies that $\omega_{\gamma^{-1}}=\omega_{\gamma}^{-1}$.

For a partition $\sigma$ of the nodes, $\Zr[\sigma]$ is defined as a weighted sum over all CRGs partitioned according to $\sigma$, 
\be
\Zr[\sigma] = \sum_{\sigma-\mathrm{CRGs}} \quad \prod_{\textrm{cycles }\gamma\in\mathrm{CRG}}\left(2-\omega_{\gamma}-\omega_{\gamma}^{-1}\right).
\ee
For a trivial connection, $\Zr[12\ldots n]$ is simply the number of spanning trees on $\G^*$ (since all cycle contributions vanish).

%\begin{figure}[t]
%\centering
%\begin{tikzpicture}[scale=0.60,font=\Large]
%\draw[dotted] (0,0) grid (10,7);
%\draw[line width=0.1cm] (3,5)--(2,5)--(2,6)--(3,6)--(4,6)--(4,5)--(3,5);
%\draw[very thick] (2,5)--(0,5)--(0,6);
%\draw[very thick] (0,5)--(0,4)--(1,4)--(1,3)--(0,3)--(0,2);
%\draw[very thick] (0,7)--(1,7)--(1,5);
%\draw[very thick] (1,7)--(4,7);
%\draw[very thick] (3,5)--(3,4)--(4,4)--(5,4)--(5,3)--(6,3);
%\draw[very thick] (3,5)--(4,5)--(4,6)--(5,6)--(5,7)--(6,7);
%\draw[very thick] (1,1)--(1,2)--(2,2)--(2,3)--(4,3)--(4,0)--(2,0)--(2,1);
%\draw[very thick] (3,1)--(3,2)--(6,2)--(6,1)--(8,1)--(8,2)--(7,2)--(7,3);
%\draw[very thick] (2,3)--(2,4);
%\draw[very thick] (1,1)--(0,1)--(0,0)--(1,0);
%\draw[very thick] (5,2)--(5,1);
%\draw[very thick] (7,1)--(7,0)--(5,0);
%\draw[very thick] (8,1)--(9,1)--(9,0);
%\draw[very thick] (8,0)--(10,0);
%\draw[very thick] (6,4)--(6,6)--(8,6)--(8,5)--(9,5)--(9,4)--(8,4)--(7,4);
%\draw[very thick] (7,5)--(7,7)--(9,7)--(9,6)--(10,6)--(10,7);
%\draw[very thick] (6,5)--(5,5);
%\draw[very thick] (9,5)--(10,5)--(10,4);
%\draw[very thick] (8,4)--(8,3)--(10,3)--(10,1);
%\draw[very thick] (10,2)--(9,2);
%\filldraw (3,2) circle (0.15cm) node[above right] {$1$};
%\filldraw (7,1) circle (0.15cm) node[below right] {$3$};
%\filldraw (1,0) circle (0.15cm) node[above right] {$4$};
%\filldraw (10,5) circle (0.15cm) node[above right] {$2$};
%\filldraw (8,4) circle (0.15cm) node[above right] {$5$};
%\end{tikzpicture}
%\caption{Cycle-rooted grove of type $\sigma=134|25$ with $n=5$ nodes.}
%\label{CRGex}
%\end{figure}

Suppose furthermore that $\sigma$ is a \emph{partial pairing}, meaning that it connects $k$ pairs of nodes and leaves the $\l$ other nodes on their own tree, i.e. $\sigma=r_1 s_1|\ldots|r_k s_k|t_1|\ldots|t_{\l}$. One defines a refined weighted sum that includes the parallel transport between paired nodes,
\be
\Zr\big[{\textstyle{{s_1}\atop{r_1}}|\ldots|{{s_k}\atop{r_k}}|t_1|\ldots|t_{\l}}\big] = \sum_{\sigma-\mathrm{CRGs}} \quad \prod_{\textrm{cycles }\gamma\in\mathrm{CRG}}\left(2-\omega_{\gamma}-\omega_{\gamma}^{-1}\right) \times \prod_{i=1}^k \: \phi_{s_i \to r_i},
\label{Zpt}
\ee
where $\phi_{s_i \to r_i}$ is the product of parallel transports along the unique path from $s_i$ to $r_i$. As a matter of notation, we write a pair of nodes $r_i,s_i$, in the argument of $\Zr$, as a column if the parallel transport $\phi_{s_i \to r_i}$ is included in the weight, as a row if it is not. When a pair $r_i,s_i$ appears as a row, the function $\Zr[\ldots|r_i s_i|\ldots]$ is symmetric in $r_i,s_i$.

The theorem below shows how to compute partition functions $\Zr[\sigma]$ when $\sigma$ is a partial pairing. To formulate it, we use the following notation. We split the set of nodes as a disjoint union $\N = R \cup S \cup T$, where $R=\left\{r_1,\ldots,r_k\right\}$, $S=\left\{s_1,\ldots,s_k\right\}$ and $T=\left\{t_1,\ldots,t_{\l}\right\}$. One of the nodes of $S$, say the $n$th node, is distinguished: indeed let $\mathbf{\Delta}_{v,w}\equiv[\mathbf{\Delta}_{\V\bs\{n\}}]_{v,w}$ be the Laplacian with Dirichlet boundary condition at $n$, and let $\Gr_{v,w}$ be its inverse, for $v,w \in \V\bs\{n\}$. Moreover, let $\widehat{\mathbf{\Delta}}$ be the matrix defined by
\begin{equation}
\begin{split}
\widehat{\mathbf{\Delta}}_{u,v}&=(\mathbf{\Delta}_{\V})_{u,v}\quad\text{for $u,v\in\V$, $u\neq n$},\quad\widehat{\mathbf{\Delta}}_{n,v}=0\quad\text{for $v\neq n$},\quad\widehat{\mathbf{\Delta}}_{n,n}=1,
\end{split}
\end{equation}
and let $\widehat{\Gr}$ be its inverse. It follows from the definition of $\widehat{\mathbf{\Delta}}$ that $\widehat{\Gr}_{u,v}=\Gr_{u,v}$ for any $u,v\in\V\bs\{n\}$.
\begin{grovethm}[\hspace{-0.025cm}\cite{KW15}\hspace{-0.025cm}]
If ${\widehat{\Gr}}\hspace{-0.05cm}\textstyle{{S\cup T}\atop {R\cup T}}$ denotes the submatrix of $\widehat{\Gr}$ labeled by rows $R\cup T$ and columns $S\cup T$, the following sum rules hold:
\begin{equation}
\Zr\det\widehat{\Gr}\hspace{-0.05cm}\textstyle{{S\cup T}\atop {R\cup T}}=\sum_{\rho \in\text{S}_k}\epsilon(\rho)\,\Zr\big[{{s_{\rho(1)}} \atop {r_1}}|\ldots|{{s_{\rho(k)}} \atop {r_k}}|t_1|\ldots|t_{\l}\big],
\label{grovethm}
\end{equation}
where 
the sum is over the symmetric group on $k$ objects, $\epsilon(\rho)$ is the parity of $\rho$, and $\Zr=\det\mathbf{\Delta}_{\V\backslash\{n\}}$. 
\end{grovethm}

Hence, linear combinations of grove partitions for partial pairings can be expressed in terms of minors of the Green matrix $\widehat{\Gr}$ of the modified Laplacian with Dirichlet boundary condition at node $n$ (the root). To simplify notations, we write $\widehat{\Gr}\hspace{-0.05cm}\textstyle{{s_1,\ldots,s_k,t_1,\ldots,t_{\l}}\atop {r_1,\ldots,r_k,t_1,\ldots,t_{\l}}}$ for $\widehat{\Gr}\hspace{-0.05cm}\textstyle{{\{s_1,\ldots,s_k\}\cup\{t_1,\ldots,t_{\l}\}}\atop {\{r_1,\ldots,r_k\}\cup\{t_1,\ldots,t_{\l}\}}}$ from now on.

On a certain type of graphs called \emph{annular-one graphs} (defined below), the grove theorem may be used to compute the function $\Zr[\sigma]$ in the limit of a trivial connection, in which case the value of $\widehat{\Gr}_{u,n}$ may be replaced with 1 \cite{KW15}.

The previous theorem will be the main tool for computing joint probabilities in the sandpile model. The next sections show how it can be used in concrete terms.

%%%%%%%%%%%%%%%%%%%%%%%%%%%%%%%%%%%%%%%%%%%%%%%%%%%%%%%%%%%%%%%%%%%%%%%%%%%%%%%%%%%%%%%%%%%%%%%%%%%%%%%%

\subsection{Annular-one graphs}

\begin{figure}[t]
\centering
\begin{tikzpicture}[scale=0.70,font=\large]

\begin{scope}[xshift=0cm]
\draw[very thick] (-3,-3) rectangle (3,3);
\draw[thick,fill=blue!15!white] (0,0) rectangle (0.5,0.5);
\draw[step=0.5cm,dotted] (-3,-3) grid (3,3);
\filldraw (0.5,0) circle (0.1cm) node[below right] {$1$};
\filldraw (0.5,0.5) circle (0.1cm) node[above right] {$2$};
\filldraw (0,0) circle (0.1cm) node[below left] {$3$};
\draw[very thick] (2,3)--(2,3.25);
\filldraw (2,3.25) circle (0.1cm) node[above right] {$4$};
\draw[thick,red] (0.25,0.25)--(0.25,-3);
\filldraw[red] (0.25,0.25) circle (0.05cm);
\draw[thick,->,red] (0.1,-1.75)--(0.5,-1.75);
\end{scope}

\begin{scope}[xshift=8cm]
\draw[very thick] (-3,-3) rectangle (3,3);
\draw[thick,fill=blue!15!white] (0,0) rectangle (0.5,0.5);
\draw[step=0.5cm,dotted] (-3,-3) grid (3,3);
\filldraw (0.5,0) circle (0.1cm) node[below right] {$1$};
\filldraw (0.5,0.5) circle (0.1cm) node[above right] {$2$};
\filldraw (0,0) circle (0.1cm) node[below left] {$3$};
\draw[very thick] (2,3)--(2,3.25);
\filldraw (2,3.25) circle (0.1cm) node[above right] {$4$};
\foreach \x in {0,0.5,...,2.75} {\draw[thick,->,red] (0,-\x)--node[below,font=\small] {$z$} (0.5,-\x);};
\end{scope}

\begin{scope}[xshift=16cm]
\draw[thick] (0,0) circle (3cm);
\draw[thick,fill=blue!15!white] (0,0) circle (1cm);
\draw[red] (0,-1)--(0,-3);
\draw[very thick,->,red] (-0.2,-2)--(0.3,-2);
\filldraw (-30:1) circle (0.1cm) node[below right] {$1$};
\filldraw (90:1) circle (0.1cm) node[above] {$2$};
\filldraw (210:1) circle (0.1cm) node[below left] {$3$};
\filldraw (50:3) circle (0.1cm) node[above right] {$4$};
\end{scope}

\end{tikzpicture}
\caption{Finite square grid with four nodes and Dirichlet boundary conditions at node 4, to which we assume all vertices on the geometric boundary of the grid are connected (this is called \emph{wired} boundary condition, as used in Section~\ref{sec2}). On the left: zipper on the dual graph going from the face adjacent to nodes $1,2,3$ to the outer face. In the middle: the edges crossed by the zipper are equipped with a parallel transport $z\in\C^*$ from left to right (and thus $z^{-1}$ from right to left). On the right: schematic representation of the nodes and the zipper on an annulus. The size of the square is eventually taken to infinity.}
\label{4nodes}
\end{figure}
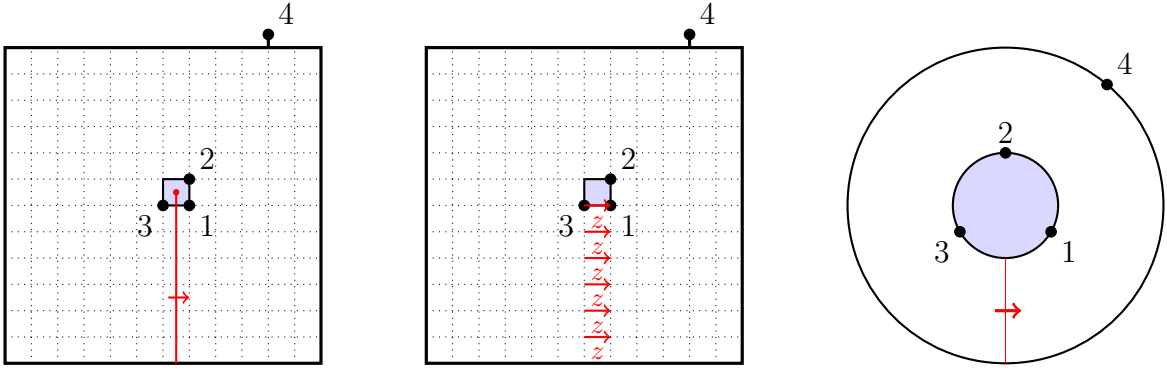

A planar graph $\G^*$ whose nodes $\left\{1,\ldots,n{-}1\right\}$ lie on the boundary of a single inner face $f$, while $n$ lies on the outer boundary, is called an \emph{annular-one} graph. We can always assume that $1,\ldots,n{-}1$ are labeled counterclockwise around $f$. We choose the connection $\Phi$ to be trivial everywhere except on the edges crossed by a \emph{zipper}, i.e. a path on the dual graph from $f$ to the outer face. The parallel transport on such edges $(k,\l)$ (called \emph{zipper edges}) is taken to be $\phi_{k,\l}=z\in\C^*$. Up to relabeling, we may assume that the zipper edge belonging to the boundary of $f$ lies between nodes $1$ and $n{-}1$ as in Fig.~\ref{4nodes}.

In the case of the square lattice $\Z^2$, we put node $n$ at infinity and choose $f$ to be the face whose lower left corner is the origin. We choose a vertical zipper down to the lower boundary, and set $\phi_{(0,m),(1,m)}=z$ for $m \le 0$.

Let us now consider a graph with $\N=\{1,2,3,4\}$ as in Fig.~\ref{4nodes} and compute the number of groves of type $12|34$. With $T=\varnothing$, and for three different choices of subsets $R,S$, the grove theorem \eqref{grovethm} yields
\begin{equation}
\begin{split}
&\Zr\det\widehat{\Gr}\textstyle{{3,4}\atop {1,2}}=\Zr[\textstyle{3 \atop 1}|\textstyle{4 \atop 2}]-\Zr[\textstyle{4 \atop 1}|\textstyle{3 \atop 2}]\,,\\
&\Zr\det\widehat{\Gr}\textstyle{{2,4}\atop {1,3}}=\Zr[\textstyle{2 \atop 1}|\textstyle{4 \atop 3}]-\Zr[\textstyle{4 \atop 1}|\textstyle{2 \atop 3}]\,,\\
&\Zr\det\widehat{\Gr}\textstyle{{1,4}\atop {2,3}}=\Zr[\textstyle{1 \atop 2}|\textstyle{4 \atop 3}]-\Zr[\textstyle{4 \atop 2}|\textstyle{1 \atop 3}]\,.
\end{split}
\label{grex1}
\end{equation}
The path joining a node $r_1\in\{1,2,3\}$ and node $4$ in $\Zr[{s_2 \atop r_2}|{4 \atop r_1}]$ can cross the zipper any number of times in both directions, so $\phi_{4\rightarrow r_1}$ varies from grove to grove. We note that these groves do not contain cycle-rooted trees, since any cycle would have a trivial monodromy. Moreover, the path between the two remaining nodes $r_2,s_2$ in $\N\bs\{r_1,4\}$ is such that $\phi_{s_2\rightarrow r_2}$ is constant over all groves in $\Zr[{s_2 \atop r_2}|{4 \atop r_1}]$, because $\G^*$ is planar. For instance, $\phi_{3\rightarrow 1} = z$ in $\Zr[{3 \atop 1}|{4 \atop 2}]$ and $\phi_{1 \rightarrow 3} = z^{-1}$ in $\Zr[{4 \atop 2}|{1 \atop 3}]$, so
\begin{equation}
\Zr[\textstyle{3 \atop 1}|{4 \atop 2}] = z \, \Zr[31|\textstyle{4 \atop 2}] = z \, \Zr[\textstyle{4 \atop 2}|13] = z^2 \, \Zr[\textstyle{4 \atop 2}|{1 \atop 3}].
%\begin{split}
%&\Zr[\textstyle{3 \atop 1}|{4 \atop 2}] = \phi_{3\rightarrow 1} \, \phi_{4\rightarrow 2} \, \Zr[1,3|2,4]=z \,\phi_{4\rightarrow 2} \, \Zr[1,3|2,4]\,,\\
%\noalign{\smallskip}
%&\Zr[\textstyle{4 \atop 2}|\textstyle{1 \atop 3}] = \phi_{1\rightarrow 3} \, \phi_{4\rightarrow 2} \, \Zr[1,3|2,4] = z^{-1}\,\phi_{4\rightarrow 2} \, \Zr[1,3|2,4]\,.
%\end{split}
\end{equation}
Likewise, we see that $\phi_{3 \to 2}=1$ in $\Zr[{4 \atop 1}|{3 \atop 2}]$ and $\phi_{2\rightarrow 1}=1$ in $\Zr[{2 \atop 1}|{4 \atop 3}]$, resulting in two further identities: 
\be
\Zr[\textstyle{4 \atop 1}|\textstyle{3 \atop 2}] = \Zr[\textstyle{4 \atop 1}|\textstyle{2 \atop 3}], \qquad \Zr[\textstyle{1 \atop 2}|\textstyle{4 \atop 3}] = \Zr[\textstyle{2 \atop 1}|\textstyle{4 \atop 3}].
\ee
This reduces the number of independent quantities to 3 and allows one, for $z \neq \pm 1$, to invert the linear system \eqref{grex1}. One obtains in particular
\begin{equation}
\begin{split}
\lim_{z\to 1}\frac{\Zr[\textstyle{2 \atop 1}|\textstyle{4 \atop 3}]}{\Zr}&=\lim_{z\to 1}\frac{\det\widehat{\Gr}\textstyle{{2,4}\atop {1,3}}+z^2\det\widehat{\Gr}\textstyle{{1,4}\atop {3,2}}+\det\widehat{\Gr}\textstyle{{3,4}\atop {2,1}}}{1-z^2}\\
&=\lim_{z\to 1}\frac{\Gr_{1,2}-z^2\,\Gr_{2,1}-\Gr_{1,3}+z^2\,\Gr_{3,1}+\Gr_{2,3}-\Gr_{3,2}}{1-z^2},
\label{grex2}
\end{split}
\end{equation}
where we used the substitution rule $\widehat{\Gr}_{i,n}\to 1$ for $1\le i\le n$, as discussed in the previous subsection. 

In the limit $z \to 1$, the weighted sum over CRGs of a given type $\sigma$ simply gives their number, and the fraction $\Zr[\sigma]/\Zr$ goes to the ratio $Z[\sigma]/Z$ of spanning forests of that type (groves are spanning forests since no cycle can contribute) to the total number of spanning trees. In this limit, both the numerator and denominator of Eq.~\eqref{grex2} vanish since $\lim_{z \to 1} \Gr_{u,v} = G_{u,v} = G_{v,u}$ is symmetric in $u,v$. The ratio however converges to
\begin{equation}
\begin{split}
\frac{Z[12|34]}{Z}&\equiv\lim_{z\rightarrow 1}\frac{\Zr[\textstyle{2 \atop 1}|\textstyle{4 \atop 3}]}{\Zr} = \lim_{z\rightarrow 1}\frac{\Gr'_{1,2} - 2z\,\Gr_{2,1} - z^2\,\Gr'_{2,1} - \Gr'_{1,3} + 2z\,\Gr_{3,1} + z^2\,\Gr'_{3,1} + \Gr'_{2,3} - \Gr'_{3,2}}{-2z}\\
&=G_{2,1}-G_{3,1}-\frac{1}{2} \Big\{[G'_{1,2}-G'_{2,1}] - [G'_{1,3}-G'_{3,1}] + [G'_{2,3}-G'_{3,2}]\Big\}\\
\noalign{\smallskip}
&=G_{1,2} - G_{1,3} - G'_{1,2} + G'_{1,3} - G'_{2,3}\,,
\end{split}
\end{equation}
where $\Gr'_{u,v}\equiv\partial_z\Gr_{u,v}$ and $G'_{u,v}=\lim_{z \to 1} \partial_z\Gr_{u,v}$, the latter of which is called the \emph{derivative of the Green function}. We see that the full dependence of the Green function $\Gr$ in the variable $z$ is not required, but only its zeroth and first order in $z{-}1$ (this remains true for a generic partition $\sigma$ \cite{KW15}). Technicalities about the way they can actually be computed are collected in the appendix, where the antisymmetry $G'_{u,v}=-G'_{v,u}$, used above in the last equality, is also proved.

\begin{figure}[t]
\centering
\begin{tikzpicture}[scale=0.5,font=\boldmath]
\tikzstyle{every node}=[draw,circle,fill=blue!25!white,scale=0.75]

\begin{scope}
\draw[thick] (0,0) circle (3cm);
\filldraw[blue!15!white] (0,0) circle (1.25cm);
\draw[thick] (0,0) circle (1.25cm);
\node (1) at (18:1.25) {$1$};
\node (2) at (90:1.25) {$2$};
\node (3) at (162:1.25) {$3$};
\node (4) at (234:1.25) {$4$};
\node (5) at (306:1.25) {$5$};
\node (6) at (90:3) {$6$};
\draw[ultra thick,blue] (2) .. controls (60:2) and (70:2.5) .. (6);
\draw[ultra thick,blue] (3) .. controls (120:2.3) .. (70:2.3);
\draw[ultra thick,blue] (1) .. controls (-45:3) and (-60:3.5) .. (4);
\draw[ultra thick,blue] (5) to[out=270,in=60] (-60:2.55);
\node[draw=none,rectangle,fill=white] at (4,0) {\huge $=$};
\end{scope}

\begin{scope}[xshift=8cm]
\draw[thick] (0,0) circle (3cm);
\filldraw[blue!15!white] (0,0) circle (1.25cm);
\draw[thick] (0,0) circle (1.25cm);
\node (1) at (18:1.25) {$1$};
\node (2) at (90:1.25) {$2$};
\node (3) at (162:1.25) {$3$};
\node (4) at (234:1.25) {$4$};
\node (6) at (90:3) {$6$};
\draw[ultra thick,blue] (2) .. controls (60:2) and (70:2.5) .. (6);
\draw[ultra thick,blue] (3) .. controls (120:2.3) .. (70:2.3);
\draw[ultra thick,blue] (1) .. controls (-45:3) and (-60:3.5) .. (4);
\node[draw=none,rectangle,fill=white] at (4,0) {\huge $=$};
\end{scope}

\begin{scope}[xshift=16cm]
\draw[thick] (0,0) circle (3cm);
\filldraw[blue!15!white] (0,0) circle (1.25cm);
\draw[thick] (0,0) circle (1.25cm);
\node (1) at (18:1.25) {$1$};
\node (2) at (90:1.25) {$2$};
\node (4) at (234:1.25) {$4$};
\node (6) at (90:3) {$6$};
\draw[ultra thick,blue] (2) .. controls (60:2) and (70:2.5) .. (6);
\draw[ultra thick,blue] (1) .. controls (-45:3) and (-60:3.5) .. (4);
\node[draw=none,rectangle,fill=white] at (4,0) {\huge $-$};
\end{scope}

\begin{scope}[xshift=24cm]
\draw[thick] (0,0) circle (3cm);
\filldraw[blue!15!white] (0,0) circle (1.25cm);
\draw[thick] (0,0) circle (1.25cm);
\node (1) at (18:1.25) {$1$};
\node (2) at (90:1.25) {$2$};
\node (3) at (162:1.25) {$3$};
\node (6) at (90:3) {$6$};
\draw[ultra thick,blue] (2) .. controls (60:2) and (70:2.5) .. (6);
\draw[ultra thick,blue] (1) .. controls (-45:3.5) and (210:3.5) .. (3);
\end{scope}

\end{tikzpicture}
\caption{Reduction of a generic partition to partial pairings for a trivial connection on an annular-one graph with six nodes.}
\label{partred}
\end{figure}
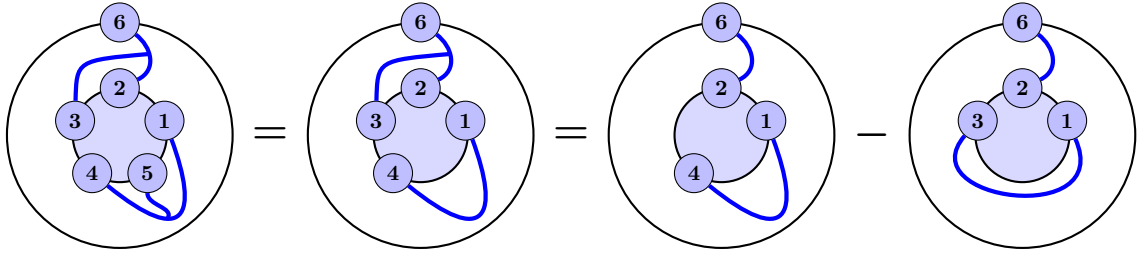

The grove theorem applies only when nodes are \emph{partially paired}, that is, when the tree components of a CRG contain either one or two nodes. When the connection is trivial, more general partitions on annular-one graphs can however be reduced to partial pairings, assuming node $n$ is not in a singleton. We illustrate the method on the partition $145|236$, shown in Fig.~\ref{partred}.

For a trivial connection, $Z[145|236]$ is the number of two-component spanning forests where $1,4,5$ lie on one tree and $2,3,6$ on the other tree. Let us first observe that in spanning forests of type $14|236$, node 5 necessarily belongs to the tree containing the nodes 1 and 4. Therefore, $Z[145|236]=Z[14|236]$, and node $5$ can be considered as an interior vertex.

Let us consider next the partition $14|26$. Node $3$ can either lie on a tree with $2$ and $6$, or with $1$ and $4$, so $Z[14|26]=Z[134|26]+Z[14|236]$. By the same argument used above, $Z[134|26]=Z[13|26]$, and we find
\begin{equation}
Z[145|236]=Z[14|236]=Z[14|26]-Z[134|26]=Z[14|26]-Z[13|26].
\label{145}
\end{equation}

%%%%%%%%%%%%%%%%%%%%%%%%%%%%%%%%%%%%%%%%%%%%%%%%%%%%%%%%%%%%%%%%%%%%%%%%%%%%%%%%%%%%%%%%%%%%%%%%%%%%%%%%

\section{Single-site probabilities on the plane}
\label{sec4}

In order to illustrate the application of the grove theorem \eqref{grovethm} to sandpile calculations, and because it forms the core of the calculations that will follow, we revisit the computation, on the infinite square lattice, of the well-known one-site probabilities $\P_a(i)$, as given by \eqref{hpred}. Their full computation by graph-theoretical methods spanned a  period of twenty years \cite{MD91,Pri94,JPR06,PP11,PPR11,KW15,CS12}, while the use of the grove theorem reduces the calculation to a few elementary steps. Because of translational invariance, the numbers $\P_a(i)=\P_a$ do not depend explicitly on $i$.

The height-one probability $\P_1=\frac{1}{4}X_0$ is the simplest one, since the definition of $X_0$ in terms of spanning trees is entirely \emph{local}: it requires that site $i$ be a leaf of the spanning tree (no predecessor among its neighbors). $\P_1$ can be computed by resorting to a new graph $\widetilde{\G}$ obtained from $\G = \Z^2$ by removing three of the adjacent edges of $i$. $\P_1$ is then simply  the ratio of the total number of spanning trees on $\widetilde{\G}^*$ to that on $\G^*$, namely the ratio of the partition functions $\widetilde{Z}$ and $Z$ pertaining respectively to $\widetilde{\G}$ and $\G$. The computation does not require a nontrivial line bundle, and reduces to a finite determinant involving the standard Green function on the plane \cite{MD91}:
\begin{equation}
\P_1 = \frac{\widetilde{Z}}{Z} = \frac{2(\pi-2)}{\pi^3} \simeq 0.0736.
\end{equation}

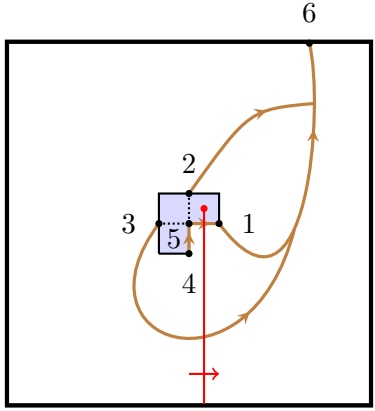
\begin{figure}[t]
\centering
\begin{tikzpicture}[scale=0.8]
\tikzstyle arrowstyle=[scale=0.9]
\tikzstyle directed=[postaction={decorate,decoration={markings,mark=at position 0.7 with {\arrow[arrowstyle]{stealth}}}}]
\draw (-3,-3) rectangle (3,3);
\draw[ultra thick] (-3.025,-3.025) rectangle (3.025,3.025);
\draw[thick,fill=blue!15!white] (-0.5,-0.5)--(-0.5,0.5)--(0.5,0.5)--(0.5,0)--(0,0)--(0,-0.5)--(-0.5,-0.5);
\draw[very thick,directed,brown] (0,-0.5)--(0,0);
\draw[very thick,directed,brown] (0,0)--(0.5,0);
\draw[very thick,directed,brown] (0.5,0) .. controls (2,-2) and (2.25,2) .. (2,3);
\draw[very thick,directed,brown] (0,0.5) .. controls (1,1.9) .. (2.10,2);
\draw[very thick,directed,brown] (-0.5,0) .. controls (-2,-2) and (1,-3) .. (1.75,-0.1);
\draw[densely dotted,thick] (0,0)--(0,0.5);
\draw[densely dotted,thick] (0,0)--(-0.5,0);
\filldraw[red] (0.25,0.25) circle (0.05cm);
\draw[thick,red] (0.25,0.25)--(0.25,-3);
\draw[thick,->,red] (0,-2.5)--(0.5,-2.5);
\filldraw (0.5,0) circle (0.05cm);
\draw (1,0) node {$1$};
\filldraw (0,0.5) circle (0.05cm);
\draw (0,1) node {$2$};
\filldraw (-0.5,0) circle (0.05cm);
\draw (-1,0) node {$3$};
\filldraw (0,-0.5) circle (0.05cm);
\draw (0,-1) node {$4$};
\filldraw (0,0) circle (0.05cm);
\draw (-0.25,-0.25) node {$5$};
\filldraw (2,3) circle (0.05cm);
\draw (2,3.5) node {$6$};
\end{tikzpicture}
\caption{Choice of nodes and edge cuts for $\P_2$. Edges belonging to the grove are schematically indicated by heavy lines. Node 5 is conventionally chosen to be the origin of the lattice.}
\label{P2}
\end{figure}

The probabilities for heights greater than one are more complicated, because they involve classes of spanning trees with nonlocal restrictions, namely that $i$ must have a fixed number of predecessors among its nearest neighbors. These restrictions can however be seen as corresponding to groves with specific node connectivities. To see this, let us denote the reference site $i$ by $5$, and its eastern, northern, western and southern neighbors by $1,2,3,4$ respectively. Node $6$ is taken as the point at infinity (i.e. the root). From \eqref{hpred}, $\P_2$ is given by
\begin{equation}
\P_2=\P_1+\frac{1}{3}X_1,
\end{equation}
where $X_1$ is the fraction of spanning trees rooted at infinity such that node $5$ has exactly one predecessor among its nearest neighbors $1,2,3,4$. By rotation invariance, we may assume without loss of generality that node $4$ is the only predecessor of node $5$ and later multiply the result by 4. The arrow going out from node $5$ can be oriented toward node $1$, $2$ or $3$, but again these three cases are equivalent. By including an extra factor 3, we may choose to orient it toward node 1, so that a typical configuration  looks like the one depicted in Fig.~\ref{P2}.

Let us consider the modified graph $\xbar{\G}$ (not to be confused with the one defined above for $\P_1$) obtained by removing the undirected edges $\{5,2\}$ and $\{5,3\}$. Then spanning forests on $\xbar{\G}$ of the type $4|12356$ are in one-to-one correspondence with those rooted spanning trees contributing to $X_1$ and containing the prescribed arrows $4 \to 5$ and $5 \to 1$, the bijection consisting in adding or removing the edge $\{5,4\}$. The fact that nodes 2 and 3 are not in the same component as 4 ensures that they are not predecessors\footnote{This nonlocal constraint, automatically accounted for by specifying the partition type $4|12356$, was handled in the old treatment by introducing $\Theta$-graphs, making the ensuing calculations much more complicated \cite{Pri94,JPR06}.} of 5. The fraction $X_1$ is therefore given by
\begin{equation}
X_1 = 12 \: \frac{\xbar{Z}[4|12356]}{Z} = 12 \, \frac{\xbar{Z}}{Z} \times \frac{\xbar{Z}[4|12356]}{\xbar{Z}},
\end{equation}
where $\xbar{Z}$ denotes the total number of spanning trees on the modified graph $\xbar{\G}$. 

%\begin{figure}[t]
%\centering
%\begin{tikzpicture}[scale=0.8]
%\tikzstyle arrowstyle=[scale=0.9]
%\tikzstyle directed=[postaction={decorate,decoration={markings,mark=at position 0.7 with {\arrow[arrowstyle]{stealth}}}}]
%\draw (-3,-3) rectangle (3,3);
%\draw[ultra thick] (-3.025,-3.025) rectangle (3.025,3.025);
%\draw[thick,fill=black!10!white] (-0.5,-0.5)--(-0.5,0.5)--(0.5,0.5)--(0.5,0)--(0,0)--(0,-0.5)--(-0.5,-0.5);
%\draw[very thick,directed] (0,-0.5)--(0,0);
%\draw[very thick,directed] (0,0)--(0.5,0);
%\draw[very thick,directed] (0.5,0) .. controls (2,-2) and (2.25,2) .. (2,3);
%\draw[very thick,directed] (0,0.5) .. controls (1,1.9) .. (2.10,2);
%\draw[very thick,directed] (-0.5,0) .. controls (-2,-2) and (1,-3) .. (1.75,-0.1);
%\draw[densely dotted,thick] (0,0)--(0,0.5);
%\draw[densely dotted,thick] (0,0)--(-0.5,0);
%\filldraw (0.25,0.25) circle (0.05cm);
%\draw[thick] (0.25,0.25)--(0.25,-3);
%\draw[thick,->] (0,-2.5)--(0.5,-2.5);
%\filldraw (0.5,0) circle (0.05cm);
%\draw (1,0) node {$1$};
%\filldraw (0,0.5) circle (0.05cm);
%\draw (0,1) node {$2$};
%\filldraw (-0.5,0) circle (0.05cm);
%\draw (-1,0) node {$3$};
%\filldraw (0,-0.5) circle (0.05cm);
%\draw (0,-1) node {$4$};
%\filldraw (0,0) circle (0.05cm);
%\draw (-0.25,-0.25) node {$5$};
%\filldraw (2,3) circle (0.05cm);
%\draw (2,3.5) node {$6$};
%\end{tikzpicture}
%\caption{Choice of nodes and edge cuts for $\P_2$. Edges belonging to the grove are schematically indicated by heavy lines. Node 5 is conventionally chosen to be the origin of the lattice.}
%\label{P2}
%\end{figure}

Since $\xbar{\G}$ is an annular-one graph, we can reduce the planar partition $\sigma = 4|12356$ to a linear combination of partial pairings, as illustrated in Fig.~\ref{partred}. Notice first that nodes 1 and 2 can be considered as interior vertices since they must necessarily belong to the same tree component as 3,5,6. Therefore we can write
\begin{equation}
\xbar{Z}[4|12356]=\xbar{Z}[4|356]=\xbar{Z}[4|56]-\xbar{Z}[34|56],
\end{equation}
since $\xbar{Z}[4|56]=\xbar{Z}[34|56]+\xbar{Z}[4|356]$. According to the grove theorem, we obtain the equations
\begin{equation}
\begin{split}
\frac{\xbar{Z}[4|56]}{\xbar{Z}}&=\lim_{z\rightarrow 1}\frac{\xbar{\Zr}[4|56]}{\xbar{\Zr}} = \xbar{G}_{4,4}-\xbar{G}_{4,5},
\end{split}
\end{equation}
\begin{equation}
\frac{\xbar{Z}[34|56]}{\xbar{Z}} = \lim_{z\rightarrow 1} \frac{\xbar{\Zr}[34|56]}{\xbar{\Zr}}=\xbar{G}_{3,4}-\xbar{G}_{3,5}-\xbar{G}'_{3,4}+\xbar{G}'_{3,5}-\xbar{G}'_{4,5},
\end{equation}
where we used $\xbar{G}_{u,v}=\xbar{G}_{v,u}$ and $\xbar{G}'_{u,v}=-\ \xbar{G}'_{v,u}$ to simplify both expressions. As illustrated in Appendix \ref{modgra}, the Woodbury formula (or the Sherman-Morrison formula applied twice, namely once for each removed edge) may be used to compute the Green function and its derivative on the modified graph $\xbar{\G}$ in terms of the same quantities on the original graph $\G$, as well as the ratio $\xbar{Z}/Z$ of partition functions. The result reads
\bea
\P_2 \egal \P_1 + \frac{12}{3} \: \frac{\xbar{Z}}{Z} \Big[(\xbar{G}_{4,4}-\xbar{G}_{4,5}) - (\xbar{G}_{3,4}-\xbar{G}_{3,5}-\xbar{G}'_{3,4}+\xbar{G}'_{3,5}-\xbar{G}'_{4,5})\Big] \label{x1} \\
\noalign{\smallskip}
\egal \frac{2(\pi-2)}{\pi^3} + 4 \cdot \frac{\pi-1}{\pi^2} \left[\frac{\pi^2-5\pi+8}{2\pi(\pi-1)} - \frac{\pi^2-10\pi+20}{16(1-\pi)}\right]\nonumber\\
\noalign{\smallskip} \egal\frac{1}{4}-\frac{1}{2\pi}-\frac{3}{\pi^2}+\frac{12}{\pi^3}\simeq 0.1739.
\eea

The computation of the height-three probability $\P_3$ is similar and makes use of the same modified graph $\xbar{\G}$ as for $\P_2$. The spanning trees contributing to $X_2$, for which the reference site (node 5) has two predecessors among its nearest neighbors, are of two types: the two predecessors form with node 5 an angle equal to $\frac{\pi}2$ or to $\pi$. We denote the corresponding two fractions\footnote{With respect to the splitting of $X_2$ in three fractions used in \cite{JPR06}, we have $X_2^{\rm (a)} = X_2^{(1)} + X_2^{(2)}$ and $X_2^{\rm (b)} = X_2^{(3)}$.} by $X_2^{\rm (a)}$ and $X_2^{\rm (b)}$. 

For type (a), we may assume that the two predecessors of node 5 are the nodes 1 and 4, and that the arrow going out of node 5 points to node 2. $X_2^{\rm (a)}$ is simply eight times this fraction of specific trees. In such a tree, the nodes 1, 4 and 5 belong to a subtree that becomes disconnected upon the removal of the edge $\{5,2\}$. After the removal, the original spanning tree breaks into two components, one containing the nodes 1, 4 and 5, the other containing 2, 3 and 6. Hence, we obtain $X_2^{\rm (a)} = 8 \, \xbar{Z}[145|236]/Z$ where $\xbar{Z}[145|236]$ is computed on the modified graph $\xbar{\G}$. Using the reduction to partial pairings shown in \eqref{145}, the same technique used above yields
\be
X_2^{\rm (a)} = 8 \, \frac{\xbar{Z}[145|236]}Z = 8 \, \Big\{\frac{\xbar{Z}[14|26]}{Z} - \frac{\xbar{Z}[13|26]}{Z}\Big\} = \frac32 - \frac2\pi + \frac4{\pi^2} - \frac{32}{\pi^3}.
\ee

For type (b), we may assume, up to a factor 4, that nodes 2 and 4 are predecessors of node 5, and that node 5 is connected to node 1. Such spanning trees are in one-to-one correspondence with the spanning forests associated with the partition $\sigma = 1356|2|4$, by removing the edges $\{2,5\}$ and $\{4,5\}$, so that $X_2^{\rm (b)} = 4  \, \xbar{Z}[1356|2|4]/Z$. This can be further simplified by observing that $\xbar{Z}[136|2|4] = \xbar{Z}[1356|2|4] + \xbar{Z}[136|2|45] = 2 \, \xbar{Z}[1356|2|4]$, because the forests of type $136|2|45$ are uniquely related to those of type $1356|2|4$ by connecting node 5 with node 1, rather than with node 4. Therefore we have
\be
X_2^{\rm (b)} = 4  \, \frac{\xbar{Z}[1356|2|4]}{Z} = 2  \, \frac{\xbar{Z}[136|2|4]}{Z} = -\frac54 + \frac5\pi + \frac2{\pi^2} - \frac{16}{\pi^3}.
\ee
Putting these results together, one finds the following expression for the height-three probability:
\be
\P_3 = \P_2 + \frac{X_2}2 = \P_2 + \frac12 (X_2^{\rm (a)} + X_2^{\rm (b)}) = \frac{3}{8}+\frac{1}{\pi}-\frac{12}{\pi^3} \simeq 0.3063.
\ee
The last probability is simply obtained by subtraction:
\be
\P_4 = 1 - \P_1 - \P_2 - \P_3 = \frac{3}{8}-\frac{1}{2\pi}+\frac{1}{\pi^2}+\frac{4}{\pi^3} \simeq 0.4462.
\ee

%%%%%%%%%%%%%%%%%%%%%%%%%%%%%%%%%%%%%%%%%%%%%%%%%%%%%%%%%%%%%%%%%%%%%%%%%%%%%%%%%%%%%%%%%%%%%%%%%%%%%%%%

\section{Multisite probabilities on the plane}
\label{sec5}

The bijection between recurrent sandpile configurations and spanning trees can in principle be used to compute multisite probabilities of an arbitrary number of heights located at sites $i_1,\ldots,i_n$. As we eventually want to compare the joint probabilities with conformal correlators, we are especially interested in the regime where all sites are mutually separated by large distances. 

As explained in Section~\ref{sec4}, handling heights 1 poses no serious problem. Counting the configurations where certain sites have a height 1 can be done by computing the total number of recurrent configurations on a locally modified graph, such that each site with height 1 has only one nearest neighbor left (thereby forcing each such site to be a leaf). Multisite height-one probabilities were computed long ago thanks to this technique (\hspace{-0.025cm}\cite{MD91} for 2-site, \cite{MR01} for up to 4-site, and \cite{Jen05a,Dur09} for general $n$-site).

The computation of a joint probability with two (or more) heights strictly larger than 1 can in principle be done by using the grove theorem. However, in order to invert the linear relations and calculate the grove fractions of interest, the grove theorem requires to work with a annular-one graph, in which the nodes---the sites where the prescribed heights are located, along with their close neighbors---are on the boundary of a single inner face, from which the zipper goes off to infinity. In case the nodes are separated by large distances, this means cutting from the original graph $\Z^2$ an unboundedly large number of edges to put all the nodes around the same face. This brings two major technical complications: $(i)$ the removal of a large number of edges defines a nonlocal (i.e. macroscopic) modification of the original graph, which makes the calculation of the modified Green function and its derivative much more complicated (for instance the Woodbury formula would require inverting a matrix of unbounded rank), and $(ii)$ the number of groves of interest on the modified graph increases exponentially with the number of removed edges. Approaches involving graphs with more than one inner face and as many zippers, or using a matrix connection, might be more suitable for these computations. However, such generalizations are currently not well enough understood. As a result, the lattice large-distance joint probabilities $\P_{a,b}(\vec r)$ of two heights $a,b>1$ remain unknown to date (note however that conformal theory predicts their correlations $\P_{a,b}(\vec r)-\P_a\P_b$ to decay like $\log^2{r}/r^4$ for large distance $r$). 

The third possible class of joint probabilities, namely those containing a single height larger than or equal to 2, is somewhat simpler. In this class, the only known results concern the three probabilities $\P_{a,1}(\vec r)$ that site $i$ has height $a=2,3$ or 4 and site $j$ has height 1, when the distance $r=|i-j|$ is large. Using graph-theoretical techniques developed earlier in \cite{Pri94}, it was shown that $\P_{a,1}(\vec r) = \P_a\P_1 + (c_a+d_a\log r)/r^4 + {\cal O}(r^{-6}\log^k r)$ for large distance $r$, with numerical constants $c_a,d_a$ explicitly known \cite{PGPR08,PGPR10}. 

In this section, we show how to compute $\P_{a,1}(\vec{r})$, $a=2,3,4$, in a more efficient way using the grove theorem and the local graph modifications explained above to handle the height 1. In addition, we extend the known results by explicitly computing the subleading contributions in $r^{-6}$. As expected, these subleading terms are not rotationally invariant. We then apply the same method to compute three-site probabilities $\P_{a,1,1}$ for $a=2,3,4$. For the purpose of comparing with field-theoretical correlation functions, it is not the joint probabilities we want to compute, but the correlators
\be
\sigma_{a,b}(i_1,i_2) = \mathbb{E}\Big[\big(\delta_{h_{i_1},a}-\P_a\big) \, \big(\delta_{h_{i_2},b}-\P_b\big)\Big] = \P_{a,b}(\vec r) - \P_a \, \P_b,
\ee
and
\be
\sigma_{a,b,c}(i_1,i_2,i_3) = \mathbb{E}\Big[\big(\delta_{h_{i_1},a}-\P_a\big) \, \big(\delta_{h_{i_2},b}-\P_b\big) \, \big(\delta_{h_{i_3},c}-\P_c\big)\Big],
\ee
here restricted to $b=c=1$. As we shall see, because of the subtractions, the calculation of three-site correlators requires the knowledge of two-site probabilities to order $r^{-6}$. Higher-order multisite probabilities $\P_{a,1,\ldots,1}$ with more heights 1 could be obtained in the same fashion.

%%%%%%%%%%%%%%%%%%%%%%%%%%%%%%%%%%%%%%%%%%%%%%%%%%%%%%%%%%%%%%%%%%%%%%%%%%%%%%%%%%%%%%%%%%%%%%%%%%%%%%%%

\subsection{Two-site probabilities}

As mentioned above, the two-site height-one correlation is well known. We simply recall the result, referring to \cite{MD91} for further details. If the two heights 1 are located at sites $i$ and $j$, and for $\vec r \equiv j-i = r {\rm e}^{{\rm i} \varphi}$, then to order 6 in the inverse distance, it is given by
\begin{equation}
\sigma_{1,1}(\vec r) = - \frac{\P_1^2}{2r^4} - \frac{4(\pi-2)}{\pi^6 r^6}\big\{1+(\pi-2)\cos 4\varphi\big\} + \O(r^{-7}),
\label{P11}
\end{equation}
where the asymptotic series for the Green function given in \eqref{Guv} has been used.

\begin{figure}[t]
\centering
\begin{tikzpicture}[scale=.8]
\tikzstyle arrowstyle=[scale=.9]
\tikzstyle directed=[postaction={decorate,decoration={markings,mark=at position 0.7 with {\arrow[arrowstyle]{stealth}}}}]
\draw (-3,-3) rectangle (9,3);
\draw[ultra thick] (-3.025,-3.025) rectangle (9.025,3.025);
\draw[thick,fill=blue!15!white] (-0.5,-0.5)--(-0.5,0.5)--(0.5,0.5)--(0.5,0)--(0,0)--(0,-0.5)--(-0.5,-0.5);
\draw[very thick,directed,brown] (0,-0.5)--(0,0);
\draw[very thick,directed,brown] (0,0)--(0.5,0);
\draw[very thick,directed,brown] (0.5,0) .. controls (2,-2) and (2.25,2) .. (2,3);
\draw[very thick,directed,brown] (0,0.5) .. controls (1,1.9) .. (2.10,2);
\draw[very thick,directed,brown] (-0.5,0) .. controls (-2,-2) and (1,-3) .. (1.75,-0.1);
\draw[densely dotted,thick] (0,0)--(0,0.5);
\draw[densely dotted,thick] (0,0)--(-0.5,0);
\filldraw[red] (0.25,0.25) circle (0.05cm);
\draw[thick,red] (0.25,0.25)--(0.25,-3);
\draw[thick,->,red] (0,-2.5)--(0.5,-2.5);
\filldraw (0.5,0) circle (0.05cm);
\draw (1,0) node {$1$};
\filldraw (0,0.5) circle (0.05cm);
\draw (0,1) node {$2$};
\filldraw (-0.5,0) circle (0.05cm);
\draw (-1,0) node {$3$};
\filldraw (0,-0.5) circle (0.05cm);
\draw (0,-1) node {$4$};
\filldraw (0,0) circle (0.05cm);
\draw (-0.25,-0.25) node {$5$};
\filldraw (2,3) circle (0.05cm);
\draw (2,3.5) node {$6$};
\draw[thick,fill=blue!15!white] (6.0,1.0)--(7.0,1.0)--(7.0,2.0)--(6.0,2.0)--(6.0,1.0);
\draw[very thick,directed,brown] (6.5,1.5)--(6.5,1.0);
\draw[densely dotted,thick] (6.0,1.5)--(7.0,1.5);
\draw[densely dotted,thick] (6.5,1.5)--(6.5,2.0);
\filldraw (6.5,1.5) circle (0.05cm);
\draw (6.77,1.25) node {$7$};
\filldraw (7.0,1.5) circle (0.05cm);
\draw (7.5,1.5) node {$8$};
\filldraw (6.0,1.5) circle (0.05cm);
\draw (6.5,2.5) node {$9$};
\filldraw (6.5,2.0) circle (0.05cm);
\draw (5.5,1.5) node {$10$};
\end{tikzpicture}
\caption{Geometrical setting used for the computation of $\P_{2,1}(\vec r)$. The dotted lines represent the removed edges. Node 5 is still at the origin, while site number 7 is located at a large distance $\vec r$.}
\label{P21}
\end{figure}
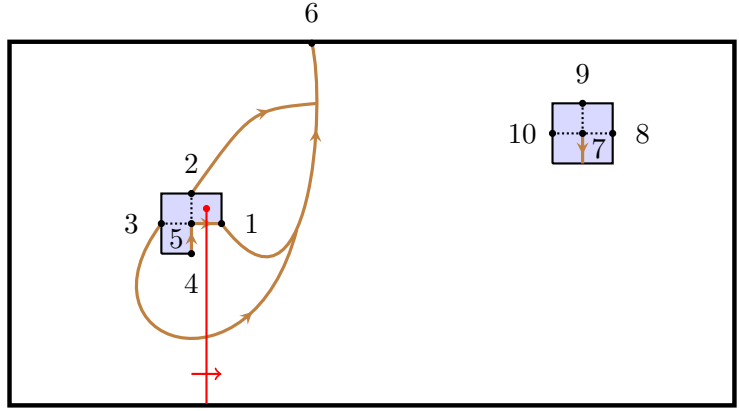

For the next case, we assume that site $i$ (chosen to be the origin) has height 2 while site $j$ has height 1, and we compute $\P_{2,1}(\vec r)$. From the discussion in Section~\ref{sec2.3}, it is given, in terms of spanning tree fractions, by
\begin{equation}
\P_{2,1}(\vec r)=\P_{1,1}(\vec r)+\frac{1}{12}X_{1,0}(\vec r),
\label{P21def}
\end{equation}
where $X_{1,0}(\vec r)$ is the fraction of spanning trees rooted at infinity such that $i$ has exactly one predecessor among its nearest neighbors, while $j$ has none. As we have seen before, the fact that $j$ is a leaf may be enforced by removing the connections with three of its nearest neighbors. There are four different ways to do so, but they are all equivalent. Choosing any specific one and multiplying by 4, we can write 
\begin{equation}
\P_{2,1}(\vec r) = \P_{1,1}(\vec r) + \frac{1}{3}\widetilde{X}_{1}(\vec r),
\label{twoone}
\end{equation}
where $\widetilde{X}_{1}(\vec r)$ denotes the fraction of spanning trees in which $i$ has exactly one predecessor among its neighbors, but on the lattice $\widetilde{\G}$ obtained from $\G=\Z^2$ by removing three edges around $j$ (in a fixed way, like those shown in Fig.~\ref{P21}). As argued in Section~\ref{sec2}, the calculation of $\P_{2,1}(\vec r)$ on $\Z^2$ amounts to computing the height-two probability at a single site but on a lattice that has been modified at a distance $r$. The modification around site $j$ however brings two complications.

The first one is that the removal of three edges at $j$ breaks the rotational invariance around $i$. As a consequence, the question of which one of its nearest neighbors, S, W, N or E, is the predecessor of $i$ matters because the four cases are no longer equivalent. They are however related by rotations if we simultaneously rotate the height 1. If we denote by $\widetilde{X}_1^{\rm S}(\vec r)$ the fraction of spanning trees (on $\widetilde{\G}$) in which the southern nearest neighbor of $i$ is its predecessor, we obtain
\be
\widetilde{X}_1(\vec r) = \widetilde{X}_1^{\rm S}(\vec r) + \widetilde{X}_1^{\rm S}({\rm e}^{{\rm i}\pi/2} \vec r) + \widetilde{X}_1^{\rm S}({\rm e}^{{\rm i}\pi} \vec r) + \widetilde{X}_1^{\rm S}({\rm e}^{3{\rm i}\pi/2} \vec r).
\label{rotate}
\ee
In each case, there are still three different possibilities for the arrow going out from $i$, but they are equivalent. Up to a factor 3, we can therefore choose to orient it toward its eastern neighbor (node 1) like we did in Section~\ref{sec4}, see Fig.~\ref{P2}.

The calculation of $\widetilde{X}_1^{\rm S}(\vec r)$ closely follows that of $X_1$ carried out in Section~\ref{sec4}. There the fraction $X_1^{\rm S} = \frac 14 X_1$ was expressed as a single grove partition on a lattice modified by the removal of the two edges $\{5,2\}$ and $\{5,3\}$. That expression is still valid in the present context, as is the ensuing formula \eqref{x1}, which here reads
\be
\widetilde{X}_1^{\rm S}(\vec r) = 3 \: \frac{\xbar{Z}}{Z} \times \Big[(\xbar{G}_{4,4}-\xbar{G}_{4,5}) - (\xbar{G}_{3,4}-\xbar{G}_{3,5}-\xbar{G}'_{3,4}+\xbar{G}'_{3,5}-\xbar{G}'_{4,5})\Big],
\label{X1S}
\ee
provided we keep in mind that the modifications referred to by the bar are made to the lattice $\widetilde{\G}$, itself a modification of $\Z^2$. Therefore, the full changes on the lattice involve the removal of five edges, two around $i$ and three around $j$, for which a total of seven sites are concerned. In addition to the three nodes 5 ($=i$), 2 and 3, we denote the other four (interior) sites by 7 ($=j$), 8, 9 and 10, as shown in Fig.~\ref{P21}.

This double modification of the lattice brings a second technical complication. According to the discussion in Appendix~\ref{modgra}, it requires computing the derivative of the Green function $G'$ for most pairs of sites among the seven sites $\{2,3,5,7,8,9,10\}$ that are concerned, in particular for sites that are far away from the zipper. When the two sites are close to the head of the zipper, the technique already used in the previous section is sufficient. If however one of the two sites is far from the zipper, one sees from the general expression \eqref{Gpdef} giving the derivative of the Green function on the plane, that an extra asymptotic analysis is needed. If both sites are far from the zipper, yet another, distinct analysis is required. It is simpler since the use of the asymptotic form of the Green function itself is permitted. Details for these two cases are given in Appendix~\ref{modgra}.

At any rate, after some algebra, we find the following expression for the function $\widetilde{X}_1^{\rm S}(\vec r)$ in polar coordinates $\vec r = r {\rm e}^{{\rm i}\varphi}$, exact to order $r^{-6}$  ($\gamma = 0.577216...$ is the Euler constant),
{\small
\begin{equation}
\begin{split}
\widetilde{X}_1^{\rm S}(\vec r)&=\frac{3 (\pi{-}2)(64{-}20\pi{-}2\pi^2{+}\pi^3)}{8 \pi^6}-\frac{3(\pi{-}2)^2}{4\pi^6} \frac{\sin\varphi}{r^3}\\
&-\frac{3(\pi{-}2)}{2\pi^6 r^4} \bigg\{(\pi{-}2) \Big[\log r + \gamma + \tfrac32 \log 2\Big] + \cos{2\varphi} + 10 - \tfrac{7\pi}{2} \bigg\}\\
&+\frac{3(\pi{-}2)}{4\pi^6} \frac{\sin \varphi}{r^5} \bigg\{4 \Big[\log r + \gamma + \tfrac32 \log 2\Big] - (4\pi{-}7) \cos{4\varphi} - (2\pi{-}3) \cos{2\varphi} - \tfrac14(\pi^2{+}16\pi{+}16) \bigg\}\\
&-\frac{1}{8\pi^6r^6} \bigg\{12(\pi{-}2) \Big[1+2(\pi{-}2)\cos 4\varphi\Big]\Big[\log r + \gamma + \tfrac32 \log 2\Big] + \tfrac32 (9\pi^3{-}30\pi^2{+}58\pi{-}62) \cos{6\varphi}\\
&-(\pi{-}2) (97\pi{-}266) \cos{4\varphi} + \tfrac32 (9\pi^3{-}36\pi^2{+}38\pi{+}2) \cos{2\varphi} + \; \tfrac32 (\pi^3{-}3\pi^2{-}60\pi{+}148) \bigg\} + \ldots\\
\end{split}
\end{equation}}
The summation in \eqref{rotate} over the four rotations eliminates all terms whose angular dependence is not a periodic function of $4\varphi$, in particular the odd powers of $r^{-1}$. The formula \eqref{twoone} yields our final result for the correlation $\sigma_{2,1}(\vec r)$, exact to order $r^{-6}$,
\bea
\sigma_{2,1}(\vec r) \egal -\: \frac{\P_1^2}{2r^4} \bigg\{\Big(\log r+\gamma+\frac{3}{2}\log 2\Big) + \frac{16-5\pi}{2(\pi-2)} \bigg\}\nonumber\\
&& -\:\frac{\pi-2}{\pi^6r^6} \bigg\{2 \Big(1+2(\pi-2)\cos 4\varphi\Big)\Big(\log r + \gamma + \tfrac32 \log 2\Big)  \nonumber\\
&& \hspace{1.5cm} - \; \tfrac16 (73\pi-218) \cos{4\varphi} + \; \frac{\pi^3-3\pi^2-44\pi+116}{4(\pi-2)} \bigg\} + \O(r^{-7}).
\label{P21exp}
\eea

The calculation of $\P_{3,1}(\vec{r})$ proceeds in exactly the same way. Referring to the decomposition of $\P_3$ in Section~\ref{sec4}, we write
\be
\P_{3,1}(\vec r) = \P_{2,1}(\vec r) + \frac18 X_{2,0}(\vec r) = \P_{2,1}(\vec r) + \frac12 \widetilde X_2(\vec r) = \P_{2,1}(\vec r) + \frac12 \Big(\widetilde X_2^{\rm (a)}(\vec r) + \widetilde X_2^{\rm (b)}(\vec r)\Big),
\ee
where the tilde refers to fractions of spanning trees on the lattice $\widetilde \G$. The fractions $\widetilde X_2^{\rm (a)}$ (resp. $\widetilde X_2^{\rm (b)}$) can be further decomposed into four (resp. two) contributions related by rotations, depending on whether the two predecessors of $i$ form the pair SE, EN, NW or WS (resp. SN or WE). The calculation then reduces to the computation of the two functions $\widetilde X_2^{\rm (a),SE}(\vec r)$ and $\widetilde X_2^{\rm (b),SN}(\vec r)$, given in terms of the same groves as in Section~\ref{sec4}, but on the modified lattice $\widetilde \G$. The required Green functions and derivatives thereof are the same as those we needed for $\P_{2,1}$.

The final result, to order $r^{-6}$, reads
\bea
\sigma_{3,1}(\vec r) \egal -\:\frac{(\pi{-}2)(8{-}\pi)}{\pi^6 r^4}\bigg\{\Big(\log r+\gamma+\frac{3}{2}\log 2\Big) - \frac{40{-}2\pi{-}\pi^2}{2(8{-}\pi)}\bigg\} \nonumber\\
&& - \: \frac{1}{\pi^6r^6} \bigg\{(8{-}\pi)\Big(1+2(\pi{-}2)\cos 4\varphi\Big)\Big(\log r + \gamma + \tfrac32 \log 2\Big)  \nonumber\\
&& \hspace{0.25cm} + \; \tfrac1{12}(\pi{-}2)(12\pi^2{+}37\pi{-}584) \cos{4\varphi} + \tfrac18 (5\pi^2{+}50\pi{-}272) \bigg\} + \O(r^{-7}).
\label{P31exp}
\eea
We do not write explicitly the last correlator $\sigma_{4,1}(\vec r)$, easily obtained by subtraction,
\be
\sigma_{4,1}(\vec r) = - \sigma_{1,1}(\vec r) - \sigma_{2,1}(\vec r) - \sigma_{3,1}(\vec r).
\ee
The dominant terms, proportional to $r^{-4}$, in the above expressions of $\sigma_{a,1}(\vec r)$ for $a>1$, reproduce the results of \cite{PGPR10}.

%%%%%%%%%%%%%%%%%%%%%%%%%%%%%%%%%%%%%%%%%%%%%%%%%%%%%%%%%%%%%%%%%%%%%%%%%%%%%%%%%%%%%%%%%%%%%%%%%%%%%%%%

\subsection{Three-site probabilities}

We now turn to three-site joint probabilities $\P_{a,1,1}$ with two heights equal to 1. While $\P_{1,1,1}$ is known \cite{MR01}, the functions $\P_{a,1,1}$ for $a\ge 2$ are new. The way they can be computed follows exactly the way $\P_{a,1}$ was computed. The core of the calculation relies on that of $\P_a$, which used the grove theorem on the modified lattice $\xbar \G$ (see Section~\ref{sec4}). A second modification around the height 1 allowed the computation of $\P_{a,1}$; now a third (similar) modification around the second height 1 is what is needed to carry out the calculation of $\P_{a,1,1}$. 

Let us denote by $i_1,i_2,i_3$ the three sites with height $a,1,1$ respectively. By translation invariance, the joint probability only depends on the two vectors $\vec r_{12} = i_2-i_1$ and $\vec r_{13} = i_3-i_1$. As the dominant term of the connected correlator $\sigma_{a,1,1}$,
\be
\sigma_{a,1,1}(\vec r_{12},\vec r_{13}) = \P_{a,1,1}(\vec r_{12},\vec r_{13}) - \P_a\,\P_{1,1}(\vec r_{23}) - \P_1\,\P_{a,1}(\vec r_{13}) - \P_1\,\P_{a,1}(\vec r_{12}) + 2\,\P_a\,\P^2_1,
\ee
is expected to be of overall order six in the distances, the knowledge of $\P_{a,1}(\vec{r})$ up to that order is required, as anticipated above. 

For the sake of simplicity, we assume the three sites $i_1,i_2,i_3$ to be aligned horizontally, so that the vectors $\vec r_{12}$ and $\vec r_{13}$ can be chosen to be along the real axis, $\vec r_{12} = (x_{21},0)$ and $\vec r_{13}=(x_{31},0)$. Up to homogeneous terms of order seven or higher in the distances, the three-point correlators read
\bea
\sigma_{1,1,1}(\vec r_{12},\vec r_{13}) \egal 0 + \ldots,\\
\noalign{\medskip}
\sigma_{2,1,1}(\vec r_{12},\vec r_{13}) \egal \frac{(\pi-2)^3}{\pi^9}\frac{1}{x_{21}^3 x_{31}^3} + \ldots,
\label{sigma211}\\
\noalign{\medskip}
\sigma_{3,1,1}(\vec r_{12},\vec r_{13}) \egal \frac{(\pi-2)^2(8-\pi)}{2\pi^9}\frac{1}{x_{21}^3 x_{31}^3} + \ldots,\\
\noalign{\medskip}
\sigma_{4,1,1}(\vec r_{12},\vec r_{13}) \egal -\frac{(\pi-2)^2(\pi+4)}{2\pi^9}\frac{1}{x_{21}^3 x_{31}^3} + \ldots
\eea
The last correlator $\sigma_{4,1,1}$ was obtained from the sum rule $\sum_a \sigma_{a,1,1} = 0$.

Two observations can be made about these results: (a) $\sigma_{1,1,1}$ vanishes identically at dominant order, and therefore also in the scaling limit, and (b) the other $\sigma_{a,1,1}$'s for $a \ge 2$ are not logarithmic, unlike the 2-correlators $\sigma_{a,1}$. As we shall argue later on, when we discuss the conformal point of view on these correlators, the property (b) is a consequence of (a) and the interpretation of the height-two, height-three and height-four variables as logarithmic partners of the height-one variable, in the scaling limit. The physical reason as to why the correlation of three heights 1 vanishes remains however unclear.

%%%%%%%%%%%%%%%%%%%%%%%%%%%%%%%%%%%%%%%%%%%%%%%%%%%%%%%%%%%%%%%%%%%%%%%%%%%%%%%%%%%%%%%%%%%%%%%%%%%%%%%%

\section{Probabilities on the upper half-plane}
\label{sec6}

Another case of interest is the lattice computation of height correlations on the upper half-plane (UHP) $\left\{(x,y)\in\Z^2|y>0\right\}$. Again we are interested in the comparison with conformal correlators, so we consider sites separated by large distances. We start by recalling the well-known one-site probabilities $\P_a(i)$ on the UHP computed in \cite{BIP93} for $a=1$ and in \cite{JPR06} for $a=2,3,4$. We compare them with the corresponding (new) probabilities on the diagonal upper half-plane (DUHP) $\left\{(x,y)\in\Z^2|y>x\right\}$.

We then compute the joint probabilities $\P_{a,1}(i,j)$ for two heights, a height $a$ at site $i$ and a height 1 at site $j$. The site $i$ is chosen to be in the bulk of the UHP (i.e. far from the boundary), while we consider two cases for $j$: in the first simpler case, $j$ is on the boundary of the UHP, and in the second case, $j$ is also in the bulk and far from $i$. The case when the two heights are located on the boundary has been completely solved, for all height values, in \cite{PR05b}.

\subsection{One-site probabilities on the upper half-plane}
\label{sec6.1}
On the upper half-plane with a horizontal boundary, the computations are very similar to those of Sections \ref{sec4} and \ref{sec5}, except that the lattice Green functions are those of the UHP. The boundary is the line of sites at $y=1$ with boundary condition either fully open or fully closed. The relevant, well-known Green functions are easily found using the image method: 
\begin{align}
G^{\textrm{op}}_{(u_1,u_2),(v_1,v_2)}&=G_{(u_1,u_2),(v_1,v_2)}-G_{(u_1,u_2),(v_1,-v_2)},\label{Gop}\\
G^{\textrm{cl}}_{(u_1,u_2),(v_1,v_2)}&=G_{(u_1,u_2),(v_1,v_2)}+G_{(u_1,u_2),(v_1,1-v_2)},\label{Gcl}
\end{align}
where $G_{u,v}$ is the Green function on the full plane $\Z^2$ (see Appendix \ref{zip}). The UHP with either boundary condition is invariant under horizontal translations, so we can choose without loss of generality the site $i$ to be located at $(0,y)$. Again we define an annular-one graph $\xbar{\G}$ obtained from $\G=\Z\times\mathbb{N}^*$ by removing edges between $i$ and two of its neighbors, so that $i$ and its four neighbors lie around a single face on $\xbar{\G}$.

The technique presented in Section~\ref{sec3} for the computation of grove probabilities works on the UHP as on the full plane, with however the new feature that the zipper can be taken to be finite or infinite, depending on whether it goes from the marked face to the boundary or off to infinity. As shown in Appendix \ref{uhp_zipper}, the two choices are equivalent but the infinite version is more convenient for practical calculations, since the derivative of the Green function on the UHP can then be written as a linear combination of that on the full plane. Therefore, we choose a zipper going upward to infinity with a nontrivial parallel transport $z$ on the horizontal edges oriented to the left, namely $\phi_{(1,y+m),(0,y+m)}=z$ for $m\ge 1$, see Fig.~\ref{zipuhp}. Then, as shown in Appendix \ref{uhp_zipper}, the derivative of the Green function on the upper half-plane reads: 
\begin{align}
G'^{\textrm{op}}_{(u_1,u_2),(v_1,v_2)} &= -\,G'_{(u_1,-u_2+y+1),(v_1,-v_2+y+1)} - G'_{(u_1,u_2+y+1),(v_1,v_2+y+1)}\nonumber\\
&\hspace{2cm} \quad+\,G'_{(u_1,-u_2+y+1),(v_1,v_2+y+1)} + G'_{(u_1,u_2+y+1),(v_1,-v_2+y+1)}, \label{Gpop}\\
G'^{\textrm{cl}}_{(u_1,u_2),(v_1,v_2)} &= -\,G'_{(u_1,-u_2+y+1),(v_1,-v_2+y+1)} - G'_{(u_1,u_2+y),(v_1,v_2+y)}
\nonumber\\
&\hspace{2cm} \quad-\,G'_{(u_1,-u_2+y+1),(v_1,v_2+y)} - G'_{(u_1,u_2+y),(v_1,-v_2+y+1)}. \label{Gpcl}
\end{align}
where $G'$ on the full plane is computed with respect to the zipper described in Section~\ref{sec3} (that is, going down).

%\begin{figure}[t]
%\centering
%\begin{tikzpicture}[scale=0.75,font=\small]
%\draw (-4,-3)--(-4,3)--(4,3)--(4,-3);
%\draw[step=0.5cm,dotted] (-4,-3) grid (4,3);
%\draw[thick] (-4,-3)--(-4,3)--(4,3)--(4,-3);
%\fill[pattern=north east lines, pattern color=gray] (-4.0125,-3.5) rectangle (4.0125,-3.1);
%\draw[thick,fill=black!10!white] (0,0)--(0.5,0)--(0.5,0.5)--(0,0.5)--(-0.5,0.5)--(-0.5,-0.5)--(0,-0.5)--(0,0);
%\filldraw[black] (0,0) circle (0.075cm);
%\draw (0.3,-0.3) node {$5$};
%\filldraw (0.5,0) circle (0.075cm);
%\draw (.9,0) node {1};
%\filldraw (0,0.5) circle (0.075cm);
%\draw (-0.25,0.9) node {2};
%\filldraw (-0.5,0) circle (0.075cm);
%\draw (-0.9,0) node {3};
%\filldraw (0,-0.5) circle (0.075cm);
%\draw (0,-0.9) node {4};
%\filldraw (3,3) circle (0.075cm) node[above right] {6};
%\foreach \x in {0.5,1,...,2.5}{\draw[thick,->] (0.5,\x) -- (0,\x);};
%\draw (2.75,-2.25) node {\small $\mathbf{i\!=\!5\!=\!(0,y)}$};
%\end{tikzpicture}
%\caption{Choice of nodes and zipper for the computation of joint height probabilities on the upper half-plane. The zipper extends up to infinity.}
%\label{zipuhp}
%\end{figure}

\begin{figure}[t]
\centering
\begin{tikzpicture}[scale=0.75,font=\small]
\draw (-4,-3)--(-4,3)--(4,3)--(4,-3);
\draw[step=0.5cm,dotted] (-4,-3) grid (4,3);
\draw[thick] (-4,-3)--(-4,3)--(4,3)--(4,-3);
\fill[pattern=north east lines, pattern color=blue] (-4.0125,-3.5) rectangle (4.0125,-3.1);
\draw[thick,fill=blue!15!white] (0,0)--(0.5,0)--(0.5,0.5)--(0,0.5)--(-0.5,0.5)--(-0.5,-0.5)--(0,-0.5)--(0,0);
\filldraw[black] (0,0) circle (0.075cm);
\draw (0.3,-0.3) node {$5$};
\filldraw (0.5,0) circle (0.075cm);
\draw (.9,0) node {1};
\filldraw (0,0.5) circle (0.075cm);
\draw (-0.25,0.9) node {2};
\filldraw (-0.5,0) circle (0.075cm);
\draw (-0.9,0) node {3};
\filldraw (0,-0.5) circle (0.075cm);
\draw (0,-0.9) node {4};
\filldraw (3,3) circle (0.075cm) node[above right] {6};
\foreach \x in {0.5,1,...,2.5}{\draw[thick,->,red] (0.5,\x) -- (0,\x);};
\draw (2.75,-2.25) node {\small $\mathbf{i\!=\!5\!=\!(0,y)}$};
\end{tikzpicture}
\caption{Choice of nodes and zipper for the computation of joint height probabilities on the upper half-plane. The zipper extends up to infinity.}
\label{zipuhp}
\end{figure}
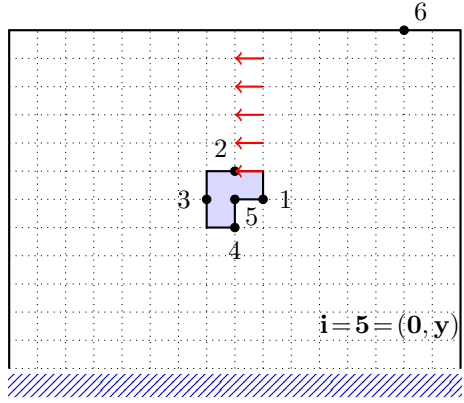

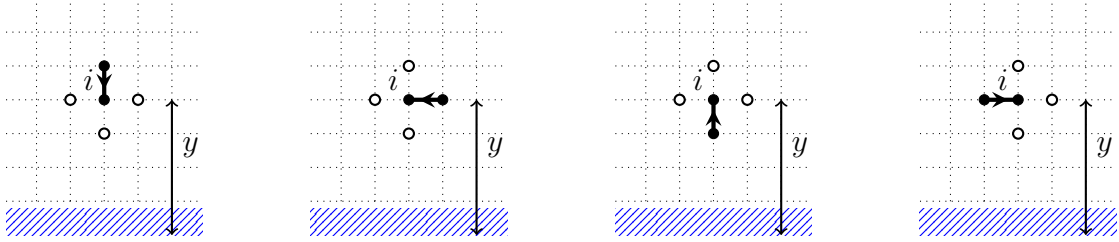
\begin{figure}[t]
\centering
\begin{tikzpicture}[scale=0.9,font=\large]
\begin{scope}[xshift=0cm]
\tikzstyle arrowstyle=[scale=1]
\tikzstyle directed=[postaction={decorate,decoration={markings,mark=at position .7 with {\arrow[arrowstyle]{stealth}}}}]
\draw[step=0.5cm,dotted] (-2.95,-1.5) grid (-0.05,1.45);
\filldraw[black] (-1.5,0) circle (0.075cm) node[above left] {$i$};
\filldraw[white] (-1,0) circle (0.075cm);
\draw[thick] (-1,0) circle (0.075cm);
\filldraw[black] (-1.5,0.5) circle (0.075cm);
\filldraw[white] (-2,0) circle (0.075cm);
\draw[thick] (-2,0) circle (0.075cm);
\filldraw[white] (-1.5,-0.5) circle (0.075cm);
\draw[thick] (-1.5,-0.5) circle (0.075cm);
\draw[directed,ultra thick] (-1.5,0.5)--(-1.5,0);
\fill[pattern=north east lines, pattern color=blue] (-2.95,-1.6) rectangle (-0.05,-2);
\draw[thick,<->] (-0.5,-2)--node[above right] {$y$}(-0.5,0);
\end{scope}
\begin{scope}[xshift=4.5cm]
\tikzstyle arrowstyle=[scale=1]
\tikzstyle directed=[postaction={decorate,decoration={markings,mark=at position .7 with {\arrow[arrowstyle]{stealth}}}}]
\draw[step=0.5cm,dotted] (-2.95,-1.5) grid (-0.05,1.45);
\filldraw[black] (-1.5,0) circle (0.075cm) node[above left] {$i$};
\filldraw[white] (-1.5,-0.5) circle (0.075cm);
\draw[thick] (-1.5,-0.5) circle (0.075cm);
\filldraw[white] (-1.5,0.5) circle (0.075cm);
\draw[thick] (-1.5,0.5) circle (0.075cm);
\filldraw[white] (-2,0) circle (0.075cm);
\draw[thick] (-2,0) circle (0.075cm);
\filldraw[black] (-1,0) circle (0.075cm);
\draw[directed,ultra thick] (-1,0)--(-1.5,0);
\fill[pattern=north east lines, pattern color=blue] (-2.95,-1.6) rectangle (-0.05,-2);
\draw[thick,<->] (-0.5,-2)--node[above right] {$y$}(-0.5,0);
\end{scope}
\begin{scope}[xshift=9cm]
\tikzstyle arrowstyle=[scale=1]
\tikzstyle directed=[postaction={decorate,decoration={markings,mark=at position .7 with {\arrow[arrowstyle]{stealth}}}}]
\draw[step=0.5cm,dotted] (-2.95,-1.5) grid (-0.05,1.45);
\filldraw[black] (-1.5,0) circle (0.075cm) node[above left] {$i$};
\filldraw[white] (-1,0) circle (0.075cm);
\draw[thick] (-1,0) circle (0.075cm);
\filldraw[white] (-1.5,0.5) circle (0.075cm);
\draw [thick] (-1.5,0.5) circle (0.075cm);
\filldraw[white] (-2,0) circle (0.075cm);
\draw [thick] (-2,0) circle (0.075cm);
\filldraw[black] (-1.5,-0.5) circle (0.075cm);
\draw[directed,ultra thick] (-1.5,-0.5)--(-1.5,0);
\fill[pattern=north east lines, pattern color=blue] (-2.95,-1.6) rectangle (-0.05,-2);
\draw[thick,<->] (-0.5,-2)--node[above right] {$y$}(-0.5,0);
\end{scope}
\begin{scope}[xshift=13.5cm]
\tikzstyle arrowstyle=[scale=1]
\tikzstyle directed=[postaction={decorate,decoration={markings,mark=at position .7 with {\arrow[arrowstyle]{stealth}}}}]
\draw[step=0.5cm,dotted] (-2.95,-1.5) grid (-0.05,1.45);
\filldraw[black] (-1.5,0) circle (0.075cm) node[above left] {$i$};
\filldraw[white] (-1,0) circle (0.075cm);
\draw[thick] (-1,0) circle (0.075cm);
\filldraw[white] (-1.5,0.5) circle (0.075cm);
\draw[thick] (-1.5,0.5) circle (0.075cm);
\filldraw[white] (-1.5,-0.5) circle (0.075cm);
\draw[thick] (-1.5,-0.5) circle (0.075cm);
\filldraw[black] (-2,0) circle (0.075cm);
\draw[directed,ultra thick] (-2,0)--(-1.5,0);
\fill[pattern=north east lines, pattern color=blue] (-2.95,-1.6) rectangle (-0.05,-2);
\draw[thick,<->] (-0.5,-2)--node[above right] {$y$}(-0.5,0);
\end{scope}
\end{tikzpicture}
\caption{The four distinct diagrams contributing to $X_1(i)$ on the upper half-plane. Neighbors of $i$ drawn as open circles are \emph{not} predecessors of $i$. The boundary of the graph, i.e. the lowest row of sites still included in the graph, is located at $y=1$.}
\label{P2uhpdiag}
\end{figure}

The presence of the boundary means that predecessor diagrams are no longer invariant under rotations. In the case of $\P_2(i)$ for example, one has to compute the fraction $X_1(i)$ of spanning trees such that $i$ has exactly one predecessor among its neighbors, either North, East, South or West, see Fig.~\ref{P2uhpdiag}. The equality $X_1^{\rm E}(i)=X_1^{\rm W}(i)$ stills holds on the UHP because of the left-right symmetry, so that there are actually three distinct diagrams to consider. In addition, the other two diagrams are related by the following identities, 
\be
X_1^{\rm N}(i)\Big|_{y\to -y} = X_1^{\rm S}(i)\quad\textrm{for open b.c.,} \qquad
X_1^{\rm N}(i)\Big|_{y\to 1-y} = X_1^{\rm S}(i)\quad\textrm{for closed b.c.},
\ee
which we prove in Appendix \ref{Diag_sym}. The combinatorial significance of these relations is not clear to us.

Although there are roughly twice as many diagrams as for the full plane, their computation procedure using the grove theorem is similar, so we simply mention the final results for one-site height probabilities. We give their expansions in terms of the distance $r$ between the reference site $i$ located at $(0,y)$ and the symmetry axes used in the method of images, namely $y=0$ (open boundary) and $y=1/2$ (closed boundary):
\begin{equation}
r=y\quad\textrm{(open b.c.)},\quad r=y-\frac{1}{2}\quad\textrm{(closed b.c.)}
\end{equation}
For the two boundary conditions, the results read:
\begin{subequations}
\begin{align}
\sigma_a^{\rm op}(r)&=\frac{1}{r^2}\Big(c_a+\frac{d_a}{2}+d_a\log r\Big)+\frac{1}{r^4}\Big(\frac{c_a}{4}+e_a+\frac{d_a}{4}\log r\Big)+\O(r^{-5}),\label{1ptop}\\
\sigma_a^{\rm cl}(r)&=-\frac{1}{r^2}\big(c_a+d_a\log r\big)-\frac{1}{r^4}\Big(\frac{c_a}{4}+f_a+\frac{d_a}{4}\log r\Big)+\O(r^{-5}),\label{1ptcl}
\end{align}
\label{Pa_uhp}%
\end{subequations}
with the various coefficients $c_a,d_a,e_a,f_a$ given in Table~\ref{UHP_coef}. The dominant terms in $r^{-2}$, depending on $c_a,d_a$ only, were first obtained in \cite{JPR06}, whereas the lower-order terms in $r^{-4}$ are new.
\vspace{5mm}
\begin{table}[htbp]
\centering
\large
\renewcommand{\arraystretch}{1.8}
\tabcolsep11pt
\begin{tabular}{|c|cccc|}
\hline
 & $c_a$ & $d_a$ & $e_a$ & $f_a$ \\
\hline
$a=1$ & $\frac{\pi-2}{2\pi^3}$ & 0 & $\frac{\pi+4}{32\pi^3}$ & $\frac{8-\pi}{32\pi^3}$ \\
$a=2$ & $\frac{\pi-2}{2\pi^3}\left(\gamma+\frac{5}{2}\text{{\small $\log 2$}}\right)+\frac{34-11\pi}{8\pi^3}$ & $\frac{\pi-2}{2\pi^3}$ & $\frac{-76-52\pi+9\pi^2}{384\pi^3}$ & $\frac{-196-28\pi+9\pi^2}{384\pi^3}$ \\
$a=3$ & $\frac{8-\pi}{4\pi^3}\left(\gamma+\frac{5}{2}\text{{\small $\log 2$}}\right)+\frac{-88+5\pi+2\pi^2}{16\pi^3}$ & $\frac{8-\pi}{4\pi^3}$ & $\frac{8+23\pi}{384\pi^3}$ & $\frac{104+17\pi}{384\pi^3}$ \\
$a=4$ & $-\frac{\pi+4}{4\pi^3}\left(\gamma+\frac{5}{2}\text{{\small $\log 2$}}\right)+\frac{36+9\pi-2\pi^2}{16\pi^3}$ & $-\frac{\pi+4}{4\pi^3}$ & $\frac{20+17\pi-9\pi^2}{384\pi^3}$ & $\frac{-4+23\pi-9\pi^2}{384\pi^3}$ \\
\hline
\end{tabular}
\caption{Numerical coefficients for one-site probabilities on the upper half-plane.}
\label{UHP_coef}
\end{table}

%%%%%%%%%%%%%%%%%%%%%%%%%%%%%%%%%%%%%%%%%%%%%%%%%%%%%%%%%%%%%%%%%%%%%%%%%%%%%%%%%%%%%%%%%%%%%%%%%%%%%

\subsection{One-site probabilities on the diagonal upper half-plane}
\label{sec6.2}
\begin{figure}[t]
\centering
\begin{tikzpicture}[scale=0.65,font=\small]
\draw (0,0) rectangle (8,8);
\draw[step=0.5cm,dotted] (0,0) grid (8,8);
\draw[thick] (0,0)--(0,8)--(8,8);
\draw[thick,fill=blue!15!white] (2,5.5)--(2.5,5.5)--(2.5,6)--(2,6)--(1.5,6)--(1.5,5)--(2,5)--(2,5.5);
\filldraw (2,5.5) circle (0.075cm);
\draw (2.3,5.2) node {$5$};
\filldraw (2.5,5.5) circle (0.075cm);
\draw (2.9,5.5) node {1};
\filldraw (2,6) circle (0.075cm);
\draw (1.75,6.4) node {2};
\filldraw (1.5,5.5) circle (0.075cm);
\draw (1.1,5.5) node {3};
\filldraw (2,5) circle (0.075cm);
\draw (2,4.6) node {4};
\filldraw (5,8) circle (0.075cm) node [above right] {6};
\foreach \x in {6,6.5,...,7.5} {\draw[thick,->,red] (2.5,\x) -- (2,\x);};
\filldraw[white] (0,0)--(8,8)--(8.2,-0.2)--(0,0);
\fill[pattern=horizontal lines, pattern color=blue,shift={(0.075cm,-0.075cm)}] (0,0)--(8,8)--(8.4,7.6)--(0.4,-0.4);
\draw (6.5,2.5) node {\small $\mathbf{i\!=\!5\!=\!(0,y)}$};
\end{tikzpicture}
\caption{Choice of nodes and zipper for the computation of height probabilities on the diagonal upper half-plane. The zipper extends up to infinity.}
\label{UDHP}
\end{figure}
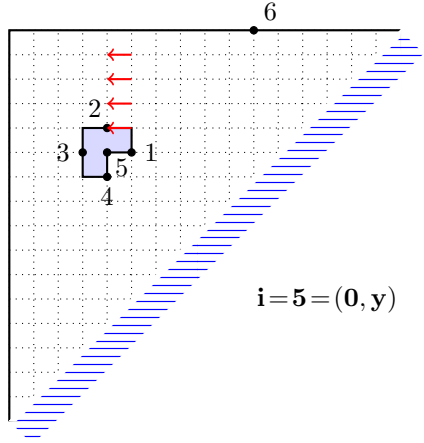

In addition to the usual upper half-plane with a horizontal boundary, we examine a new form of semi-infinite lattice, namely the diagonal upper half-plane (DUHP) $\left\{(x,y)\in\Z^2|y>x\right\}$. Indeed, we note that the Green functions for open and closed boundary conditions can be obtained quite simply by the method of images, and read:
\begin{align}
G^{\mathrm{op}}_{(u_1,u_2),(v_1,v_2)}&=G_{(u_1,u_2),(v_1,v_2)}-G_{(u_1,u_2),(v_2,v_1)}, \label{Gop_diag}\\
G^{\mathrm{cl}}_{(u_1,u_2),(v_1,v_2)}&=G_{(u_1,u_2),(v_1,v_2)}+G_{(u_1,u_2),(v_2-1,v_1+1)}. \label{Gcl_diag}
\end{align}
Without loss of generality, we can choose the same nodes and zipper as the ones on the UHP (see Fig.~\ref{UDHP}). The derivatives of the Green functions on the DUHP are then given by:
\begin{align}
\begin{split}
G'^{\textrm{op}}_{(u_1,u_2),(v_1,v_2)}&=- \, G^{\prime}_{(u_1,y+1-u_2),(v_1,y+1-v_2)}+G^{\prime}_{(u_1,y+1-u_2),(v_2,y+1-v_1)}\\
&\hspace{2cm} \quad + G^{\prime}_{(u_2,y+1-u_1),(v_1,y+1-v_2)}-G^{\prime}_{(u_2,y+1-u_1),(v_2,y+1-v_1)},
\label{Gpop_D}
\end{split}\\
\begin{split}
G'^{\textrm{cl}}_{(u_1,u_2),(v_1,v_2)}&=-\,G^{\prime}_{(u_1,y+1-u_2),(v_1,y+1-v_2)}-G^{\prime}_{(u_1,y+1-u_2),(v_2-1,y-v_1)}\\
&\hspace{2cm} \quad - \, G^{\prime}_{(u_2-1,y-u_1),(v_1,y+1-v_2)}-G^{\prime}_{(u_2-1,y-u_1),(v_2-1,y-v_1)},
\label{Gpcl_D}
\end{split}
\end{align}
where $G'$ on the full plane is computed with respect to the zipper described in Section~\ref{sec3}.

%\begin{figure}[t]
%\centering
%\begin{tikzpicture}[scale=0.65,font=\small]
%\draw (0,0) rectangle (8,8);
%\draw[step=0.5cm,dotted] (0,0) grid (8,8);
%\draw[thick] (0,0)--(0,8)--(8,8);
%\draw[thick,fill=black!10!white] (2,5.5)--(2.5,5.5)--(2.5,6)--(2,6)--(1.5,6)--(1.5,5)--(2,5)--(2,5.5);
%\filldraw (2,5.5) circle (0.075cm);
%\draw (2.3,5.2) node {$5$};
%\filldraw (2.5,5.5) circle (0.075cm);
%\draw (2.9,5.5) node {1};
%\filldraw (2,6) circle (0.075cm);
%\draw (1.75,6.4) node {2};
%\filldraw (1.5,5.5) circle (0.075cm);
%\draw (1.1,5.5) node {3};
%\filldraw (2,5) circle (0.075cm);
%\draw (2,4.6) node {4};
%\filldraw (5,8) circle (0.075cm) node [above right] {6};
%\foreach \x in {6,6.5,...,7.5} {\draw[thick,->] (2.5,\x) -- (2,\x);};
%\filldraw[white] (0,0)--(8,8)--(8.2,-0.2)--(0,0);
%\fill[pattern=horizontal lines, pattern color=gray,shift={(0.075cm,-0.075cm)}] (0,0)--(8,8)--(8.4,7.6)--(0.4,-0.4);
%\draw (6.5,2.5) node {\small $\mathbf{i\!=\!5\!=\!(0,y)}$};
%\end{tikzpicture}
%\caption{Choice of nodes and zipper for the computation of height probabilities on the diagonal upper half-plane. The zipper extends up to infinity.}
%\label{UDHP}
%\end{figure}

For the two boundary conditions, we write the results in terms of the Euclidean distance $r$ between the reference site $i=(0,y)$ and the reflection axes used in the method of images, namely $y=x$ (open boundary) and $y=x+1$ (closed boundary), so that
\begin{equation}
r = \frac{y}{\sqrt{2}}\quad\textrm{(open b.c.)},\quad r=\frac{y-1}{\sqrt{2}}\quad\textrm{(closed b.c.)}.
\end{equation}
We find that one-site probabilities on the DUHP read:
\begin{subequations}
\begin{align}
\sigma_a^{\rm op}(r)&=\frac{1}{r^2}\Big(c_a+\frac{d_a}{2}+d_a\log r\Big) - \frac{1}{r^4}\Big(\frac{c_a}{4} - \widetilde e_a +\frac{d_a}{4}\log r\Big)+\O(r^{-5}),\\
\sigma_a^{\rm cl}(r)&=-\frac{1}{r^2}\big(c_a+d_a\log r\big) + \frac{1}{r^4}\Big(\frac{c_a}{4}- \widetilde f_a+\frac{d_a}{4}\log r\Big)+\O(r^{-5}). 
\end{align}
\end{subequations} 
The dominant terms in $r^{-2}$ of $\sigma_a^{\mathrm{op,cl}}$ are identical on the UHP and the DUHP, as expected. However, the subleading terms in $r^{-4}$ differ for $a > 1$, in the numerical values of the coefficients $\widetilde e_a,\widetilde f_a$, listed in Table~\ref{UDHP_coef}, and in the overall sign, which, strangely enough, gets swapped.

\begin{table}[htbp]
\centering
\large
\renewcommand{\arraystretch}{1.8}
\tabcolsep12pt
\begin{tabular}{|c|cccc|}
\hline
& $a=1$ & $a=2$ & $a=3$ & $a=4$ \\
\hline
$\widetilde{e}_a$ & $\frac{\pi+4}{32\pi^3}$ & $\frac{-224+22\pi+9\pi^2}{384\pi^3}$ & $\frac{152-7\pi}{192\pi^3}$ & $\frac{-128-20\pi-9\pi^2}{384\pi^3}$ \\
$\widetilde{f}_a$ & $\frac{8-\pi}{32\pi^3}$ & $\frac{-368+58\pi+9\pi^2}{384\pi^3}$ & $\frac{224-13\pi}{192\pi^3}$ & $\frac{-176-20\pi-9\pi^2}{384\pi^3}$ \\
\hline
\end{tabular}
\caption{Numerical coefficients for one-site probabilities on the diagonal upper half-plane.}
\label{UDHP_coef}
\end{table}

%%%%%%%%%%%%%%%%%%%%%%%%%%%%%%%%%%%%%%%%%%%%%%%%%%%%%%%%%%%%%%%%%%%%%%%%%%%%%%%%%%%%%%%%%%%%%%%%%%%%%

\subsection{Mixed bulk-boundary two-site correlations}
\label{sec6.3}

In this first case, we compute the two-point probabilities $\P_{a,1}$, when the height $a$ is in the bulk of the UHP, far from the boundary, and the height 1 is on the boundary. For simplicity, we have considered the situation where the two heights are vertically aligned, namely $h_i=a$ at $i=(0,y)$ and $h_j=1$ at $j=(0,1)$. As before, we actually compute the correlations,
\be
\sigma_{a,1}^{\rm bound}(y) = \mathbb{E}\Big[\big(\delta_{h_{i},a}-\P_a^{}\big) \, \big(\delta_{h_{j},1}-\P_1^{\rm bound}\big)\Big] = \P_{a,1}(y) - \P_a^{}(y) \, \P_1^{\rm bound},
\ee
where the probabilities $\P_1^{\rm bound}$ that a site on the boundary of the UHP has height 1 are given by \cite{BIP93}:
\be
\P_1^{\rm op} = \frac{9}{2}-\frac{42}{\pi}+\frac{320}{3\pi^2}-\frac{512}{9\pi^3} \simeq 0.1038, \qquad 
\P_1^{\rm cl} = \frac{3}{4}-\frac{2}{\pi} \simeq 0.1134.
\ee

We skip the details of the calculations and merely give the final results:
\begin{align}
\sigma_{1,1}^{\rm op}(y)&=-\frac{2(\pi-2)(16-3\pi)(32-9\pi)}{9\pi^6 y^4}+\ldots,\\
\sigma_{2,1}^{\rm op}(y)&=-\frac{(16-3\pi)(32-9\pi)}{9\pi^6 y^4}\left\{18-6\pi+2(\pi-2)\Big(\log y+\gamma+\frac{5}{2}\log 2\Big)\right\}+\ldots,\\
\sigma_{3,1}^{\rm op}(y)&=-\frac{(16-3\pi)(32-9\pi)}{18\pi^6 y^4}\left\{-48+3\pi+\pi^2+2(8-\pi)\Big(\log y+\gamma+\frac{5}{2}\log 2\Big)\right\}+\ldots,\\
\sigma_{1,1}^{\rm cl}(y)&=-\frac{2(\pi-2)(3\pi-8)}{\pi^5 y^4}+\ldots,\\
\sigma_{2,1}^{\rm cl}(y)&=-\frac{3\pi-8}{\pi^5 y^4}\left\{20-7\pi+2(\pi-2)\Big(\log y+\gamma+\frac{5}{2}\log 2\Big)\right\}+\ldots,\\
\sigma_{3,1}^{\rm cl}(y)&=-\frac{3\pi-8}{2\pi^5 y^4}\left\{-56+4\pi+\pi^2+2(8-\pi)\Big(\log y+\gamma+\frac{5}{2}\log 2\Big)\right\}+\ldots
\end{align}

%%%%%%%%%%%%%%%%%%%%%%%%%%%%%%%%%%%%%%%%%%%%%%%%%%%%%%%%%%%%%%%%%%%%%%%%%%%%%%%%%%%%%%%%%%%%%%%%%%%%%

\subsection{Bulk two-site correlations}
\label{sec6.4}

Here we consider the two-site probabilities $\P_{a,1}(i,j)$ when both heights $h_i=a$ and $h_j=1$ are in the bulk of the UHP, far from the boundary and far from each other. Again, for simplicity, we choose the two sites to be aligned vertically, $i=(0,y_1)$ and $j=(0,y_2)$, with $y_2>y_1$. We have computed, to total order 4 in the inverse distances, the correlations defined as
\be
\sigma_{a,1}(y_1,y_2) = \mathbb{E}\Big[\big(\delta_{h_{i},a}-\P_a^{}\big) \, \big(\delta_{h_{j},1}-\P_1\big)\Big] 
= \P_{a,b}(y_1,y_2)-\P_a(y_1)\P_b-\P_a\P_b(y_2)+\P_a\P_b,
\ee
and for the two boundary conditions, open and closed. In terms of the two functions
\be
P(y_1,y_2) = \frac{1}{8 y_1^2 y_2^2}-\frac{1}{(y_1-y_2)^4}-\frac{1}{(y_1+y_2)^4}, \qquad Q(y_1,y_2) = \frac{1}{(y_1-y_2)^4}-\frac{1}{(y_1+y_2)^4},
\ee
we have found the following results,
{\small
\begin{align}
\sigma_{1,1}^{\rm op}(y_1,y_2)&=\frac{2(\pi-2)^2}{\pi^6} P(y_1,y_2) + \ldots,
\label{P11uhpop}\\
\sigma_{2,1}^{\rm op}(y_1,y_2)&=\frac{2(\pi-2)^2}{\pi^6}\left[P(y_1,y_2) \Big(\log y_1+\gamma+\frac{5}{2}\log 2\Big) + Q(y_1,y_2)\log\left|\frac{y_2+y_1}{y_2-y_1}\right|\right]
\nonumber \\
&\quad - \frac{\pi-2}{16\pi^6 y_1^2 y_2^2 (y_1^2-y_2^2)^4}\Big[3(3\pi-10)(y_1^8+y_2^8)+12(58-19\pi)y_1^6 y_2^2\nonumber\\
&\hspace{4cm}+6(482-151\pi)y_1^4 y_2^4+4(142-41\pi)y_1^2 y_2^6\Big]+\ldots,
\label{P21uhpop}\\
\sigma_{3,1}^{\rm op}(y_1,y_2)&=\frac{(\pi-2)(8-\pi)}{\pi^6}\left[P(y_1,y_2) \Big(\log y_1+\gamma+\frac{5}{2}\log 2\Big) + Q(y_1,y_2)\log\left|\frac{y_2+y_1}{y_2-y_1}\right|\right] \nonumber\\
&\quad-\frac{\pi-2}{32\pi^6 y_1^2 y_2^2 (y_1^2-y_2^2)^4}\Big[(2\pi^2+3\pi-72)(y_1^8+y_2^8) - 4(10\pi^2+27\pi-456)y_1^6 y_2^2 \nonumber \\
&\hspace{3cm}-6(30\pi^2+61\pi-1208)y_1^4 y_2^4-4(10\pi^2+11\pi-328)y_1^2 y_2^6\Big] + \ldots
\end{align}
}
for the open boundary condition, and
{\small
\begin{align}
\sigma_{1,1}^{\rm cl}(y_1,y_2)&=\frac{2(\pi-2)^2}{\pi^6} P(y_1,y_2) + \ldots,
\label{P11uhpcl}
\\
\sigma_{2,1}^{\rm cl}(y_1,y_2)&=\frac{2(\pi-2)^2}{\pi^6}\left[P(y_1,y_2) \Big(\log y_1+\gamma+\frac{5}{2}\log 2\Big) + Q(y_1,y_2)\log\left|\frac{y_2+y_1}{y_2-y_1}\right|\right]
\nonumber \\
&\quad -\frac{\pi-2}{16\pi^6 y_1^2 y_2^2 (y_1^2-y_2^2)^4}\Big[(11\pi-34)(y_1^8+y_2^8) + 4(194-67\pi)y_1^6 y_2^2\nonumber\\
&\hspace{4cm}+2(1498-479\pi)y_1^4 y_2^4 + 4(98-19\pi)y_1^2 y_2^6\Big]+\ldots,
\label{P21uhpcl}
\displaybreak[0]\\
\sigma_{3,1}^{\rm cl}(y_1,y_2)&=\frac{(\pi-2)(8-\pi)}{\pi^6}\left[P(y_1,y_2) \Big(\log y_1+\gamma+\frac{5}{2}\log 2\Big) + Q(y_1,y_2)\log\left|\frac{y_2+y_1}{y_2-y_1}\right|\right] \nonumber\\
&\quad-\frac{\pi-2}{32\pi^6 y_1^2 y_2^2 (y_1^2-y_2^2)^4}\Big\{(\pi+8)(2\pi-11)(y_1^8+y_2^8) - 4(-536+37\pi+10\pi^2)y_1^6 y_2^2 \nonumber\\
&\hspace{3cm} -2(90\pi^2+209\pi-3832)y_1^4 y_2^4 - 4(10\pi^2-11\pi-152)y_1^2 y_2^6 \Big] + \ldots
\end{align}
}%
for the closed boundary condition. As usual, $\sigma_{4,1}=-\sigma_{1,1}-\sigma_{2,1}-\sigma_{3,1}$ for both boundary conditions.

%%%%%%%%%%%%%%%%%%%%%%%%%%%%%%%%%%%%%%%%%%%%%%%%%%%%%%%%%%%%%%%%%%%%%%%%%%%%%%%%%%%%%%%%%%%%%%%%%%%%%

\section{Conformal field theory}
\label{sec7}

The Abelian sandpile model is believed to be described in the scaling limit by a conformal field theory. The reference \cite{MD92} was the first one to suggest the value of its central charge, namely $c=-2$. Since then, this value as well as many more refined aspects have been examined, largely confirming the validity of the conformal point of view. A recent review devoted to this question can be found in \cite{Rue13}.

According to this view, the various degrees of freedom of the discrete model go over to specific conformal fields in such a way that the continuum limit of the discrete correlations yield the corresponding field-theoretical correlations. In the favorable cases, the fundamental or most natural microscopic variables of the discrete model are associated with primary conformal fields. A number of identifications of this type have been proposed for the sandpile model, and these include the height variables, relevant for the lattice correlations computed in the previous sections.

The height variable at every nonboundary site is a microscopic random variable taking the four integer values from 1 to 4. As suggested by the calculations of the previous sections, one does not consider the random height variables themselves, but, at each site, the indicator functions $\delta_{h(i),a}, \, a=1,2,3,4$, of the four fixed-height events. For reasons explained above, one subtracts by their own expectation value on the infinite discrete plane, and therefore one considers
\be
h_a(i) \equiv \delta_{h(i),a} - \P_a \qquad \hbox{for }a=1,2,3,4.
\label{fixedheight}
\ee
On the lattice, they define four ``fixed-height variables'', not all independent since they satisfy the obvious linear relation $\sum_a h_a(i) = 0$ at each site. This decomposition into fixed values is not usual in lattice models, but in the present case, it reveals components of fundamentally different nature and therefore provides an enriched perspective. The various correlators computed in previous sections are simply equal to the expectation values of products of fixed-height variables, $\sigma_{a_1,a_2,\ldots}(i_1,i_2,\ldots) = {\mathbb E}\big[h_{a_1}(i_1) \, h_{a_2}(i_2) \ldots\big].$ 

The basic definition \eqref{fixedheight} of the fixed-height variables can easily be adapted to boundary sites (with specific boundary conditions). It turns out that the bulk conformal fields $h_a(z,\bar z)$ associated with the height variables $h_a(i)$ far from boundaries are more complicated (reflecting in a way the difficulty of computing their joint probabilities on the lattice). In light of the results for one-point functions on the upper half-plane, the nature of the four fields was first conjectured in \cite{PR05b}, and then completed in \cite{JPR06}; this conjecture has been found to be consistent with all subsequent calculations. It can be formulated as follows.

\bigskip \noindent
{\it
The bulk fixed-height fields $h_a(z,\bar z)$ can be identified with specific fields in a nonchiral indecomposable staggered Virasoro module at $c=-2$, containing a logarithmic (Jordan) pair of conformal fields $(\phi,\psi)$ of weights $(1,1)$. The field $\phi$ is primary and left- and right-degenerate at level 2, while its logarithmic partner $\psi$ has the following infinitesimal conformal transformations, which involve two additional fields $\rho$ and $\bar \rho$,
\begin{subequations}
\be
L_0 \psi=\bar{L}_0 \psi = \psi+\lambda \phi,\quad L_1 \psi=\rho,\quad \bar{L}_1 \psi=\bar{\rho},\quad L_{n>1} \psi=\bar{L}_{n>1} \psi=0.
\label{Lopsi}
\ee
The fields $\rho$ and $\bar\rho$ are respectively left-primary and right-primary of weights $(0,1)$ and $(1,0)$,
\be
L_{n\ge 0}\,\rho=\bar{L}_{n\ge 0}\,\bar{\rho}=0,\quad\bar{L}_0\rho=\rho,\quad L_0\bar{\rho}=\bar{\rho},\quad\bar{L}_{n\ge 1}\,\rho=L_{n\ge 1}\,\bar{\rho}=\kappa\mathbb{I}\,\delta_{n,1}.
\label{Lnrho}
\ee
Two additional relations further characterize the module:
\be
L_{-1}\rho=\bar{L}_{-1}\bar{\rho}=\beta\lambda \phi, \qquad\quad \beta=\frac12.
\label{betaeq}
\ee
\label{conj}%
\end{subequations}
Up to normalizations, the height-one field $h_1$ can be identified with $\phi$, each of the other three $h_2, h_3, h_4$ with (a specific choice of) $\psi$.
}

\bigskip\medskip
The constants $\lambda$ and $\kappa$ are related to the normalizations of $\phi$ and $\psi$. The nature of the parameter $\beta$, however, is quite different, see below. Beyond the choice of normalization of $\phi$ and $\psi$, one has the freedom to redefine $\psi$ by adding to it an arbitrary multiple of $\phi$, since the above structural relations will be preserved. It is in this sense that the last statement in the conjecture should be understood: $h_2,h_3,h_4$ are proportional to each other up to a multiple of $\phi$.

In the following, we use the notation $h_a(z,\bar z)$ for the {\it normalized} fields arising from the scaling limit of the lattice height variables $h_a(i)$, their normalizations being directly inherited from those of the lattice variables. To relate them to the fields belonging to the representation described above, we arbitrarily fix the normalizations of $\phi$ and $\psi$ by requiring the strict identities 
\begin{subequations}
\be
\phi(z,\bar z) \equiv h_1(z,\bar z), \qquad \psi(z,\bar z) \equiv h_2(z,\bar z),
\label{h12}
\ee
which also fix the values of $\lambda$ and $\kappa$ to 
\be
\lambda=-\frac12\, \qquad \kappa = -\frac{\P_1}4.
\ee 
The other two height variables are then (conjecturally) given by the following combinations \cite{JPR06},
\bea
h_3(z,\bar z) \egal \frac{8-\pi}{2(\pi-2)} \; \psi(z,\bar z) + \frac{\pi^3-5\pi^2+12\pi-48}{4(\pi-2)^2} \; \phi(z,\bar z),\label{h3}\\
\noalign{\medskip}
h_4(z,\bar z) \egal -\frac{\pi+4}{2(\pi-2)} \; \psi(z,\bar z) + \frac{32+4\pi+\pi^2-\pi^3}{4(\pi-2)^2} \; \phi(z,\bar z).
\label{h4}
\eea
\label{hs}%
\end{subequations}
Let us note that the values of these coefficients imply that the height field, defined as the scaling limit of the random height variable, and equal to $h(z,\bar z) = h_1 + 2h_2 + 3h_3 + 4h_4 = h_2 + 2h_3 + 3h_4$, is itself a logarithmic field of type $\psi$.

We see that the fixed-height variables belong to the conformal representation that contains the identity. It was shown \cite{PR04} that the identity in the bulk has itself a logarithmic partner $\omega$, satisfying
\be
L_{n \ge 1} \, \omega = \bar L_{n \ge 1} \, \omega = 0, \qquad L_0 \, \omega = \bar L_0 \, \omega = -\frac1{4\pi} {\mathbb I}.
\label{omega}
\ee
It is normalized by requiring that $\la \omega(z,\bar z) \ra = 1$. On the lattice, $\omega(z,\bar z)$ was identified as corresponding to the insertion of dissipation at $z$; in fact, as we shall briefly recall, it plays a central role in the understanding of fixed-height correlators. More details (multisite correlators of $\omega$, fusion and dissipation field $\omega_b$ at a closed boundary site) can be found in \cite{PR04}.

In turn $\omega$ is related to the fields discussed above; indeed, one might suspect that $\rho$ and $\bar\rho$ are actually its two descendants at level 1. Earlier calculations on the lattice \cite{PR04} are consistent with the following relations,
\be
\rho = \delta \, \bar L_{-1}\omega, \quad \bar\rho = \delta \, L_{-1}\omega, \quad \phi = -4\delta \,L_{-1}\bar L_{-1} \omega, \qquad \delta = \frac{\pi-2}{\pi^2} = \frac\pi2 \P_1.
\ee
Therefore, the height-one field $h_1(z,\bar z)$ would be a level-2 descendant of the dissipation field, $h_1 = -4\delta \, \partial\bar \partial \omega$. Some of the lattice calculations of two-point correlators presented in Section~\ref{sec6} confirm these relations.

{\it Chiral} staggered modules of the kind discussed above were studied in \cite{GK96}, where it was noted that different values of $\beta$ correspond to different equivalence classes of such modules (see \cite{KR09} for a mathematical analysis attempting a classification of staggered modules). For nonchiral modules, no classification result is known; one however knows that the bosonic sector of the free symplectic fermions \cite{GK99} realizes a nonchiral representation\footnote{That representation is in fact a module for a W-algebra, and as such, is larger. For instance, it contains four logarithmic pairs of fields with weights $(1,1)$, whereas here only one has been identified so far (of course, this does not rule out the possibility to find more). However, if one restricts to the action of the Virasoro modes only, one finds the relations \eqref{conj}.}  satisfying the same structural relations \eqref{conj} but with a different value of $\beta$, namely $\beta=-1$ (\hspace{-0.025cm}\cite{JPR06} to see the explicit realization). Since the chiral restrictions of the symplectic representation and of the module discussed in the above conjecture are not isomorphic, it is expected that they are not isomorphic either as nonchiral modules.

The scaling limit of fixed-height variables on a boundary has also been discussed. In this case, it depends on the boundary condition and was studied in \cite{PR05a} for the open and closed boundary conditions. The results are simpler than in the bulk. It was found that in the continuum limit, the boundary fixed-height fields are chiral fields of scaling dimension 2, which are neither logarithmic nor primary. In particular, the height-one field, which is the only one involved in the calculations of Section~\ref{sec6}, is proportional to the stress-energy tensor for either boundary condition,
\be
h_1^{\rm op}(x) = \frac{(16-3\pi)(32-9\pi)}{9\pi^3}\,T(x), \qquad h_1^{\rm cl}(x) = -\frac{3\pi-8}{\pi^2}\,T(x).
\label{h1bord}
\ee
We refer to \cite{PR05a} for more details about these identifications and those of the other height fields.

What we want to do in the rest of this section is to focus on the bulk height fields to see whether the new correlations functions computed in Section~\ref{sec6} are consistent with the conjectured identifications stated in \eqref{hs}. 

One part of the conjecture can be easily verified, namely the fact that the field $h_3$ is a linear combination of $h_1$ and $h_2$ (the same then follows for $h_4$ in view of the relation $\sum_a h_a = 0$). Indeed, it is a simple matter to see by inspection that the correlators that involve the height 3 satisfy this linear relation {\it at dominant order}. This can be explicitly checked on the two-site and three-site  correlators on the plane in Section~\ref{sec5}, and on the one-site and two-site correlators on the upper half-plane in Section~\ref{sec6}. The lattice one-site functions on the upper half-plane, given in \eqref{Pa_uhp}, make it clear that the linear relation does not hold beyond the dominant order.

The part of the conjecture concerned with the conformal transformations of the fields $\phi$ and $\psi$ is much deeper and implies strong constraints on the functional form of the lattice correlators that involve the heights 1 and 2. The one-point functions $\sigma_a^{\rm op,cl}(y)$ on the upper half-plane were at the basis of the conjecture. For each boundary condition taken separately, these functions are clearly consistent with $\psi$ and $\phi$ being a logarithmic pair. However, the specific way $\sigma_a^{\rm op}(y)$ and $\sigma_a^{\rm cl}(y)$ are related at dominant order---they depend on the same coefficients, see \eqref{1ptop} and \eqref{1ptcl}---can actually be computed from conformal field theory, by using the field switching between the open and closed boundary conditions, identified as a primary field of weight $-1/8$ \cite{Rue02}. It was done in \cite{JPR06} and provides a highly nontrivial and convincing check of the conjecture. 

The two-point correlators $\sigma_{1,1}(\vec r)$ and $\sigma_{2,1}(\vec r)$ on the plane were discussed in \cite{PGPR08,PGPR10} and, in the scaling limit, confirm the logarithmic partnership of $\phi$ and $\psi$. We shall briefly rediscuss these two-point functions to understand how their form should be understood, on the conformal side, not as two- but as three-point correlators where the extra background field $\omega(\infty)$ is to be inserted. This feature will be crucial to understand the rather unusual form of the lattice three-point correlator $\sigma_{2,1,1}(\vec r_{12},\vec r_{13})$. Interpreted as a conformal nonchiral four-point function, we shall show that its unexpected functional form follows naturally from the above conjecture {\it and} the logarithmic nature of the background field $\omega(\infty)$. 

Being related to chiral four-point functions, the lattice two-point correlators $\sigma_{a,1}^{\rm op,cl}(y)$ on the upper half-plane computed in Section~\ref{sec6} offer further opportunities to verify the consistency of the conformal framework provided by the conjecture.

%%%%%%%%%%%%%%%%%%%%%%%%%%%%%%%%%%%%%%%%%%%%%%%%%%%%%%%%%%%%%%%%%%%%%%%%%%%%%%%%%%%%%%%%%%%%%%%%%%%%%

\subsection{Correlations on the plane}
\label{sec7.1}

At dominant order, the lattice two-point correlators on the plane reported in Section~\ref{sec5}, 
\be
\sigma_{1,1}(\vec r) = \frac{a}{r^4} + \ldots, \qquad \sigma_{2,1}(\vec r) = \frac{1}{r^4} \, (a \log{r} + b) + \ldots
\label{latt2}
\ee
look familiar, including the distinctive presence of the same coefficient $a$ in the two correlators. It is however well known \cite{Flo03} that the self-correlations of the primary field of a logarithmic pair are all identically zero, implying in particular $\la \phi(1)\phi(2) \ra \equiv 0$. As the coefficient $a$ computed on the lattice is not zero ($a=-\P_1^2/2$), the correlators $\sigma_{1,1}$ and $\sigma_{2,1}$ cannot correspond to conformal two-point functions. 

The way out was discussed in \cite{PR04} and has a clear physical interpretation. The probabilities computed on the infinite lattice are limits of similar quantities formulated at finite volume. The formulation of the sandpile dynamics is well defined at finite volume provided it involves dissipation of sand (to the sink), here chosen to be located on the boundaries of the grid (the dissipative sites). The infinite volume limit of multisite probabilities remains well defined, but the dissipation is sent off to infinity. On the other hand, the conformal field theory is defined right away on the infinite plane, and has no trace of the necessary dissipation. Therefore, dissipation has to be inserted at infinity by hand, in the form of a background insertion $\omega(\infty)$. Its conformal weights (0,0) do not alter the dimensions of the correlators.

Therefore, the lattice correlators in \eqref{latt2} should correspond respectively to the following three-point functions,
\begin{equation*}
\la \phi(z_1,\bar z_1) \phi(z_2,\bar z_2) \omega(\infty) \ra \qquad {\rm and} \qquad \la \psi(z_1,\bar z_1) \phi(z_2,\bar z_2) \omega(\infty) \ra.
\end{equation*}
Indeed, one finds that these three-point functions reproduce the form given in \eqref{latt2} provided the two-point functions $\la \phi(z_1,\bar z_1) \phi(z_2,\bar z_2) \ra$ and $\la \psi(z_1,\bar z_1) \phi(z_2,\bar z_2) \ra$ vanish identically \cite{PR05b}. This is again very natural from the sandpile point of view, since these two-point functions would correspond to fractions of recurrent configurations with fixed heights at sites 1 and 2 in a model with no dissipation; such models are known to have no recurrent configurations at all \cite{Dha06}. In what follows, we use this argument to set all correlations involving only $\phi$'s and $\psi$'s to zero.

Let us now come to the three-point correlators. When the three insertion points are aligned (horizontally or vertically), we recall that these are given at dominant order (terms of dimension 6) by
\be
\sigma_{1,1,1}(\vec r_{12},\vec r_{13}) = 0 + \ldots, \qquad \sigma_{2,1,1}(\vec r_{12},\vec r_{13}) = \frac{(\pi-2)^3}{\pi^9} \, \frac1{r_{12}^3 r_{13}^3} + \ldots
\label{latt3}
\ee

It is surprising, in the sandpile model, that the correlator of three heights 1 vanishes (at dominant order). To a large extent, this can be understood in the conformal picture. The four-point function $\la \phi(z_1,\bar z_1) \phi(z_2,\bar z_2) \phi(z_3,\bar z_3) \omega(\infty) \ra$ should be associated with the scaling limit of the correlator $\sigma_{1,1,1}$. From the vanishing of $\la \phi(z_1,\bar z_1) \phi(z_2,\bar z_2) \phi(z_3,\bar z_3) \ra$, the four-point function can be computed as if $\omega$ is primary of weights (0,0) (not assumed to be degenerate at level 1 though, like the identity). However, together the Ward identities and the degeneracy conditions of $\phi$ at level 2 in the theory with central charge $c=-2$, namely $(L_{-1}^2 - 2L_{-2})\phi = (\bar L_{-1}^2 - 2\bar L_{-2})\phi = 0$, offer no solution that is fully symmetric under the permutations of sites 1, 2 and 3.

Let us now examine the other three-point correlator $\sigma_{2,1,1}$ in light of the nonchiral four-point function $\la \psi(z_1,\bar z_1) \phi(z_2,\bar z_2) \phi(z_3,\bar z_3) \omega(\infty)\ra$. As is usually the case for a correlation involving logarithmic fields, the calculation proceeds in several steps \cite{Flo03}. Finite regular conformal transformations of $\omega$ involve $\omega$ itself and the identity, whereas those of $\psi$ involve $\psi$ itself, $\phi$, $\rho$, $\bar\rho$ and the identity. This makes a total of 9 four-point functions to be computed before the last one $\la \psi(1)\phi(2)\phi(3)\omega(\infty)\ra$ can be obtained. However, many of them either vanish or are already known:
\begin{subequations}
\bea
&& \hspace{-20mm}\la \psi(1)\phi(2)\phi(3) \ra = \la \phi(1)\phi(2)\phi(3) \ra = \la \rho(1)\phi(2)\phi(3) \ra = \la \bar\rho(1)\phi(2)\phi(3) \ra = \la \phi(2)\phi(3) \ra = 0, \\
\noalign{\medskip} 
&& \hspace{-20mm} \la \phi(1)\phi(2)\phi(3) \omega(\infty) \ra = 0, \qquad \la \phi(2)\phi(3)\omega(\infty) \ra = \frac{C_{\phi\phi}}{|z_2-z_3|^4}, \quad C_{\phi\phi} = -\frac{\P_1^2}2.
\eea
\end{subequations}
In what follows, we summarize the calculations of the remaining three correlations
\begin{equation*}
\la \rho(1)\phi(2)\phi(3)\omega(4)\ra,\,\,\la \bar\rho(1)\phi(2)\phi(3)\omega(4)\ra\,\,\text{and}\,\,\la \psi(1)\phi(2)\phi(3)\omega(4)\ra
\end{equation*}
for general positions. To simplify the analysis, we retain only the most general solutions that are symmetric under the exchange of sites 2 and 3 {\it in the limit where the site 4 goes to infinity}.

The integration of the infinitesimal transformations given in \eqref{Lopsi} and \eqref{Lnrho} yields the following transformation rules of $\rho,\phi$ and $\omega$ under $w \to z(w), \, \bar w \to \bar z(\bar w)$ \cite{JPR06},
\begin{subequations}
\bea
\omega(z,\bar z) \egal \omega(w,\bar w) - \frac1{4\pi} \log\Big|\frac{{\rm d}w}{{\rm d}z}\Big|^2, \qquad 
\phi(z,\bar z) = \Big|\frac{{\rm d}w}{{\rm d}z}\Big|^2 \, \phi(w,\bar w),\\
\noalign{\medskip}
\rho(z,\bar z) \egal \frac{{\rm d}\bar w}{{\rm d}\bar z} \, \rho(w,\bar w) + \frac\kappa2 \Big(\frac{{\rm d}^2 \bar w}{{\rm d}\bar z^2}\Big/\frac{{\rm d} \bar w}{{\rm d}\bar z}\Big).
\eea
\end{subequations}
The M\"obius transformations (with $z_{ij}\equiv z_i-z_j$)
\be
w(z) = \frac{(z_1 - z)z_{34}}{z_{13}(z - z_4)}, \qquad \bar w(\bar z) = \frac{(\bar z_1 - \bar z)\bar z_{34}}{\bar z_{13}(\bar z - \bar z_4)},
\ee
map the four points $z_1,z_2,z_3$ and $z_4$ to $0,x =(z_{12}z_{34})/(z_{13}z_{24}),1$ and $\infty$ respectively (and likewise for the conjugate variables). The above transformation rules yield the following form:
\be
\la \rho(1)\phi(2)\phi(3)\omega(4) \ra = -\frac{\bar z_{34}}{\bar z_{13}\bar z_{14}} \,
\frac1{|z_{23}|^4} \: \Big\{F(x,\bar x) + \kappa C_{\phi\phi} \, \frac{\bar z_{13}}{\bar z_{34}}  \Big\},
\label{rhophiphi}
\ee
with $F(x,\bar x) = |1-x|^4 \:\la \rho(0,0)\phi(x,\bar x)\phi(1,1)\omega(\infty) \ra$.

Further constraints on $F(x,\bar x)$ come from the left and right degeneracy conditions of $\phi$ at level 2. The left degeneracy condition of $\phi(2)$ (or of $\phi(3)$) yields a homogeneous differential equation in $x$, namely
\be
x(1-x) \,\partial^2 F + 2 \,\partial F = 0.
\ee
It can be strengthened in the following way. The correlator $\la (L_{-1}\rho)(1)\phi(2)\phi(3)\omega(4) \ra$ is proportional to $\la \phi(1)\phi(2)\phi(3)\omega(4) \ra$, which is identically zero. Therefore $\la \rho(1)\phi(2)\phi(3)\omega(4) \ra$ cannot depend on $z_1$, and so $F(x,\bar x)$ cannot depend on $x$, $\partial F = 0$.

The right degeneracy of $\phi$ actually delivers two independent, inhomogeneous equations, depending on whether we write it for $\phi(2)$ or for $\phi(3)$. They combine to give a first-order equation,
\be
\overline{\cal D} F(x,\bar x) = -\kappa C_{\phi\phi}\frac{1-\bar x}{\bar x^2}\,, \qquad \overline{\cal D} \equiv \bar \partial + \frac{\bar x+2}{\bar x(1-\bar x)}.
\label{DR}
\ee
The general solution is obtained from a particular solution of the inhomogeneous equation, for instance $-\kappa C_{\phi\phi} (1-\bar x)^2/\bar x^2$, and an arbitrary element of $\ker \, \overline{\cal D}$, given by $A (1-\bar x)^3/\bar x^2$. Inserting this general form into \eqref{rhophiphi}, taking the limit $z_4,\bar z_4 \to \infty$ and requiring that the result be symmetric under $2 \leftrightarrow 3$ fix $A=\kappa C_{\phi\phi}/2$ and lead to the unique solution,
\be
F(x,\bar x) = -\kappa C_{\phi\phi} \, \frac{(1+\bar x)(1-\bar x)^2}{2\bar x^2}.
\label{F}
\ee

The correlation involving $\bar\rho$ instead of $\rho$ is similar since the transformations of $\bar \rho$ are those of $\rho$ with left and right exchanged. The corresponding correlation is simply obtained by exchanging the variables with their conjugates,
\be
\la \bar\rho(1)\phi(2)\phi(3)\omega(4) \ra = \kappa C_{\phi\phi} \, \frac{z_{34}}{z_{13} z_{14}} \,
\frac1{|z_{23}|^4} \: \Big\{\frac{(1+x)(1-x)^2}{2x^2} - \frac{z_{13}}{z_{34}}  \Big\}.
\ee

Finally, the same procedure may be used to compute the last correlator. To do this, we use the finite transformation law of $\psi$ under $w \to z(w)$, $\bar w \to \bar z(\bar w)$ \cite{JPR06},
\bea
\psi(z,\bar z) \egal \Big|\frac{{\rm d}w}{{\rm d}z}\Big|^2 \, \Big[\psi(w,\bar w) + \log\Big|\frac{{\rm d}w}{{\rm d}z}\Big|^2 \: \phi(w,\bar w) \Big] + \frac12 \Big(\frac{{\rm d}^2 w}{{\rm d}z^2}\Big/\frac{{\rm d}w}{{\rm d}z}\Big) \, \frac{{\rm d}\bar w}{{\rm d}\bar z} \, \rho(w,\bar w) \nonumber\\
\noalign{\medskip}
&& \hspace{7mm} + \; \frac12 \frac{{\rm d}w}{{\rm d}z} \, \Big(\frac{{\rm d}^2 \bar w}{{\rm d}\bar z^2}\Big/\frac{{\rm d} \bar w}{{\rm d}\bar z}\Big)\, \bar\rho(w,\bar w) + \frac{\kappa}4 \Big(\frac{{\rm d}^2 w}{{\rm d}z^2}\Big/\frac{{\rm d}w}{{\rm d}z}\Big)\Big(\frac{{\rm d}^2 \bar w}{{\rm d}\bar z^2}\Big/\frac{{\rm d} \bar w}{{\rm d}\bar z}\Big).
\label{finitepsi}
\eea
The same M\"obius transformation as before yields
\be
\la \psi(1)\phi(2)\phi(3)\omega(4) \ra = \frac1{|z^{}_{14}z_{23}^2|^2} \, \Big\{\Big|\frac{z_{34}}{z_{13}}\Big|^2 G(x,\bar x) + \frac{\bar z_{34}}{\bar z_{13}} \, F(x,\bar x) + \frac{z_{34}}{z_{13}} \, F(\bar x,x) + \kappa C_{\phi\phi} \Big\},
\label{psiphiphi}
\ee
with $G(x,\bar x) = |1-x|^4 \:\la \psi(0,0)\phi(x,\bar x)\phi(1,1)\omega(\infty) \ra$ and for the function $F(x,\bar x)$ given in \eqref{F}.

The left and right degeneracy conditions of $\phi(2)$ and $\phi(3)$ each give two differential equations for $G(x,\bar x)$; so four equations in total, two in $x$ and two in $\bar x$. They combine to give two first-order equations, very close to the one encountered above, as they involve the same differential operator (in $x$ or $\bar x$),
\begin{subequations}
\bea
{\cal D} G(x,\bar x) \egal -\frac{1-x}{x^2} \, F(x,\bar x),\\
\noalign{\smallskip}
\overline{\cal D} G(x,\bar x) \egal -\frac{1-\bar x}{\bar x^2}\,F(\bar x,x),
\eea
\end{subequations}
Their compatibility is readily established on account of the right equation \eqref{DR} satisfied by $F(x,\bar x)$.

The general solution depends on a free parameter $B$, related to the choice of an arbitrary element in $\ker \{{\cal D},\overline{\cal D}\}$,
\be
G(x,\bar x) = \frac12 \kappa C_{\phi\phi} \, (x + \bar x) \, \Big|\frac{1-x}{x}\Big|^4 + B \, \Big|\frac{(1-x)^3}{x^2}\Big|^2.
\ee
Inserted in \eqref{psiphiphi}, the limit $z_4,\bar z_4 \to \infty$ yields a result that is symmetric in $2 \leftrightarrow 3$, for any value of $B$,
\be
\la \psi(1)\phi(2)\phi(3)\omega(\infty) \ra = \frac12 \kappa C_{\phi\phi} \, \frac1{|z_{12}z_{13}|^2} \, \Big[ \frac1{z_{13}\bar z_{12}} + \frac1{z_{12}\bar z_{13}} \Big] + B \, \Big|\frac{z_{23}}{z_{12}^2 z_{13}^2}\Big|^2.
\label{psipp}
\ee

We note that the term proportional to $B$ has a singularity in $|z_{12}|^{-4}$, namely the most singular term possible in the fusion $\psi(1)\phi(2)$. The only fields that can enter the fusion at this order are the identity and $\omega$. The vanishing of $\la \psi(1)\phi(2) \ra$ forbids the presence of $\omega$, while the correlator $\la \psi(1)\phi(2)\omega(\infty) \ra = (a \log{|z_{12}|} + b)/|z_{12}|^4$ shows that the identity enters the fusion with a logarithmic term. Since the last term in \eqref{psipp} contains no logarithm, we conclude that $B=0$.

When the three fields are aligned horizontally, the variables $z_{ij}$ are real, so that the previous correlator reduces to
\be
\la h_2(\vec r_1) \, h_1(\vec r_2) \, h_1(\vec r_3) \, \omega(\infty) \ra = \kappa C_{\phi\phi} \, \frac1{r_{12}^3 r_{13}^3}.
\ee
With $\kappa = -\P_1/4$ and $C_{\phi\phi} = -\P_1^2/2$ (see above), we obtain $\kappa C_{\phi\phi} = \P_1^3/8 = (\pi-2)^3/\pi^9$, which exactly reproduces the dominant order of $\sigma_{2,1,1}(\vec r_{12},\vec r_{13})$, given in \eqref{sigma211}.

%%%%%%%%%%%%%%%%%%%%%%%%%%%%%%%%%%%%%%%%%%%%%%%%%%%%%%%%%%%%%%%%%%%%%%%%%%%%%%%%%%%%%%%%%%%%%%%%%%%%%

\subsection{Bulk-boundary correlations}

Section~\ref{sec6.3} gave the results of our lattice computations of the two-point correlations $\sigma_{a,1}^{\rm op,cl}$ of two heights, the height $a$ being in the bulk of the upper half-plane, the height 1 on the boundary, taken to be either open or closed. For simplicity, the height $a$ was located right above the height 1, at a distance $y$ from the boundary.

For general positions $x$ (real) and $z$, these mixed bulk-boundary correlators should correspond to the following conformal correlators, at dominant order,
\be
\sigma_{a,1}^{\rm op}(z,\bar z;x) = \la h_a(z,\bar z) \, h_1^{\rm op}(x) \ra_{\rm op} + \ldots, \quad 
\sigma_{a,1}^{\rm cl}(z,\bar z;x) = \la h_a(z,\bar z) \, h_1^{\rm cl}(x) \, \omega(\infty) \ra_{\rm cl} + \ldots
\ee
Dissipation at infinity has been inserted in the second correlator; no such insertion is needed when the boundary is open, since each boundary site is dissipative. As before, one may concentrate on $a=1,2$.

The boundary height-one field being proportional to the stress-energy tensor, the $\sigma_{a,1}^{\rm op,cl}(z,\bar z;x)$'s are related to the one-point functions $\sigma_{a}^{\rm op,cl}(z,\bar z)$. These are known from \cite{JPR06} and have been recalled in (the dominant terms of) \eqref{1ptop} and \eqref{1ptcl}. When the boundary is open, the UHP stress-energy tensor can be thought of as the sum of the left and right chiral tensors \cite{Car84}. Using this property together with the chiral conformal transformations in \eqref{conj}, we find that for $a=1,2$,
\bea
\hspace{-1cm}\langle T(x) \, h_a(z,\bar z) \rangle_{\rm op} \egal \Big\{\frac{1}{(x-z)^2} + \frac{1}{x-z}\partial_z +\frac{1}{(x-\bar{z})^2} + \frac{1}{x-\bar{z}}\partial_{\bar{z}}\Big\} \la h_a(z,\bar{z}) \ra_{\rm op} \nonumber\\
\noalign{\medskip}
\plus \Big\{ \frac{\la \rho(z,\bar z) \ra_{\rm op}}{(x-z)^3} + \frac{\la \bar\rho(z,\bar z) \ra_{\rm op}}{(x-\bar z)^3} - \frac{\la h_1(z,\bar z) \ra_{\rm op}}{2(x-z)^2} - \frac{\la h_1(z,\bar z) \ra_{\rm op}}{2(x-\bar z)^2}\Big\} \, \delta_{a,2}.
\label{Tha}
\eea

To complete the calculation, we need to evaluate $\la \rho(z,\bar z) \ra_{\rm op}$ and $\la \bar\rho(z,\bar z) \ra_{\rm op}$. The quickest way to compute them is by using the relations $\rho = \delta \bar L_{-1}\omega$, $\bar\rho = \delta L_{-1}\omega$ as well as 
\be
\la \omega(z,\bar z) \ra_{\rm op} = \frac1{2\pi} \log{|z-\bar z|} + \gamma_0,
\ee 
with $\gamma_0=\frac1{2\pi}(\gamma+\frac32 \log{2})+1$ and $\gamma=0.577216...$ the Euler constant \cite{PR04}. Hence, we have
\begin{subequations}
\bea
&&\hspace{-1cm} \la \rho(z,\bar z) \ra_{\rm op} = -\frac{d_2}{z-\bar z}, \quad \la \bar\rho(z,\bar z) \ra_{\rm op} = \frac{d_2}{z-\bar z},\\
\noalign{\smallskip}
&&\hspace{-1cm} \la h_1(z,\bar z) \ra_{\rm op} = -\frac{4d_2}{(z-\bar z)^2}, \quad \la h_2(z,\bar{z}) \ra_{\rm op} = -\frac{4}{(z-\bar{z})^2} \Big\{c_2 + \frac{d_2}2 + d_2 \log\Big|\frac{z-\bar{z}}{2}\Big|\Big\},
\label{oneopen}
\eea
\end{subequations}
where the coefficients $c_2,d_2$ have been given explicitly in Section~\ref{sec6}. Plugging these in \eqref{Tha} yields
\begin{subequations}
\bea
\la T(x) \, h_1(z,\bar z) \ra_{\rm op} \egal -\frac{4d_2}{|x-z|^4},\\
\la T(x) \, h_2(z,\bar z) \ra_{\rm op} \egal -\frac{4}{|x-z|^4}\Big\{c_2 + \frac{3d_2}4 + \frac{d_2}4 \frac{(z-\bar z)^2}{|x-z|^2} + d_2 \log\Big|\frac{z-\bar{z}}{2}\Big|\Big\}.
\eea
\label{open}
\end{subequations}

When the boundary is closed, one has, from \eqref{omega},
\bea
\hspace{-0.5cm}\langle T(x) \, h_a(z,\bar z) \, \omega(\infty)\rangle_{\rm cl} \egal \Big\{\frac{1}{(x-z)^2} + \frac{1}{x-z}\partial_z +\frac{1}{(x-\bar{z})^2} + \frac{1}{x-\bar{z}}\partial_{\bar{z}} \Big\} \la h_a(z,\bar{z}) \omega(\infty) \ra_{\rm cl} \nonumber\\
\noalign{\medskip}
&& \hspace{-4.5cm} + \: \Big\{ \frac{\la \rho(z,\bar z) \omega(\infty) \ra_{\rm cl}}{(x-z)^3} + \frac{\la \bar\rho(z,\bar z) \omega(\infty) \ra_{\rm cl}}{(x-\bar z)^3} - \frac{\la h_1(z,\bar z) \omega(\infty) \ra_{\rm cl}}{2(x-z)^2} - \frac{\la h_1(z,\bar z) \omega(\infty) \ra_{\rm cl}}{2(x-\bar z)^2}\Big\} \, \delta_{a,2} \nonumber\\
\noalign{\smallskip}
&& \hspace{-4.5cm} + \lim_{w,\bar w \to \infty} \Bigg[ \Big\{\frac{1}{x-w}\partial_{w} + \frac{1}{x-\bar w}\partial_{\bar w}\Big\} \la h_a(z,\bar{z}) \omega(w,\bar w) \ra_{\rm cl} - \frac{\la h_a(z,\bar z) \ra_{\rm cl}}{4\pi(x-w)^2} - \frac{\la h_a(z,\bar z) \ra_{\rm cl}}{4\pi(x-\bar w)^2} \Bigg].
\eea
The terms on the last line give however no contribution: $\la h_a(z,\bar z) \ra_{\rm cl}$ vanishes identically (no dissipation), and $\la h_a(z,\bar{z}) \omega(w,\bar w) \ra_{\rm cl}$ does not depend on $w,\bar w$ (the precise location of the dissipation is immaterial for this two-point function).

Repeating the same steps as before, this time using \cite{PR04}
\be
\la \omega(z_{1},\bar z_1) \omega(z_{2},\bar z_2) \ra_{\rm cl}  = {1 \over \pi} \log{|z_{12}|} 
+ 2\gamma_{0} + {1 \over 2\pi} \log{|z_{1}-\bar z_{2}|^{2} \over |z_{1}-\bar
z_{1}| |z_{2}-\bar z_{2}|},
\ee
we obtain
\begin{subequations}
\bea
&&\hspace{-1cm} \la \rho(z,\bar z) \omega(\infty) \ra_{\rm cl} = \frac{d_2}{z-\bar z}, \quad \la \bar\rho(z,\bar z)  \omega(\infty)\ra_{\rm cl} = -\frac{d_2}{z-\bar z},\\
\noalign{\smallskip}
&&\hspace{-1cm} \la h_1(z,\bar z)  \omega(\infty)\ra_{\rm cl} = \frac{4d_2}{(z-\bar z)^2}, \quad \la h_2(z,\bar{z})  \omega(\infty) \ra_{\rm cl} = \frac{4}{(z-\bar{z})^2} \Big\{c_2 + d_2 \log\Big|\frac{z-\bar{z}}{2}\Big|\Big\},
\eea
\end{subequations}
with the same coefficients $c_2,d_2$. In turn, this leads to
\begin{subequations}
\bea
\la T(x) \, h_1(z,\bar z) \, \omega(\infty) \ra_{\rm cl} \egal \frac{4d_2}{|x-z|^4},\\
\la T(x) \, h_2(z,\bar z) \, \omega(\infty) \ra_{\rm cl} \egal \frac{4}{|x-z|^4}\Big\{c_2 + \frac{d_2}4 + \frac{d_2}4 \frac{(z-\bar z)^2}{|x-z|^2} + d_2 \log\Big|\frac{z-\bar{z}}{2}\Big|\Big\}.
\eea
\label{closed}%
\end{subequations}

The correlators \eqref{open} and \eqref{closed}, evaluated when the height $a$ is right above the boundary height 1, namely for $x-z=-{\rm i}y$, simplify to give
\begin{subequations}
\bea
\hspace{-1cm}\la T(x) \, h_1(z,\bar z) \ra_{\rm op} = -\frac{4d_2}{y^4},&& 
\la T(x) \, h_2(z,\bar z) \ra_{\rm op} = -\frac{4}{y^4}\Big\{c_2 - \frac{d_2}4 + d_2 \log y \Big\},\\
\hspace{-1cm}\la T(x) \, h_1(z,\bar z) \, \omega(\infty) \ra_{\rm cl} = \frac{4d_2}{y^4},&& 
\la T(x) \, h_2(z,\bar z) \, \omega(\infty) \ra_{\rm cl} = \frac{4}{y^4}\Big\{c_2 - \frac{3d_2}4 + d_2 \log y \Big\}.
\eea
\end{subequations}
Inserting the explicit values of $c_2,d_2$ and the proportionality constants between $h_1^{\rm op,cl}(x)$ and $T(x)$, given in \eqref{h1bord}, exactly yields the dominant orders of the lattice correlators $\sigma_{a,1}^{\rm op,cl}(y)$ computed in Section~\ref{sec6.3}.

%%%%%%%%%%%%%%%%%%%%%%%%%%%%%%%%%%%%%%%%%%%%%%%%%%%%%%%%%%%%%%%%%%%%%%%%%%%%%%%%%%%%%%%%%%%%%%%%%%%%

\subsection{Two-point bulk correlations on the upper half-plane}

The last case concerns the lattice correlators $\sigma_{a,1}^{\rm op,cl}$ for two heights vertically aligned in the bulk of the upper half-plane. As before, we focus on $a=1,2$ and compute the corresponding conformal correlation function for arbitrary positions, expected to describe the dominant order of the lattice correlators,
\be
\begin{split}
\sigma_{a,1}^{\rm op}(\vec r_1,\vec r_2) &= \la h_a(z_1,\bar z_1) h_1(z_2,\bar z_2) \ra_{\rm op} + \ldots,\\
\sigma_{a,1}^{\rm cl}(\vec r_1,\vec r_2) &= \la h_a(z_1,\bar z_1) h_1(z_2,\bar z_2) \omega(\infty) \ra_{\rm cl} + \ldots
\label{2ptuhp}
\end{split}
\ee
Since the field $h_1 = \phi$ is degenerate at level 2, the previous correlators satisfy second-order differential equations for $a=1$ and for $a=2$. However, for $a=1$, it is much quicker to start from the self-correlators of $\omega$ and use the relations giving the nonchiral fields $\rho, \bar \rho$ and $\phi$ in terms of $\omega$. This avoids the solving of differential equations and circumvents the problem of fixing the integration constants. In addition, it yields the correlators involving $\rho$ and $\bar \rho$, which are in any case required for the case $a=2$, at least for the open boundary condition. We start with $a=1$.

The correlators with two (resp. three) dissipation fields in presence of an open (resp. closed) boundary are easily computed (they are given by $2 \times 2$ (resp. $3 \times 3$) determinants \cite{PR04}),
\begin{subequations}
\bea
&& \hspace{-1cm} \la \omega(z_1,\bar z_1) \omega(z_2,\bar z_2) \ra_{\rm op} = -{1 \over 4\pi^{2}} \log^{2}\Big|{z_1-z_2 \over z_1-\bar z_2}\Big| + \Big ({1 \over 2\pi} \log{|z_1- \bar z_1|} + \gamma_{0}\Big)\nonumber\\
&& \hspace{0.5cm} \times\Big({1 \over 2\pi} \log{|z_2 - \bar z_2|} + \gamma_{0}\Big),\\
\noalign{\medskip}
&& \hspace{-1cm} \la \omega(z_1,\bar z_1) \omega(z_2,\bar z_2) \omega(z_3,\bar z_3) \ra_{\rm cl} = 
-{1 \over 4\pi^{2}} \log\big|(z_1-z_2)(z_1-\bar z_2)\big| \cdot \log{\frac{|(z_1-z_2)(z_1-\bar z_2)|}{|(z_2-z_3)(z_2-\bar z_3)|^2}} \nonumber\\
\noalign{\smallskip}
&& \hspace{0.5cm} + \: \Big(\gamma_0 - {1 \over 2\pi} \log{|z_1- \bar z_1|}\Big) \Big(\gamma_0 - {1 \over 2\pi} \log{\frac{|z_3- \bar z_3|}{|(z_2-z_3)(z_2-\bar z_3)|^2}}\Big) + {\rm cyclic}.
\eea
\end{subequations}
Upon derivation, one easily gets the correlators involving $\rho, \bar \rho$ and $\phi$. For the open boundary condition, they can be written in the following way:
\begin{subequations}
\small
\begin{align}
\begin{split}
\la \rho(1) \phi(2) \ra_{\rm op} &= -\frac{\P_1^2}8 \Big\{ \frac1{(z_1\!-\!\bar z_2)(\bar z_1\!-\!z_2)^2} - \frac2{(z_1\!-\!\bar z_1)(z_2\!-\!\bar z_2)^2} + \frac{(z_1\!-\!\bar z_1)^2 - (z_1\!-\!z_2)(z_1\!-\!\bar z_2)}{(z_1\!-\!z_2)(\bar z_1\!-\!z_2)^2(\bar z_1\!-\!\bar z_2)^2}\Big\}, 
\end{split}\\
\begin{split}
\la \bar \rho(1) \phi(2) \ra_{\rm op} &= -\frac{\P_1^2}8 \Big\{ \frac1{(z_1\!-\!\bar z_2)^2(\bar z_1\!-\!z_2)} + \frac2{(z_1\!-\!\bar z_1)(z_2\!-\!\bar z_2)^2} + \frac{(z_1\!-\!\bar z_1)^2 - (\bar z_1\!-\!z_2)(\bar z_1\!-\!\bar z_2)}{(z_1\!-\!z_2)^2(z_1\!-\!\bar z_2)^2(\bar z_1\!-\!\bar z_2)}\Big\},
\end{split}\\
\begin{split}
\la \phi(1) \phi(2) \ra_{\rm op} &= \frac{\P_1^2}2 \Big\{\frac2{(z_1-\bar z_1)^2(z_2-\bar z_2)^2} - \frac1{|z_1-z_2|^4} - \frac1{|z_1-\bar z_2|^4} \Big\}.
\end{split}
\end{align}
\label{twoopen}%
\end{subequations}
The third equation in particular, evaluated when $z_1 = x+{\rm i}y_1$ and $z_2 = x+{\rm i}y_2$ are vertically aligned, reproduces exactly the dominant terms of $\sigma_{1,1}^{\rm op}(y_1,y_2)$ reported in Section~\ref{sec6.4}.

Similar calculations when the boundary is closed leads to slightly different results:
\begin{subequations}
\begin{align}
\begin{split}
\la \rho(1) \phi(2) \omega(\infty) \ra_{\rm cl} &= \la \rho(1) \phi(2) \ra_{\rm op} - \frac{\P_1^2}{4}\frac{(\bar z_1-z_2)+(\bar z_1-\bar z_2)}{(\bar z_1-z_2)^2(\bar z_1-\bar z_2)^2}, 
\end{split}\\
\begin{split}
\la \bar\rho(1) \phi(2) \omega(\infty) \ra_{\rm cl} &= \la \bar\rho(1) \phi(2) \ra_{\rm op} - \frac{\P_1^2}{4}\frac{(z_1-z_2)+(z_1-\bar z_2)}{(z_1-z_2)^2(z_1-\bar z_2)^2}, 
\end{split}\\
\noalign{\medskip}
\begin{split}
\la \phi(1) \phi(2) \omega(\infty) \ra_{\rm cl} &= \la \phi(1) \phi(2) \ra_{\rm op}\,.% = \frac{\P_1^2}2 \Big\{\frac2{(z_1\!-\!\bar z_1)^2(z_2\!-\!\bar z_2)^2} - \frac1{|z_1\!-\!z_2|^4} - \frac1{|z_1\!-\!\bar z_2|^4} \Big\}.
\end{split}
\end{align}
\end{subequations}
The correlators involving $\rho$ and $\bar\rho$ for the closed boundary differ from their open analogues by terms independent of $z_1$ and $\bar z_1$ respectively, with the consequence that the closed correlator with two $\phi$'s in general positions is exactly equal to the one corresponding to the open boundary, in agreement with the lattice results \eqref{P11uhpop} and \eqref{P11uhpcl} when the two insertion points are vertically aligned.

The case $a=2$ for the open boundary condition requires to compute the nonchiral two point-function $\la \psi(1) \phi(2) \ra_{\rm op}$. We first perform a M\"obius transformation, taken to be real to preserve the boundary, so as to bring the two insertion points to positions that only depend on the anharmonic ratio of $z_1^{},\bar z_1,z_2^{},\bar z_2$ given by
\be
x=\frac{(z_1-\bar z_1)(z_2-\bar z_2)}{(z_1-z_2)(\bar z_1-\bar z_2)} = -\frac{4y_1y_2}{|z_1-z_2|^2},
\qquad \text{(real negative)}
\ee
where $y_1,y_2>0$ are the imaginary parts of $z_1,z_2$. A convenient choice is to map the two points $z_1,z_2$ onto respectively $w_1={\rm i} t$ and $w_2=2+{\rm i}t$ with $t = \sqrt{-x}>0$. The proper map takes the usual form $w(z)=(az+b)/(cz+d)$ with the following values of the parameters,
\begin{subequations}
\bea
&& a = t \, (x_1-x_2) - 2y_2, \qquad b = 2x_1y_2 - t \, (x_1^2 - x_1x_2 + y_1^2 - y_1y_2),\\
\noalign{\medskip}
&& c = y_1 - y_2, \qquad d = x_1y_2 - x_2y_1 - 2 \frac{y_1y_2}t,
\eea
\end{subequations}
where $x_1,x_2$ are the real parts of $z_1,z_2$.

The transformation law \eqref{finitepsi} of $\psi$ yields the following identity,
\bea
\la \psi(z_1^{},\bar z_1) \phi(z_2^{},\bar z_2) \ra_{\rm op} \egal \frac{x^2}{y_1^2y_2^2} \, \Big\{ \la \psi(w_1^{},\bar w_1) \phi(w_2^{},\bar w_2) \ra_{\rm op} + \lambda \log{\Big(\frac{-x}{y_1^2}\Big)} \la \phi(w_1^{},\bar w_1) \phi(w_2^{},\bar w_2) \ra_{\rm op}\Big\} \nonumber\\
&& - \frac{x^2(y_1-y_2)}{y_1^2 y_2^2} \Bigg\{\frac{\la \rho(w_1^{},\bar w_1) \phi(w_2^{},\bar w_2) \ra_{\rm op}}{t(\bar z_1-\bar z_2)-2y_2} + \frac{\la \bar\rho(w_1^{},\bar w_1) \phi(w_2^{},\bar w_2) \ra_{\rm op}}{t(z_1^{}-z_2^{})-2y_2}\nonumber\\
&& - \frac{\kappa (y_1-y_2)}{4y_2} \frac{\la \phi(w_2^{},\bar w_2) \ra_{\rm op}}{y_1+y_2-t(x_1-x_2)}\Bigg\}. \nonumber\\
\eea
Of the five correlators appearing on the right-hand side, the last four are known functions of $x$, and given in \eqref{oneopen} and \eqref{twoopen}. Further using $\lambda=-\frac12$ and $\kappa=-\frac{\P_1}4$, one may simplify the previous expression to
\begin{align}
\la \psi(z_1^{},\bar z_1) \phi(z_2^{},\bar z_2) \ra_{\rm op}& = \frac{x^2}{y_1^2y_2^2} \, \Big[ H(x) + \frac{\P_1^2}{64} \, \frac{x^4-2x^3+4x-2}{(1-x)^2} \, \log{\Big(\frac{-x}{y_1^2}\Big)}\Big]\nonumber\\
&\qquad+\frac{\P_1^2}{128} \, \frac{y_1-y_2}{y_1^2y_2^3} \, \frac{x^3(x-2)}{(1-x)^2},
\end{align}
where $H(x) = \la \psi(w_1^{},\bar w_1) \phi(w_2^{},\bar w_2) \ra_{\rm op}$. The degeneracy of $\phi$ implies that the function $H(x)$ satisfies the following differential equation,
\be
x(1-x) H''(x) + 2(1-2x) H'(x) - \frac{2H(x)}{x(1-x)} = -\frac{\P_1^2}{64} \frac{x^4-14x^3+24x^2-20x+6}{x^3(1-x)^3}.
\ee
The general solution depends on two integration constants, which can be easily fixed by considering two limiting cases. Indeed, for a large horizontal separation of $\psi$ and $\phi$ (i.e. $|x_1-x_2| \to \infty$ and $y_1,y_2$ finite), and for a large distance to the boundary (i.e. $y_1,y_2 \to \infty$ and $|z_1-z_2|$ finite), the correlator $\la \psi(z_1^{},\bar z_1) \phi(z_2^{},\bar z_2) \ra_{\rm op}$ must go respectively to $\la \psi(z_1^{},\bar z_1) \ra_{\rm op} \, \la \phi(z_2^{},\bar z_2) \ra_{\rm op}$ and to the bulk correlator $\la \psi(z_1^{},\bar z_1) \phi(z_2^{},\bar z_2) \ra$ on the full plane.

The final result reads
\bea
\la \psi(z_1^{},\bar z_1) \phi(z_2^{},\bar z_2) \ra_{\rm op} \egal \frac{\P_1}{32y_1^2y_2^2} \, \frac{x^4-2x^3+4x-2}{(1-x)^2} \, \Bigg[ \frac{3(3\pi-10)}{2\pi^3} - \P_1 \Big(\log{y_1} + \gamma + \frac52 \log{2}\Big) \Bigg] \nonumber\\
&& \hspace{1cm} + \: \frac{\P_1^2}{64y_1^2y_2^2} \, \Bigg[\frac{x^3(x-2)}{(1-x)^2} \, \Big(\log{(1-x)} + \frac{y_1}{2y_2}\Big) - \frac{x^2}{1-x} \Bigg].
\label{psiphiopen}
\eea

As a first check, one can verify that when the two fields are vertically aligned, $x_1=x_2$, the previous form exactly reproduces the dominant term of the lattice correlator $\sigma_{2,1}^{\rm op}(y_1,y_2)$ given in \eqref{P21uhpop}. A second and independent check concerns the limit $y_2 \to 0$, which brings the height-one field $\phi$ to the boundary. In this limit, the bulk $\phi$ is expected to expand on open-boundary fields of its own conformal family, to which the identity and the stress-energy tensor (the open-boundary height-one field)  belong, namely
\be
\phi(z_2^{},z_2^*) \simeq \frac{C_{-2}}{y_2^2}\, {\mathbb I} + C_0 \, T(x_2) + \ldots
\ee
The expansion of the correlator \eqref{psiphiopen} should then reproduce, up to constants, the correlators $\la \psi(z_1^{},\bar z_1)\ra_{\rm op}$ and $\la \psi(z_1^{},\bar z_1) T(x_2) \ra_{\rm op}$ at order $y_2^{-2}$ and $y_2^0$ respectively, and give a vanishing term at order $y_2^{-1}$. This is exactly what we find; in addition, the two fusion coefficients read $C_{-2}=\P_1/4$ and $C_0=-4\P_1$.

Although the lattice result for $\sigma_{2,1}^{\rm cl}(y_1,y_2)$, given in \eqref{P21uhpcl}, is very close to $\sigma_{2,1}^{\rm op}(y_1,y_2)$, the situation is more complicated on the conformal side. It requires to compute the equivalent of a chiral six-point correlator, due to the extra background field $\omega$ (a chiral five-point function if one uses the boundary dissipation field). So a satisfactory argument to write the functional form of $\sigma_{2,1}^{\rm cl}(y_1,y_2)$ for general positions is still lacking at the moment.

%%%%%%%%%%%%%%%%%%%%%%%%%%%%%%%%%%%%%%%%%%%%%%%%%%%%%%%%%%%%%%%%%%%%%%%%%%%%%%%%%%%%%%%%%%%%%%%%%%%%%%%%

\section{Conclusion}

Applied to the sandpile model, the technique proposed by Kenyon and Wilson has considerably eased the calculation of height correlations. The computation of the next-to-leading order of $\P_{a,1}(i,j)$ for $a \ge 2$ or of $\P^{\rm op,cl}_{a,1}(i,j)$ on the upper half-plane, for instance, would have been extremely tedious with traditional graph-theoretical techniques. These new ideas would presumably allow one to compute correlators with more heights 1, at a linear or at most polynomial time cost, but explicit results of that sort would be of little intrinsic interest. 

The real challenge, as far as lattice height correlations are concerned, is to find a suitable method for handling two or more heights larger than or equal to 2, which are separated by large distances. Correlations of heights larger than 2 at short distances can in principle be computed using the formalism of Kenyon and Wilson, and even this simpler case is bound to lead to complicated combinatorics as many different predecessor diagrams contribute (as mentioned in the introduction, David Wilson did exactly that for arbitrary heights at a few adjacent sites). But with a nontrivial connection localized on a single zipper, the correlation of two heights 2 at large distance is beyond what is currently tractable.

On the conformal side, little is known of the specific logarithmic conformal theory underlying the scaling limit of the sandpile model. What is presently known or assumed of that conformal theory is perfectly consistent with the lattice calculations, but only offers a very partial view of it. Therefore, more data from the lattice are needed to gain insight into what the conformal theory looks like. In particular, knowing whether the representation of the identity resembles that of the symplectic triplet theory, with its multiple copies of the $(\phi,\psi)$ Jordan pair for instance, would be a major progress.

%%%%%%%%%%%%%%%%%%%%%%%%%%%%%%%%%%%%%%%%%%%%%%%%%%%%%%%%%%%%%%%%%%%%%%%%%%%%%%%%%%%%%%%%%%%%%%%%%%%%%

\subsection*{Acknowledgments}
This work is supported by the Belgian Interuniversity Attraction Poles Program P7/18 through the network DYGEST (Dynamics, Geometry and Statistical Physics). PR is Senior Research Associate of the Belgian Fonds National de la Recherche Scientifique (FNRS).

%%%%%%%%%%%%%%%%%%%%%%%%%%%%%%%%%%%%%%%%%%%%%%%%%%%%%%%%%%%%%%%%%%%%%%%%%%%%%%%%%%%%%%%%%%%%%%%%%%%%%

\newpage
\appendix
\section{Green functions}
In this appendix, we briefly review the Green function in presence of a zipper, following \cite{KW15},  and in particular the way the first derivative can be computed. As required by the calculations presented in the text, we consider the Green function and its derivative on the infinite planar graph $\Z^2$, the half-infinite planar graph $\Z \times \mathbb N^*$, as well as local modifications thereof by which a finite number of edges are removed.

\subsection{Zipper on the plane}
\label{zip}

The type of spanning tree probabilities we want to compute requires the knowledge of the Green function $\Gr_{u,v}(z) = (\mathbf{\Delta}^{-1}(z))_{u,v}$ on the plane, in presence of a (semi-infinite) zipper. The Green function depends on the parameter $z$ carried by the zipper, however the full dependence on $z$ is not needed; only its zeroth and first orders around $z=1$ are required. By writing the Laplacian $\mathbf{\Delta}(z)$ with the zipper as a perturbation of the standard Laplacian $\Delta\equiv\mathbf{\Delta}(1)$, one obtains \cite{KW15} 
\begin{equation}
\Gr_{u,v}(z)=G_{u,v}-(z-1)\sum_{(k,\l):\,\phi_{k,\l}=z}\Big(G_{u,k} \, G_{\l,v}-G_{u,\l} \, G_{k,v}\Big)+\O(z-1)^2.
\label{fullGr}
\end{equation}
The zeroth order $G_{u,v} = \Gr_{u,v}(1)$ is the standard Green function on the lattice $\Z^2$, given by
\be
G(v-u) \equiv G_{u,v} = \int_{-\pi}^{\pi}\:\frac{{\rm d}\alpha\,{\rm d}\beta}{8\pi^2} \; \frac{{\rm e}^{{\rm i} (\alpha p+\beta q)}}{2 - \cos\alpha - \cos\beta} = \int_{-\pi}^{\pi}\:\frac{{\rm d}\alpha}{4\pi} \; \frac{t^{|q|}\,{\rm e}^{{\rm i} \alpha p}}{\sqrt{y^2 - 1}}, \qquad v-u=(p,q),
\label{Gsq} 
\ee
with $y(\alpha)\equiv 2-\cos\alpha$ and $t(\alpha)\equiv y(\alpha)-\sqrt{y(\alpha)^2 - 1}$. Although the integral is divergent, the difference $G_{u,v}-G_{0,0}$ is finite for any $u,v$. The leading asymptotic behavior for large distances $|u-v|$ as well as finite-distance values are well known \cite{Spi76}. Subleading terms in the large-distance expansion of $G_{u,v}$ in inverse powers of $r\equiv|u-v|$ have been computed in \cite{GPP09}. With $v-u = r {\rm e}^{{\rm i}\varphi}$, the first few terms read
\bea
\hspace{-0.5cm} G_{u,v} - G_{0,0} \egal - \frac{1}{2\pi} \Big(\log r + \gamma + \frac32 \log 2\Big) + \frac{\cos{4\varphi}}{24\pi \, r^2} + \frac1{r^4} \Big(\frac{3 \cos{4\varphi}}{80\pi} + \frac{5\cos{8\varphi}}{96\pi}\Big)\nonumber\\
\noalign{\medskip}
&& \hspace{-2.5cm} + \; \frac1{r^6} \Big(\frac{51 \cos{8\varphi}}{224\pi} + \frac{35\cos{12\varphi}}{144\pi}\Big) + 
\frac1{r^8} \Big(\frac{217 \cos{8\varphi}}{640\pi} + \frac{45\cos{12\varphi}}{16\pi} + \frac{1925\cos{16\varphi}}{768\pi}\Big) + \ldots 
\label{Guv}
\eea

From \eqref{fullGr}, the first derivative $G'_{u,v} \equiv \partial_z\Gr_{u,v}(z)\big|_{z=1}$ is given by
\begin{equation}
G'_{u,v}=\sum_{(k,\l):\,\phi_{k,\l}=z}\big(G_{u,\l} \, G_{k,v}-G_{u,k} \, G_{\l,v}\big).
\label{Gpdef}
\end{equation}
It satisfies the useful identity
\begin{equation}
\left(\Delta_0 G'\right)_{u,v}=-\left(G'\Delta_0\right)_{v,u}=\sum_{(k,\l):\,\phi_{k,\l}=z}\Big[\delta_{u,\l}\,G_{k,v}-\delta_{u,k}\,G_{\l,v}\Big].
\label{Gpharm}
\end{equation}
We note that $G'_{u,v}$ has the same singularity, proportional to $G_{0,0}$, as $G_{u,v}$ itself, with however a main difference: the coefficient of $G_{0,0}$ is constant for $G_{u,v}$, but is a complicated function of $u,v$ for $G'_{u,v}$.

On $\Z^2$, the summation in \eqref{Gpdef} is infinite, but can be reduced to a finite sum by combining three ingredients: the antisymmetry of $G'_{u,v}$, the rotation and translation invariance, and  deformations of the zipper. The symmetry of the Green function implies the antisymmetry of its first derivative, $G'_{u,v}=-G'_{v,u}$. Rotation and translation invariance of $G'_{u,v}$ are manifest provided $u,v$ {\it and} the zipper are all rotated or translated simultaneously. 

Finally, the derivative $G'_{u,v}$ only depends weakly on the location of the zipper. For fixed $u,v$,  the zipper can in fact be freely deformed---keeping the endpoints fixed---without changing the value of $G'_{u,v}$, except when the zipper line is moved over $u$ or $v$, in which case the move brings an extra contribution. 

%\begin{figure}[t]
%\centering
%\begin{tikzpicture}[scale=1.15,>=stealth,font=\normalsize]
%\draw[densely dashed] (-1.25,1.75)--(1.25,1.75)--(1.25,-0.25)--(-0.25,-0.25)--(-0.25,-0.75)--(-1.75,-0.75)--(-1.75,1.25)--(-1.25,1.25)--(-1.25,1.75);
%\foreach \x in {-1,-0.5,...,1} {\draw[->] (\x,1.5)--(\x,2);}
%\foreach \x in {0,0.5,...,1.5} {\draw[->] (1,\x)--(1.5,\x);}
%\foreach \x in {0,0.5,...,1} {\draw[->] (\x,0)--(\x,-0.5);}
%\draw[->] (-0.5,-0.5)--(0,-0.5);
%\foreach \x in {-0.5,-1,-1.5} {\draw[->] (\x,-0.5)--(\x,-1);}
%\foreach \x in {-0.5,0,...,1} {\draw[->] (-1.5,\x)--(-2,\x);}
%\draw[->] (-1.5,1)--(-1.5,1.5);
%\draw[->] (-1,1.5)--(-1.5,1.5);
%\draw[step=0.5cm,dotted] (-2.4,-1.4) grid (1.9,2.4);
%\foreach \x in {-1.5,-1,...,1}{\foreach \y in {0,0.5,1}{\filldraw (\x,\y) circle (0.04cm);}}
%\foreach \x in {-1.5,-1,-0.5}{\filldraw (\x,-0.5) circle (0.04cm);}
%\foreach \x in {-1,-0.5,0,0.5,1}{\filldraw (\x,1.5) circle (0.04cm);}
%\filldraw (1.25,1.75) circle (0.04cm);
%\filldraw (-1.75,-0.75) circle (0.04cm);
%\draw (-0.1,-0.75) node {$L$};
%\draw (-0.25,0.75) node {$S_L$};
%\draw (-1.95,-0.85) node {$A$};
%\draw (1.48,1.85) node {$B$};
%\end{tikzpicture}
%\caption{Zipper forming a closed loop $L$. The arrows indicate the oriented edges on which the parallel transport equals $z$.}
%\label{ziploop}
%\end{figure}

\begin{figure}[t]
\centering
\begin{tikzpicture}[scale=1.15,>=stealth,font=\normalsize]
\draw[densely dashed,red] (-1.25,1.75)--(1.25,1.75)--(1.25,-0.25)--(-0.25,-0.25)--(-0.25,-0.75)--(-1.75,-0.75)--(-1.75,1.25)--(-1.25,1.25)--(-1.25,1.75);
\foreach \x in {-1,-0.5,...,1} {\draw[->] (\x,1.5)--(\x,2);}
\foreach \x in {0,0.5,...,1.5} {\draw[->] (1,\x)--(1.5,\x);}
\foreach \x in {0,0.5,...,1} {\draw[->] (\x,0)--(\x,-0.5);}
\draw[->] (-0.5,-0.5)--(0,-0.5);
\foreach \x in {-0.5,-1,-1.5} {\draw[->] (\x,-0.5)--(\x,-1);}
\foreach \x in {-0.5,0,...,1} {\draw[->] (-1.5,\x)--(-2,\x);}
\draw[->] (-1.5,1)--(-1.5,1.5);
\draw[->] (-1,1.5)--(-1.5,1.5);
\draw[step=0.5cm,dotted] (-2.4,-1.4) grid (1.9,2.4);
\foreach \x in {-1.5,-1,...,1}{\foreach \y in {0,0.5,1}{\filldraw[red] (\x,\y) circle (0.04cm);}}
\foreach \x in {-1.5,-1,-0.5}{\filldraw[red] (\x,-0.5) circle (0.04cm);}
\foreach \x in {-1,-0.5,0,0.5,1}{\filldraw[red] (\x,1.5) circle (0.04cm);}
\filldraw[blue] (1.25,1.75) circle (0.04cm);
\filldraw[blue] (-1.75,-0.75) circle (0.04cm);
\draw (-0.1,-0.75) node[red] {$L$};
\draw (-0.25,0.75) node[red] {$S_L$};
\draw (-1.95,-0.85) node[blue] {$A$};
\draw (1.48,1.85) node[blue] {$B$};
\end{tikzpicture}
\caption{Zipper forming a closed loop $L$. The arrows indicate the oriented edges on which the parallel transport equals $z$.}
\label{ziploop}
\end{figure}

To show this, we consider the derivative $G'^{\,\rm loop}_{u,v}$ as given by \eqref{Gpdef} for a closed zipper loop $L$ as in Fig.~\ref{ziploop}. Let us denote by $S_L$ the subset of vertices that lie inside the contour $L$. Since the contributions of an oriented edge $(k,\l)$ and its opposite $(\l,k)$ cancel each other out in $G'$, one can extend the summation to all edges $(k,\ell)$ such that $k$ is in $S_L$:
\begin{align}
G'^{\,\rm loop}_{u,v} &= -\sum_{(k,\l)\in L}\Big(G_{u,k}\,G_{\l,v}-G_{u,\l}\,G_{k,v}\Big) = -\sum_{(k,\l): \, k \in S_L}\Big(G_{u,k}\,G_{\l,v}-G_{u,\l}\,G_{k,v}\Big)\nonumber\\
&=-\sum_{k\in S_L}\left[G_{u,k}\Big(4G_{k,v}-(\Delta_0 G)_{k,v}\Big)-\Big(4G_{u,k}-(\Delta_0 G)_{u,k}\Big)G_{k,v}\right]\nonumber\\
&=G_{u,v}\sum_{k\in S_L}\big(\delta_{k,v}-\delta_{k,u}\big).
\label{Gloop}
\end{align}
With respect to a pair of points $A,B$ on the loop $L$, one can compare the values of $G'_{u,v}$ for the two zippers with endpoints $A$ and $B$, one going through the lower-right of $S_L$, the other going through the upper-left of $S_L$. The two stretches can be considered as deformations of each other, except for the orientation of the arrows. Since reversing the arrows merely changes the sign of the corresponding $G'_{u,v}$, the full loop contribution $G'^{\,\rm loop}_{u,v}$ may be written as $G'^{\,\rm loop}_{u,v} = G'^{\,\rm new}_{u,v} - G'^{\,\rm old}_{u,v}$ for a proper choice of orientation of the arrows. The identity \eqref{Gloop} then shows that
\be
G'^{\,\rm new}_{u,v} - G'^{\,\rm old}_{u,v} = 
\begin{cases}
G_{u,v} & \hbox{if the zipper has crossed $v$ only}, \\
-G_{u,v} & \hbox{if the zipper has crossed $u$ only}, \\
0 & \hbox{if the zipper has crossed none or both of $u,v$}. 
\end{cases}
\label{newold}
\ee
According to our convention of which is the new zipper and which is the old one, the vertices $u,v$ are crossed in the direction of the arrows attached to the moving zipper. If they are crossed in the opposite direction, the difference \eqref{newold} changes sign.

\begin{figure}[t]
\centering
\begin{tikzpicture}[scale=0.7,>=stealth]

\begin{scope}[xshift=0cm]
\draw[step=0.5cm,dotted] (-3.45,-1.95) grid (3.45,1.95);
\filldraw[blue] (-0.5,0.5) circle (0.075cm);
\draw[blue] (-0.625,0.375) rectangle (-0.375,0.625);
\filldraw[blue] (2.5,-1) circle (0.075cm);
\filldraw[red] (0.25,0.25) circle (0.075cm);
\draw[->,red] (0.25,0.25) -- (0.25,-2);
\draw[->,red] (0,-0.75) -- (0.5,-0.75);
\draw[xshift=0.25cm,thick,->] (0:2cm) arc (0:180:2cm);
\draw[thick,font=\small] (-2.75,-1.25) node {(a)};
\end{scope}

\begin{scope}[xshift=8cm]
\draw[step=0.5cm,dotted] (-3.45,-1.95) grid (3.45,1.95);
\filldraw[blue] (1,0) circle (0.075cm);
\draw[blue] (0.875,-0.125) rectangle (1.125,0.125);
\filldraw[blue] (-2,1.5) circle (0.075cm);
\filldraw[red] (0.25,0.25) circle (0.075cm);
\draw[->,red] (0.25,0.25) -- (0.25,2);
\draw[->,red] (0.5,1.25) -- (0.,1.25);
\draw[xshift=0.25cm,thick,->] (90:1cm) arc (90:270:0.75cm);
\draw[thick,font=\large] (-2.75,-1.25) node {\small (b)};
\end{scope}

\begin{scope}[xshift=16cm]
\draw[step=0.5cm,dotted] (-3.45,-1.95) grid (3.45,1.95);
\filldraw[blue] (1,0) circle (0.075cm);
\draw[blue] (0.875,-0.125) rectangle (1.125,0.125);
\filldraw[blue] (-2,1.5) circle (0.075cm);
\filldraw[red] (0.25,0.25) circle (0.075cm);
\draw[->,red] (0.25,0.25) -- (0.25,-2);
\draw[->,red] (0,-0.75) -- (0.5,-0.75);
\draw[thick,font=\large] (-2.75,-1.25) node {\small (c)};
\end{scope}

\begin{scope}[xshift=0cm,yshift=-5cm]
\draw[step=0.5cm,dotted] (-3.45,-1.95) grid (3.45,1.95);
\filldraw[blue] (1,0) circle (0.075cm);
\draw[blue] (0.875,-0.125) rectangle (1.125,0.125);
\filldraw[blue] (-2,1.5) circle (0.075cm);
\filldraw[red] (0.25,0.25) circle (0.075cm);
\draw[->,red] (-1.25,1.25) -- (0.25,1.25) -- (0.25,-2);
\draw[->,red] (0,-0.75) -- (0.5,-0.75);
\filldraw[red] (0.25,0.75) circle (0.075cm);
\filldraw[red] (0.25,1.25) circle (0.075cm);
\filldraw[red] (-0.25,1.25) circle (0.075cm);
\filldraw[red] (-0.75,1.25) circle (0.075cm);
\filldraw[red] (-1.25,1.25) circle (0.075cm);
\draw[->,red] (0,0.5) -- (0.5,0.5);
\draw[->,red] (0,1) -- (0.5,1);
\draw[->,red] (0,1) -- (0,1.5);
\draw[->,red] (-0.5,1) -- (-0.5,1.5);
\draw[->,red] (-1,1) -- (-1,1.5);
\draw[thick,font=\large] (-2.75,-1.25) node {\small (d)};
\end{scope}

\begin{scope}[xshift=8cm,yshift=-5cm]
\draw[step=0.5cm,dotted] (-3.45,-1.95) grid (3.45,1.95);
\filldraw[blue] (1,0) circle (0.075cm);
\draw[blue] (0.875,-0.125) rectangle (1.125,0.125);
\filldraw[blue] (-2,1.5) circle (0.075cm);
\draw[->,red] (-1.25,1.25) -- (0.25,1.25) -- (0.25,-2);
\draw[->,red] (0,-0.75) -- (0.5,-0.75);
\filldraw[red] (-1.25,1.25) circle (0.075cm);
\draw[->] (0.25,0.5) -- (-1.25,-0.75);
\draw[thick,font=\large] (-2.75,-1.25) node {\small (e)};
\end{scope}

\begin{scope}[xshift=16cm,yshift=-5cm]
\draw[step=0.5cm,dotted] (-3.45,-1.95) grid (3.45,1.95);
\filldraw[blue] (-0.5,0.5) circle (0.075cm);
\draw[xshift=3cm,yshift=-1.5cm,blue] (-0.625,0.375) rectangle (-0.375,0.625);
\filldraw[blue] (2.5,-1) circle (0.075cm);
\filldraw[red] (0.25,0.25) circle (0.075cm);
\draw[->,red] (0.25,0.25) -- (0.25,-2);
\draw[->,red] (0,-0.75) -- (0.5,-0.75);
\draw[thick,font=\large] (-2.75,-1.25) node {\small (f)};
\end{scope}

\end{tikzpicture}
\caption{Computation of $G'_{u,v}$ for $u=(-1,1)$ and $v=(5,-2)$. The boxed vertex corresponds to the first argument of the derivative of the Green function $G'$. The six panels (a)--(f) illustrate the transformations and deformations explained in the text.}
\label{zipdefex}
\end{figure}
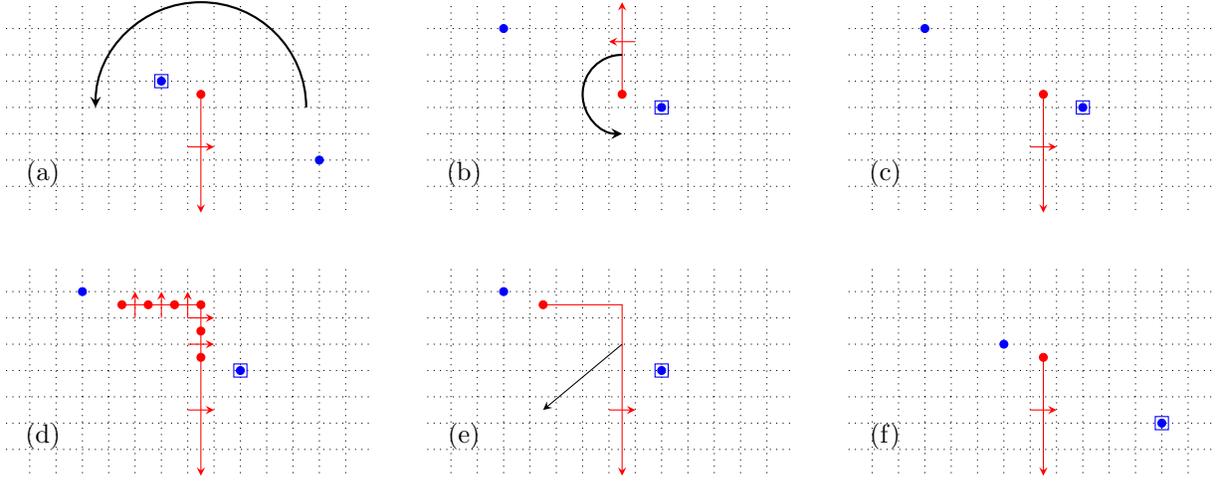

Let us illustrate how $G'_{u,v}$ can be computed in closed form for the particular case of the zipper $\left\{(0,m),(1,m)\right\}_{m\le 0}$ on the square lattice, using the previous observations \cite{KW15}. We shall compute the derivative $G'_{u,v}$ for $u=(-1,1)$ and $v=(5,-2)$, following the steps pictured in Fig.~\ref{zipdefex}.

We begin by rigidly rotating the whole lattice (zipper and marked points) by $180^\circ$ around the black dot to which the zipper is attached. Under this rotation, a vertex $x$ is mapped to $x'=(1,1)-x$, so the two reference points $(-1,1)$ and $(5,-2)$ are mapped to $u'=(2,0)$ and $v'=(-4,3)$ respectively, while the zipper is now pointing upward (indicated by an up arrow in the middle equation below). To put the zipper back in the original position, we rotate it by $180^\circ$ to the left, this time keeping the marked points fixed. In the process, the zipper goes over $v'=(-4,3)$ in the direction of the arrows, producing an additional term:
\begin{equation}
G'_{(-1,1),(5,-2)} = G'^{\,\uparrow}_{(2,0),(-4,3)} = G'_{(2,0),(-4,3)} - G_{(2,0),(-4,3)}.
\end{equation}

In the next step, we add a finite number of zipper edges so that the endpoint is in the same relative position with respect to $v'$ as it was with respect to $u$ in the original configuration. In our example, one needs to add the following set of five edges:
\begin{equation}
E = \Big\{\big((0,1),(1,1)\big)\,,\,\big((0,2),(1,2)\big)\,,\,\big((0,2),(0,3)\big)\,,\,\big((-1,2),(-1,3)\big)\,,\,\big((-2,2),(-2,3)\big)\Big\}.
\end{equation}
Deforming the zipper again to make it go straight down, we recover the original relative positions of the vertices with respect to the zipper. A rigid translation puts the whole configuration in the original position except for the inversion of $u$ and $v$ (the square box has changed place). In the example, one finds:
\begin{align}
G'_{(-1,1),(5,-2)} &= G'_{(2,0),(-4,3)} - G_{(2,0),(-4,3)}\nonumber\\
&= G'_{(5,-2),(-1,1)} - G_{(2,0),(-4,3)} + \sum_{(k,\ell) \in E} \;[G_{(2,0),k}\,G_{\ell,(-4,3)} - G_{(2,0),\ell}\,G_{k,(-4,3)}]\nonumber\\
&= - \frac12 G_{(2,0),(-4,3)} + \frac12 \sum_{(k,\ell) \in E} \;[G_{(2,0),k}\,G_{\ell,(-4,3)} - G_{(2,0),\ell}\,G_{k,(-4,3)}]\nonumber\\
&= G_{0,0} \, \left(-\frac{15}{4}+\frac{152}{15\pi}\right) - \frac{677}{32} + \frac{1559}{21\pi} - \frac{1939}{90\pi^2},
\label{Gpcompex}
\end{align}
where $G_{0,0}$ is the divergent part of the Green function.

Proceeding this way, one obtains the following values of the derivative $G_{u,v}'$ for the origin and its four nearest neighbors, with the same zipper location as in the previous calculation,
%%%%%%%%
{\scriptsize
\begin{equation}
\bordermatrix{G'_{u,v} & {\scriptstyle(1{,}0)} & {\scriptstyle(0{,}1)} & {\scriptstyle({-}1{,}0)} & {\scriptstyle(0{,}{-}1)} & {\scriptstyle(0{,}0)}\cr
{\scriptstyle (1{,}0)} & 0 & \frac{1}{2}G_{0,0}{-}\frac{1}{2\pi} & G_{0,0}\hspace{-0.75mm}\left(1{-}\frac{1}{\pi}\right){-}\frac{5}{8}{+}\frac{5}{4\pi} & G_{0,0}\hspace{-0.75mm}\left(\frac{1}{2}{+}\frac{1}{\pi}\right){-}\frac{3}{4\pi} & \frac{3}{4}G_{0,0}{-}\frac{5}{32}\cr
{\scriptstyle(0{,}1)} & -\frac{1}{2}G_{0,0}{+}\frac{1}{2\pi} & 0 & G_{0,0}\hspace{-0.75mm}\left(\frac{1}{2}{-}\frac{1}{\pi}\right){-}\frac{1}{4\pi} & \frac{1}{\pi}G_{0,0}{-}\frac{3}{8}{+}\frac{3}{4\pi} & \frac{1}{4}G_{0,0}{-}\frac{3}{32}\cr
{\scriptstyle({-}1{,}0)} & G_{0,0}\hspace{-0.75mm}\left(-1{+}\frac{1}{\pi}\right){+}\frac{5}{8}{-}\frac{5}{4\pi} & G_{0,0}\hspace{-0.75mm}\left(-\frac{1}{2}{+}\frac{1}{\pi}\right){+}\frac{1}{4\pi} & 0 & G_{0,0}\hspace{-0.75mm}\left(-\frac{1}{2}{+}\frac{2}{\pi}\right){+}\frac{1}{4}{-}\frac{1}{\pi} & G_{0,0}\hspace{-0.75mm}\left(-\frac{1}{4}{+}\frac{1}{\pi}\right){-}\frac{1}{32}\cr
{\scriptstyle(0{,}{-}1)} & -G_{0,0}\hspace{-0.75mm}\left(\frac{1}{2}{+}\frac{1}{\pi}\right){+}\frac{3}{4\pi} & -\frac{1}{\pi}G_{0,0}{+}\frac{3}{8}{-}\frac{3}{4\pi} & G_{0,0}\hspace{-0.75mm}\left(\frac{1}{2}{-}\frac{2}{\pi}\right){-}\frac{1}{4}{+}\frac{1}{\pi} & 0 & G_{0,0}\hspace{-0.75mm}\left(\frac{1}{4}{-}\frac{1}{\pi}\right){+}\frac{1}{32}\cr
{\scriptstyle(0{,}0)} & -\frac{3}{4}G_{0,0}{+}\frac{5}{32} & -\frac{1}{4}G_{0,0}{+}\frac{3}{32} & G_{0,0}\hspace{-0.75mm}\left(\frac{1}{4}{-}\frac{1}{\pi}\right){+}\frac{1}{32} & G_{0,0}\hspace{-0.75mm}\left(-\frac{1}{4}{+}\frac{1}{\pi}\right){-}\frac{1}{32} & 0\cr}
\label{gprime}
\end{equation}}%
These are the only values of $G'_{u,v}$ needed to complete the computation of the single-height probabilities of Section~\ref{sec4}.

%%%%%%%%%%%%%%%%%%%%%%%%%%%%%%%%%%%%%%%%%%%%%%%%%%%%%%%%%%%%%%%%%%%%%%%%%%%%%%%%%%%%%%%%%%%%%%%%%%%%%%%%

\subsection{Zipper on the upper half-plane}
\label{uhp_zipper}

On the upper half-plane, a zipper going out from an inner face can be of essentially two different kinds, either going off to infinity or terminating somewhere on the boundary, as illustrated in Fig.~\ref{zip_inf}. It is semi-infinite in the former case, finite in the latter. Despite this seemingly substantial difference, the derivatives of the Green function associated with them are closely related, as were those associated with two different zippers on the full plane. As shown in Appendix~\ref{zip}, two zippers on the full plane starting from the same point and going to infinity in different directions are deformable into each other; the associated Green function derivatives are equal or differ by $\pm G_{u,v}$. The same relation holds on the UHP.

Let us first consider the zippers $\rm Z_2$ and $\rm Z_3$. Each one involves a nontrivial parallel transport on a boundary edge. According to \eqref{Gpdef}, the Green function derivatives $G'^{\rm op/cl}_{u,v}({\rm Z_2})$ and $G'^{\rm op/cl}_{u,v}({\rm Z_3})$ are given by finite sums over two different sets of edges, where the Green function $G$ is to be replaced with the form appropriate to the boundary condition. Let us now observe that either sum may be extended to include all or some of those edges depicted at the bottom of Fig.~\ref{zip_inf}. Indeed, any of them brings a zero contribution to the sum, since
\be
G_{u,(x,1)} \, G_{(x,0),v} - G_{u,(x,0)} \, G_{(x,1),v}
\ee
vanishes identically for both boundary conditions. For the open boundary, the relevant Green function $G^{\rm op}_{u,v}$ vanishes when one of its arguments is on the line $y=0$; while for the closed boundary, it satisfies the identities $G^{\rm cl}_{u,(x,1)} = G^{\rm cl}_{u,(x,0)}$ and $G^{\rm cl}_{(x,1),v} = G^{\rm cl}_{(x,0),v}$ for any sites $u$ and $v$.

Instead of using the zipper $\rm Z_3$, we can complement it with the segment lying between the two endpoints of $\rm Z_2$ and $\rm Z_3$. We then see that the so-extended $\rm Z_3$ line together with the $\rm Z_2$ line form a closed circuit. The result of Appendix \ref{zip} implies that the Green function derivatives $G'^{\rm op/cl}_{u,v}({\rm Z_2})$ and $G'^{\rm op/cl}_{u,v}({\rm Z_3})$ are equal or differ by a factor $\pm G^{\rm op/cl}_{u,v}$ depending on whether $u$ and/or $v$ are in the region delimited by the zippers $\rm Z_2$ and $\rm Z_3$ (including the boundary sites). We note that the fact that the edges at the bottom are outside the domain of interest does not alter the argument.

\begin{figure}[bt]
\centering
\begin{tikzpicture}[scale=0.8]
\draw (-4,-3)--(-4,3)--(5,3)--(5,-3);
\draw[step=0.5cm,help lines,semithick,dotted] (-4,-3) grid (5,3);
\fill[pattern=north east lines, pattern color=blue] (-4.0125,-3.8) rectangle (5.0125,-3.1);
\draw[thick,fill=blue!15!white] (0,0)--(0.5,0)--(0.5,0.5)--(0,0.5)--(-0.5,0.5)--(-0.5,-0.5)--(0,-0.5)--(0,0);
\filldraw (0.25,0.25) circle (0.075cm);
\filldraw (0.25,-3.25) circle (0.075cm);
\filldraw (2.75,-3.25) circle (0.075cm);
\draw[very thick,red] (0.25,0.25)--(0.25,2.25);
\draw[very thick,dotted,red] (0.25,2.25)--(0.25,3);
\draw[very thick,red] (0.25,0.25)--(0.25,-3.25);
\draw[very thick,red] (0.25,0.25)--(1.25,0.25)--(1.25,-1.25)--(2.75,-1.25)--(2.75,-3.25);
\draw[thick,->,red] (0,-3)--(0.5,-3);
\draw[thick,->,red] (0,-1.5)--(0.5,-1.5);
\draw[thick,->,red] (1,-0.5)--(1.5,-0.5);
\draw[thick,->,red] (2,-1.5)--(2,-1);
\draw[thick,->,red] (2.5,-3)--(3,-3);
\draw[thick,->,red] (0.5,1.5)--(0,1.5);
\draw[very thick,blue] (-3.25,-3.25)--(2.75,-3.25);
\draw[very thick,dotted,blue] (-4,-3.25)--(-3.25,-3.25);
\draw[thick,font=\small,red] (-0.25,2.25) node {$\rm Z_1$};
\draw[thick,font=\small,red] (-0.25,-2.25) node {$\rm Z_2$};
\draw[thick,font=\small,red] (3.25,-1.25) node {$\rm Z_3$};
\foreach \x in {-3.0,-2.5,...,2.5}  {\draw[thick,->,blue] (\x,-3)--(\x,-3.5);}
\filldraw[red] (0.25,0.25) circle (0.075cm);
\filldraw[red] (0.25,-3.25) circle (0.075cm);
\filldraw[red] (2.75,-3.25) circle (0.075cm);
\end{tikzpicture}
\caption{Three possible zippers on a semi-infinite annular-one graph, starting from the same point, the black dot inside the inner grey face, and terminating at infinity or on the boundary. The stretch at the bottom does not contribute to the Green function derivative, but helps to close the zipper lines.}
\label{zip_inf}
\end{figure}

The same argument allows one to relate the Green function derivatives associated with $\rm Z_1$ and $\rm Z_2$. Extending $\rm Z_2$ by the line going from the boundary end point of $\rm Z_2$ to the left infinity shows that this extended zipper line is deformable to that of $\rm Z_1$ (through the point at infinity). Hence, the corresponding Green function derivatives are equal or differ by $\pm G^{\rm op/cl}_{u,v}$ depending on whether $u$ and/or $v$ are crossed when the lines are moved. 

Therefore, any choice of zipper, be it finite or infinite, is equally good as any other; if the Green function derivatives are not equal, though closely related, they should yield the same results when computing grove probabilities. Let us  consider the specific zipper $\rm Z_1$ in Fig.~\ref{zip_inf}, starting on the edge $\big((1,y+1),(0,y+1)\big)$ and going upward, which is the zipper we used in the calculations of Section~\ref{sec6}, see Fig.~\ref{zipuhp}. For the open boundary condition, the corresponding Laplacian is noted $\mathbf{\Delta}^{\rm op}(z)$, while $\Gr^{\rm op}(z)$ is its inverse.

It is not difficult to see that $\Gr^{\rm op}(z)$ can be related, by the method of images, to an inverse Laplacian on the full plane. Because the method involves a reflection through the real axis, the relevant Laplacian on the full plane must be defined relative to two semi-infinite zippers, namely the original zipper Z$^{}_1$ in the UHP and its reflected version Z$^*_1$, starting on the edge $\big((1,-y-1),(0,-y-1)\big)$ and going downward. We shall denote by $\mathbf{\Delta}^*(z)$ and $\Gr^*(z)$ the corresponding Laplacian and its inverse. The double zipper Z$^{}_1 \cup \,{\rm Z}^*_1$ on the plane ensures the following symmetry $\Gr^*_{u,v}(z) = \Gr^*_{u,v^*}(z)$ whenever $u=(u_1,0)$ is on the real axis,  where $v^*=(v_1,-v_2)$ is the reflected site of $v=(v_1,v_2)$. It then follows that $\Gr^{\rm op}(z)$ is equal to
\be
\Gr^{\rm op}_{u,v}(z) = \Gr^*_{u,v}(z) - \Gr^*_{u,v^*}(z), \qquad u,v \in {\rm UHP}.
\label{images}
\ee
At order 0 in $(1{-}z)$, it yields the usual relation \eqref{Gop}, while at order 1, we obtain,
\be
G'^{\rm op}_{u,v} = G'^{\uparrow}_{u,v} + G'^{\downarrow}_{u,v} - G'^{\uparrow}_{u,v^*} - G'^{\downarrow}_{u,v^*}\,,
\qquad v^* = (v_1,-v_2),
\ee
where the up (resp. down) arrow refers to ${\rm Z}^{}_1$ (resp. ${\rm Z}^*_1$).

A translation and a change of orientation, combined with a reflection for ${\rm Z}^{}_1$, bring the two zippers ${\rm Z}^{}_1$ and ${\rm Z}^*_1$ onto the zipper we have used on the plane in Sections \ref{sec4} and \ref{sec5} (namely, starting at the edge $((0,0),(1,0))$ and going downward). We therefore find the following relations:
\be
G'^{\uparrow}_{u,v} = -G'_{u^*+(0,y+1),v^*+(0,y+1)}, \qquad G'^{\downarrow}_{u,v} = -G'_{u+(0,y+1),v+(0,y+1)},
\ee
where $G'$ is computed relative to the zipper we used on the plane. Combining the previous two equations, we obtain Eq.~\eqref{Gpop}:
\begin{align}
G'^{\textrm{op}}_{(u_1,u_2),(v_1,v_2)} &= -\,G'_{(u_1,-u_2+y+1),(v_1,-v_2+y+1)} - G'_{(u_1,u_2+y+1),(v_1,v_2+y+1)}\nonumber\\
&\hspace{2cm} \quad+\,G'_{(u_1,-u_2+y+1),(v_1,v_2+y+1)} + G'_{(u_1,u_2+y+1),(v_1,-v_2+y+1)}.
\end{align}

The arguments are easily adapted to the closed boundary condition by taking into account the appropriate reflection, leading to
\be
G'^{\rm cl}_{u,v} = G'^{\uparrow}_{u,v} + G'^{\downarrow}_{u,v} + G'^{\uparrow}_{u,v^*} + G'^{\downarrow}_{u,v^*}\,,
\qquad v^* = (v_1,1-v_2).
\ee
Paying attention to the way the zipper ${\rm Z}^{}_1$ must be reflected readily gives the expression quoted in the text, in \eqref{Gpcl}. Similarly, one finds that the corresponding derivatives on the diagonal upper half-plane are given by Eqs.~\eqref{Gpop_D} and \eqref{Gpcl_D}.

%%%%%%%%%%%%%%%%%%%%%%%%%%%%%%%%%%%%%%%%%%%%%%%%%%%%%%%%%%%%%%%%%%%%%%%%%%%%%%%%%%%%%%%%%%%%%%%%%%%%%%%%

\subsection{Modified graphs}
\label{modgra}

Most of the spanning tree computations presented in the text involve the removal of one or more edges from the graph $\G=\Z^2$ or $\G=\Z\times\mathbb{N}^*$. The resulting Laplacian $\xbar{\mathbf{\Delta}}(z)$ on the modified graph $\xbar{\G}$ is a local perturbation of the original Laplacian $\mathbf{\Delta}(z)$ on the full graph by a matrix of finite rank. The inverse $\xbar{\Gr} = [\xbar{\mathbf{\Delta}}(z)]^{-1}$ and the determinant of $\xbar{\mathbf{\Delta}}(z)$ can be computed in terms of the corresponding quantities for $\mathbf{\Delta}(z)$ by making use of the Woodbury formula. As $\mathbf{\Delta}(z)^{\rm t}=\mathbf{\Delta}(z^{-1})$ and $\xbar{\mathbf{\Delta}}(z)^{\rm t}=\xbar{\mathbf{\Delta}}(z^{-1})$, the modified Laplacian can be written as $\xbar{\mathbf{\Delta}} = \mathbf{\Delta} - U^{\rm t}U$ if the removed edges have a trivial parallel transport. In which case the Woodbury formula implies
\bea
&& \xbar{\mathbf{\Delta}}^{-1} = \mathbf{\Delta}^{-1} + [\mathbf{\Delta}^{-1} U^{\rm t} \, ({\mathbb I} - U \mathbf{\Delta}^{-1} U^{\rm t})^{-1} \, U \mathbf{\Delta}^{-1}], \label{wood} \\
\noalign{\medskip}
&& \det \,\xbar{\mathbf{\Delta}} = \det \mathbf{\Delta} \times \det ({\mathbb I} - U \mathbf{\Delta}^{-1} U^{\rm t}).
\label{wood2}
\eea
When the perturbation $U^{\rm t}U$ has finite rank $r$, the matrix $U$ may be taken as an $r \times \infty$ rectangular matrix. Then the matrix to be inverted, ${\mathbb I} - U \mathbf{\Delta}^{-1} U^{\rm t}$, is $r$-dimensional, as is the last determinant on the second line. If $U^{\rm t}U$ has rank 1, the Woodbury formula reduces to the Sherman-Morrison formula.

Let us first illustrate the use of the Woodbury formula when two edges are removed, as was the case for single-site probabilities on the plane, reviewed in Section~\ref{sec4}. The computations required to remove the two edges $\{5,2\}$ and $\{5,3\}$. The only nonzero entries of the perturbation $U^{\rm t}U$ have row and column indices in the set $\{5,2,3\}$, and are given by
\be
U^{\rm t}U\Big|_{\{5,2,3\}} = 
\begin{pmatrix} 
2 & -1 & -1 \\ -1 & 1 & 0 \\ -1 & 0 & 1
\end{pmatrix}.
\ee
This matrix being of rank 2, a convenient choice is to take 
\be
U = \begin{pmatrix}
\ \cdots & \!\!\!\begin{matrix}
1 & -1 & 0 \\ 1 & 0 & -1 
\end{matrix} & \!\!\cdots\ 
\end{pmatrix},
\label{firstU}
\ee
where the columns shown are labeled by the vertices 5, 2 and 3, all the others being identically zero.

The matrices $\xbar{\mathbf{\Delta}}$ and $\mathbf{\Delta}$ in \eqref{wood} depend on $z$, but for the purpose of computing $\xbar{G}$ and $\xbar{G}'$, only the zeroth and first orders in $z-1$ are required. By using the following explicit values of $G,G'$ on the plane, see \eqref{gprime},
\bea
&& \textstyle G_{2,5} = G_{3,5} = G_{0,0}-\frac14, \quad G_{2,3} = G_{0,0}-\frac1\pi,\\
\noalign{\medskip}
&& \textstyle G'_{2,5} = \frac14 G_{0,0} - \frac{3}{32}, \quad G'_{2,3} = (\frac12-\frac1\pi)G_{0,0} - \frac1{4\pi}, \quad G'_{5,3} = (\frac14-\frac1\pi)G_{0,0} + \frac1{32},
\eea
one finds the expansion of $({\mathbb I} - U \mathbf{\Delta}^{-1} U^{\rm t})^{-1}$ to first order,
\be
({\mathbb I} - U \mathbf{\Delta}^{-1} U^{\rm t})^{-1} = \frac\pi{2(\pi-1)}\begin{pmatrix}
\pi & \pi-2 \\ \pi-2 & \pi
\end{pmatrix}
+ \frac{\pi(4-\pi)}{16(\pi-1)} \begin{pmatrix}
0 & 1 \\ -1 & 0 
\end{pmatrix} (z-1) + {\cal O}(z-1)^2.
\ee
It leads to the following expression for $\xbar{\Gr}$, valid to first order,
\bea
\xbar{\Gr}_{u,v} \egal \Gr_{u,v} + \frac\pi2 \big(2\Gr_{u,5}{-}\Gr_{u,2}{-}\Gr_{u,3}\big)\big(2\Gr_{5,v}{-}\Gr_{2,v}{-}\Gr_{3,v}\big) + \frac\pi{2(\pi{-}1)} \big(\Gr_{u,2}{-}\Gr_{u,3}\big)\big(\Gr_{2,v}{-}\Gr_{3,v}\big) \nonumber\\
\noalign{\smallskip}
&& \hspace{-1.4cm} + \; \frac{\pi(4{-}\pi)}{32(\pi{-}1)} (z{-}1) \Big\{\big(G_{u,2}{-}G_{u,3}\big)\big(2G_{5,v}{-}G_{2,v}{-}G_{3,v}\big) - \big(2G_{u,5}{-}G_{u,2}{-}G_{u,3}\big)\big(G_{2,v}{-}G_{3,v}\big)\Big\} + \ldots \nonumber\\
\eea
from which the explicit formulas for $\xbar{G}$ and $\xbar{G}'$ needed in Section~\ref{sec4} are easily derived. The ratio of partition functions, needed in the same calculations of Section~\ref{sec4}, is straightforward to compute in the limit $z \to 1$:
\be
\frac{\xbar{Z}}{Z} = \frac{\det \xbar{\Delta}}{\det \Delta} = \det ({\mathbb I} - U \Delta^{-1} U^{\rm t}) = \frac{\pi-1}{\pi^2}.
\ee

In the calculation of multisite probabilities presented in Section~\ref{sec5}, extra lattice changes were to be considered, namely the removal of three edges in the neighborhood of each height 1. For simplicity, let us focus on $\P_{2,1}(\vec r)$, which was shown to be given in terms of essentially the same grove fractions as for $\P_2$, but on a lattice $\widetilde \G$ obtained from $\Z^2$ by cutting three edges around the height 1. One may therefore proceeds in two steps.

The first step relates the Green function $\xbar \Gr(z)$ on the fully modified lattice $\xbar{\G}$ to the function $\widetilde \Gr(z)$ pertaining to $\widetilde \G$. For this, we use the Woodbury formula \eqref{wood} with the matrix given in \eqref{firstU}. Because the entries $\xbar \Gr_{u,v}(z)$ are only required for $u,v$ close to site $i$ (where the height 2 is located), see the expression \eqref{X1S}, we similarly need the entries of $\widetilde \Gr(z)$ for sites close to $i$.

The second step is to relate $\widetilde \Gr(z)$ to $\Gr(z)$, the $z$-dependent Green function on the usual square lattice $\Z^2$. For this, we use a second matrix $V$ implementing the removal of the three edges between site $7$ and sites 8, 9 and 10, as pictured in Figure \ref{P21}. Again, a convenient choice is to set
\be
V = \begin{pmatrix}
\ \cdots & \!\!\!\begin{matrix}
1 & -1 & 0 & 0\\ 1 & 0 & -1 & 0\\ 1 & 0 & 0 & -1
\end{matrix} & \!\!\cdots\ 
\end{pmatrix}.
\label{removal1}
\ee
However, since $V$ is located around the distant site $j$ (with height 1), the Woodbury identity shows that the calculation of $\widetilde \Gr_{u,v}(z)$ for $u,v$ close to site $i$ requires the knowledge of $\Gr_{u,v}(z)$ for $u$ and/or $v$ close to site $j$, namely far from the head of the zipper. The zeroth order in $z-1$, namely $G_{u,v}$, is well known and has been recalled in \eqref{Guv}. To compute the first order $G'_{u,v}$, the technique reviewed in \ref{zip} is no longer helpful, as one would need to add an arbitrarily large number of new zipper edges. Indeed, we found it more convenient to resort to the defining expression \eqref{Gpdef} for $G'_{u,v}$ in which we use the integral representation \eqref{Gsq} of $G_{u,v}$. In this way, the infinite summation over the edges of the zipper can easily be carried out. 

As an illustration, let us review the calculation of $G'_{u,v}$ where $u=(0,0)$ is the origin and $v=\vec r = (p,q)$ is the site 7 (the height 1). As before, the edges crossed by the zipper form the set $\{(k,\ell) = (k,k+(1,0))\}$ where $k$ runs over the sites $(0,-m)$ with $m \ge 0$. For simplicity, we assume $q$ to be positive and large, and show later on how to deal with other cases, i.e. $q$ small and/or negative. Since calculations of two-site probabilities in Section~\ref{sec5} are carried out to order $r^{-6}$, our purpose here is likewise to obtain the asymptotic expansion of $G'_{0,\vec r}$ to that order. According to \eqref{Gpdef}, and for $q$ positive, it is given by 
\bea
G'_{0,\vec r} \egal  \sum_k \: (G_{0,\l}\,G_{k,\vec r} - G_{0,k}\,G_{\l,\vec r})  = 
\sum_{m \ge 0} \Big(G(1,m)G(p,q{+}m) - G(0,m)G(p{-}1,q{+}m) \Big) \nonumber\\
\egal \sum_{m=0}^\infty \; \int\!\!\!\!\int_0^\pi \frac{{\rm d}\alpha_1 {\rm d}\alpha_2}{4\pi^2} \; \frac{\big[\cos{p\alpha_1} \cos{\alpha_2} - \cos{(p-1)\alpha_1}\big]}{\sqrt{y_1^2-1}\sqrt{y_2^2-1}} \: t_1^{q+m} \, t_2^m \nonumber\\
\noalign{\smallskip}
\egal \int_0^\pi \frac{{\rm d}\alpha_1}{2\pi} \: \frac{t_1^q}{\sqrt{y_1^2-1}} \; \int_0^\pi \frac{{\rm d}\alpha_2}{2\pi} \; \frac{\big[\cos{p\alpha_1} \cos{\alpha_2} - \cos{(p-1)\alpha_1}\big]}{\sqrt{y_2^2-1}} \: \frac{1}{1-t_1t_2},
\label{Gor}
\eea
where $y_i\equiv y(\alpha_i)$ and $t_i \equiv t(\alpha_i)$, defined right after \eqref{Gsq}. 

The principle underlying the asymptotic evaluation of this double integral is simple \cite{JPR06}. One first observes that the function $t_1$ of $\alpha_1$ decreases away from the origin. From $t_1 \simeq 1-\alpha_1+\ldots$ for small $\alpha_1$, it follows that for large $q$, $t_1^q \simeq {\rm e}^{-q\alpha_1}$ decreases exponentially (with polynomial corrections, see below). This suggests to expand the rest of the integrand in powers of $\alpha_1$ and simply integrate term by term. A simple dimensional analysis shows that the integral of ${\rm e}^{-q\alpha_1}\alpha_1^k$ contributes to order $r^{-(k+1)}$, so that the expansion of the integrand to order $\alpha_1^5$ is sufficient. In fact, the only integrals we shall need are the following one,
\be
\int_0^\infty \: {{\rm d}\alpha_1} \, {\rm e}^{-q\alpha_1} \frac{\sin{p\alpha_1}}{\alpha_1} (A \log\alpha_1 + B) = \frac1{2} \Big\{2B - A \big[\!\log{(p^2+q^2)} + 2\gamma \big]\Big\} \arctan{\frac pq},
\label{elem}
\ee
and its $p$- and $q$-derivatives (in order to bring higher powers of $\alpha_1$ in the integrand). The extension of the integration domain from $[0,\pi]$ to $[0,\infty)$ is valid up to exponentially small corrections.

The idea just explained is simple but requires extra care for two reasons. First, the naive expansion  of the integrand in powers of $\alpha_1$, before doing the integral over $\alpha_2$, is not allowed because it yields increasingly singular functions of $\alpha_2$, which are not integrable. And second, as noted above, $G'_{0,\vec r}$ is expected to contain a divergent piece proportional to $G_{0,0}$, which needs to be properly identified. 

In order to handle these two difficulties, we split the expression \eqref{Gor} into three pieces, 
\bea
G'_{0,\vec r} \egal \int_0^\pi \frac{{\rm d}\alpha_1}{4\pi^2} \: \frac{t_1^q}{\sqrt{y_1^2-1}} \; \int_0^\pi {\rm d}\alpha_2 \; \frac{\cos{p\alpha_1} (\cos{\alpha_2} - 1)}{\sqrt{y_2^2-1}} \: \frac{1}{1-t_1t_2} \\
\noalign{\smallskip}
\plus \int_0^\pi \frac{{\rm d}\alpha_1}{4\pi^2} \: \frac{t_1^q}{\sqrt{y_1^2-1}} \; \int_0^\pi {\rm d}\alpha_2 \; \frac{\cos{p\alpha_1} - \cos{(p-1)\alpha_1}}{\sqrt{y_2^2-1}} \: \Big(\frac{1}{1-t_1t_2} - \frac1{1-t_1}\Big) \label{g2} \\
\noalign{\smallskip}
\plus \int_0^\pi \frac{{\rm d}\alpha_1}{4\pi^2} \: \frac{t_1^q}{\sqrt{y_1^2-1}} \; \int_0^\pi {\rm d}\alpha_2 \; \frac{\cos{p\alpha_1} - \cos{(p-1)\alpha_1}}{\sqrt{y_2^2-1}} \: \frac1{1-t_1},
\label{g3}
\eea
which we call respectively $G_1$, $G_2$ and $G_3$.

\medskip\noindent
{\bf First contribution.} We rewrite the function of $\alpha_2$ involved in $G_1$ as
\be
\left\{\frac{\cos{\alpha_2} - 1}{\sqrt{y_2^2-1} (1-t_1t_2)} - \frac{\alpha_1 + \frac{\alpha_1^3}{12} + \frac{\alpha_1^5}{120}}{2\sqrt{y_3^2-1}} \right\} + \frac{\alpha_1 + \frac{\alpha_1^3}{12} + \frac{\alpha_1^5}{120}}{2\sqrt{y_3^2-1}},
\label{G1_det}
\ee
where $y_3 \equiv y(\alpha_1+\alpha_2)$. In the first term inside the brackets, the subtracted term is such that the expansion in $\alpha_1$ to order 6 produces coefficients that are regular functions of $\alpha_2$, which can be integrated exactly,
\be
\int_0^\pi {\rm d}\alpha_2 \left\{\frac{\cos{\alpha_2} - 1}{\sqrt{y_2^2-1} (1-t_1t_2)} - \frac{\alpha_1 + \frac{\alpha_1^3}{12} + \frac{\alpha_1^5}{120}}{2\sqrt{y_3^2-1}} \right\} = -\frac{3\pi}4 - \frac{\alpha_1^2}{4\sqrt{2}} - \frac{7\alpha_1^4}{192\sqrt{2}} - \frac{137\alpha_1^6}{30720\sqrt{2}} + \ldots
\ee
The function in the second term of Eq.~\eqref{G1_det} can also be integrated exactly and then expanded for small $\alpha_1$,
\begin{align}
\int_0^\pi \frac{{\rm d}\alpha_2}{\sqrt{y_3^2-1}} &= \Big[-{\rm arcth}\frac{\sqrt{2} \cos\frac{\alpha_2}2}{\sqrt{3-\cos{\alpha_2}}}\Big]^{\pi+\alpha_1}_{\alpha_1} \nonumber\\
&= -\log{\alpha_1} + \tfrac 32 \log 2 +\frac{\alpha_1}{2\sqrt{2}} + \frac{\alpha_1^2}{24} + \frac{\alpha_1^3}{32\sqrt{2}} \nonumber\\
&\qquad-\frac{43\alpha_1^4}{5760} + \frac{11\alpha_1^5}{5120\sqrt{2}} + \frac{949\alpha_1^6}{725760} + \ldots
\label{arcth}
\end{align}
The two terms together and the further expansion of $(y_1^2-1)^{-1/2}$ yield 
\be
G_1 = -\frac38 G_{0,\vec r} + \int_0^\pi \frac{{\rm d}\alpha_1}{8\pi^2} \: t_1^q \, \cos{p\alpha_1} \Big\{\big(-\log{\alpha_1}+\frac 32 \log{2}\big)\big(1 + \frac{\alpha_1^4}{32}\big) + \frac{\alpha_1^2}{24} - \frac{43 \alpha_1^4}{5760} + \ldots \Big\}.
\ee

\medskip\noindent
{\bf Second contribution.} We use the same subtraction trick to rewrite the function in \eqref{g2} as 
\begin{equation*}
\left\{\frac{1}{\sqrt{y_2^2-1}} \: \Big(\frac{1}{1-t_1t_2} - \frac1{1-t_1}\Big) + \frac{1 + \frac{\alpha_1^2}4 + \frac{\alpha_1^4}{96} + \frac{19\alpha_1^6}{5760}}{\sqrt{y_1^2-1}\sqrt{y_3^2-1}}\right\} - \frac{1 + \frac{\alpha_1^2}4 + \frac{\alpha_1^4}{96} + \frac{19\alpha_1^6}{5760}}{\sqrt{y_1^2-1}\sqrt{y_3^2-1}}
\end{equation*}
and apply the same method as for $G_1$. The integral of the function in curly brackets yields
\begin{align}
&\int_0^\pi {\rm d}\alpha_2 \left\{\frac{1}{\sqrt{y_2^2-1}} \: \Big(\frac{1}{1-t_1t_2} - \frac1{1-t_1}\Big) + \frac{1 + \frac{\alpha_1^2}4 + \frac{\alpha_1^4}{96} + \frac{19\alpha_1^6}{5760}}{\sqrt{y_1^2-1}\sqrt{y_3^2-1}}\right\} \nonumber\\
&\quad =\frac{1}{2\sqrt{2}} + \frac{11\alpha_1^2}{96\sqrt{2}} + \frac{787\alpha_1^4}{46080\sqrt{2}} + \ldots
\end{align}
Together with the previous result \eqref{arcth}, simple trigonometric identities and a few more expansions in $\alpha_1$, it leads to the following expression for $G_2$,
\begin{align}
G_2 &= -\int_0^\pi \frac{{\rm d}\alpha_1}{8\pi^2} \: t_1^q \, \cos{p\alpha_1} \Big\{\big(-\log{\alpha_1}+\frac 32 \log{2}\big)\big(1 + \frac{\alpha_1^4}{32}\big) + \frac{\alpha_1^2}{24} - \frac{43 \alpha_1^4}{5760} + \ldots \Big\}  \nonumber\\
&\quad + \int_0^\pi \frac{{\rm d}\alpha_1}{8\pi^2} \: t_1^q \, \sin{p\alpha_1} \Big\{\big(-\log{\alpha_1}+\frac 32 \log{2}\big)\big(\frac2{\alpha_1} - \frac{\alpha_1}{6} + \frac{43\alpha_1^3}{720} - \frac{949\alpha_1^5}{60480}\big)\nonumber\\
&\hspace{4cm} + \frac{\alpha_1}{12} - \frac{7\alpha_1^3}{320} + \frac{4607\alpha_1^5}{725760} + \ldots \Big\}.
\end{align}
Curiously, the first integral is exactly the opposite of that in $G_1$, so that the two cancel out.

\medskip\noindent
{\bf Third contribution.} The last contribution \eqref{g3} is the simplest one since the two integrals are decoupled, the one over $\alpha_2$ simply giving a multiple of $G_{0,0}$. A few expansions yield
\bea
G_3 \egal -G_{0,0} \int_0^\pi \frac{{\rm d}\alpha_1}{2\pi} \: t_1^q \Big\{\sin{p\alpha_1}\:  \big(\frac{1}{\alpha_1} + \frac12 - \frac{\alpha_1}{12} - \frac{\alpha_1^2}8 -\frac{\alpha_1^3}{720} + \frac{5\alpha_1^4}{192} - \frac{\alpha_1^5}{30240} + \ldots\big)\nonumber\\
\noalign{\smallskip}
&& \hspace{4cm} + \: \cos{p\alpha_1} \: \big(\frac12 + \frac{\alpha_1}{4} - \frac{\alpha_1^3}{24} + \frac{19\alpha_1^5}{1920} + \ldots\big)\Big\}.
\eea

\medskip
The last step before doing the remaining integrals is to recast $t_1^q$ into a more workable function. The expansion of $\log{t_1}$,
\be
\log{t_1} = -\alpha_1 + \frac{\alpha_1^3}{12} - \frac{\alpha_1^5}{96} + \frac{79\alpha_1^7}{40320} + \ldots,
\ee
shows that the following expansion is sufficient to finish the calculations to the required order,
\be
t_1^q = {\rm e}^{-q\alpha_1} \: \Big(1 + \frac{q\alpha_1^3}{12} - \frac{q\alpha_1^5}{96} + \frac{q^2\alpha_1^6}{288} + \frac{79\alpha_1^7}{40320} - \frac{q^2\alpha_1^8}{1152} + \frac{q^3 \alpha_1^9}{10368} + \ldots\Big).
\ee

By using the integral \eqref{elem}, the rest of the computation is straightforward. For completeness, we give the final result, more conveniently expressed in polar coordinates, $p=r\cos\varphi$, $q=r\sin\varphi$ with $0<\varphi<\pi$, so $p^2+q^2=r^2$ and $\arctan\left(p/q\right)=\pi/2-\varphi$:
\small
\bea
G'_{0,\vec r} \egal G_{0,0} \Big(\frac{\varphi}{2 \pi } - \frac{5}{8} - \frac{\cos\varphi{-}\sin \varphi}{4 \pi  r} - \frac{3 \cos 2 \varphi{-}\sin 4 \varphi}{24 \pi  r^2} - \frac{\cos 3 \varphi{+}\sin 3 \varphi{+}\cos 5\varphi{-}\sin 5 \varphi}{16 \pi  r^3} \nonumber\\
\noalign{\smallskip}
&& \hspace{2cm} -\: \frac{27 \sin 4 \varphi{+}60 \cos 6 \varphi{-}25 \sin 8 \varphi}{480\, \pi  r^4} - \frac{5 [\cos 7 \varphi{+}\sin 7 \varphi{+}\cos 9 \varphi{-}\sin 9 \varphi]}{32 \pi  r^5} \nonumber\\
\noalign{\smallskip}
&& \hspace{2cm} -\:  \frac{189 \cos 6 \varphi{+}972 \sin 8 \varphi{+}2205 \cos 10 \varphi{-}980 \sin 12\varphi}{4032\, \pi  r^6}+ \ldots \Big) \nonumber\\
\noalign{\smallskip}
&&\hspace{-1cm}-\,\big(\log r{+}\gamma{+}\tfrac32 \log 2\big) \Big(-\frac{5\pi{-}4\varphi}{16\pi^2} + \frac{\sin 4\varphi}{48\pi^2r^2} + \frac{18\sin 4\varphi{+}25 \sin 8\varphi}{960\,\pi^2r^4} + \frac{459\sin 8\varphi {+}490 \sin 12\varphi}{4032\,\pi^2r^6} + \ldots\Big) \nonumber\\
\noalign{\smallskip}
&&\hspace{-1cm} -\,\frac14 (5\pi{-}4\varphi) \Big(\frac{\cos 4\varphi}{48\pi^2r^2} + \frac{18\cos 4\varphi{+}25 \cos 8\varphi}{960\,\pi^2r^4} + \frac{459\cos 8\varphi{+}490 \cos 12\varphi}{4032\,\pi^2r^6} + \ldots \Big)\nonumber\\
\noalign{\smallskip}
&&\hspace{-1cm} +\,\frac{\sin 4\varphi}{32\pi^2r^2} + \frac{90\sin 4\varphi{+}137 \sin 8\varphi}{2304\,\pi^2r^4} + \frac{3483\sin 8\varphi{+}3805\sin 12\varphi}{11520 \, \pi^2r^6} + \ldots
\eea
\normalsize
The expression, albeit complicated, remarkably simplifies for diagonal positions $\vec r=(p,p)$ $\left(\varphi=\frac{\pi}4\right)$ and vertical positions $\vec r=(0,q)$ $\left(\varphi=\frac{\pi}{2}\right)$:
\begin{align}
G'_{0,(p,p)}&=-\frac12 G_{0,0} + \frac1{4\pi} \big(\log r + \gamma + \tfrac32 \log 2\big) +\frac1{48\pi r^2} - \frac{7}{960\,\pi r^4} + \frac{31}{4032\,\pi r^6} + \ldots\\
G'_{0,(0,q)}&=G_{0,0}\left(-\frac{3}{8}+\frac{1}{4\pi r}+\frac{1}{8\pi r^2}+\frac{1}{8\pi r^3}+\frac{1}{8\pi r^4}+\frac{5}{16\pi r^5}+\frac{19}{32\pi r^6}\right)\nonumber\\
&\quad+\frac{3}{16\pi}\big(\log r+\gamma+\tfrac{3}{2}\log 2\big)-\frac{1}{64\pi r^2}-\frac{43}{1280\pi r^4}-\frac{949}{5376\pi r^6}+\ldots
\end{align}
This is best understood using the technique reviewed in Appendix~\ref{zip}, which enables us to write $G'_{0,\vec{r}}$ as a finite sum whose number of terms grows linearly with $p$ or $q$. In case $p=q$ or $p=0$, huge cancellations happen, and we are left with only a few terms:
\begin{align}
G'_{0,(p,p)}&=-\frac{1}{2}G(p,p),\\
G'_{0,(0,q)}&=-\frac{1}{2}G(0,q)+\frac{1}{2}G(0,0)G(0,q-1)-\frac{1}{2}G(1,0)G(0,q).
\end{align}

Next we discuss the extension of $G'_{0,\vec{r}}$ with $r\gg 1$ to $\pi\le\varphi\le 2\pi$. Remember indeed that the development presented above only holds if $q=r\sin\varphi\gg 1$. If $q$ is small (i.e. $\varphi$ close to $0$, $\pi$ or $2\pi$), $G'_{0,\vec{r}}$ can still be evaluated using transformation properties of $G'$ under zipper deformations:
\begin{align}
G'_{0,(p,q)}&=G'_{0,(-q,p)}-G(0,0)G(p-1,q)+G(1,0)G(p,q)\quad\textrm{for $|p|\gg 1$ and $p>0$,}\\
G'_{0,(p,q)}&=G'_{0,(q,-p)}+G(0,0)G(p,q-1)-G(1,0)G(p,q)\quad\textrm{for $|p|\gg 1$ and $p<0$}.
\end{align}
If rather $|q|\gg 1$ but $q<0$, the asymptotic expression of the derivative of the Green function may be obtained from the following relation:
\be
G'_{0,\vec r}=G'_{0,-\vec r}+G_{0,0}\big[G(p,q-1)-2 G(p,q)+ G(p+1,q)\big]-\frac12 {\rm sign}(p-\tfrac12) \, G(p,q).
\ee

%%%%%%%%%%%%%%%%%%%%%%%%%%%%%%%%%%%%%%%%%%%%%%%%%%%%%%%%%%%%%%%%%%%%%%%%%%%%%%%%%%%%%%%%%%%%%%%%%%%%%%%%

\section{Symmetries and maps between predecessor diagrams on half-planes}
\label{Diag_sym}

In Section~\ref{sec6}, we computed one-site probabilities $\P_a(i)$ on (horizontal and diagonal) upper half-planes, and several two-site probabilities $\P_{a,1}(i,j)$ on the horizontal half-plane. The boundary conditions considered were either fully open or fully closed. Contrary to their full-plane analogues for one-site probabilities, predecessor diagrams on half-planes are not invariant under rotations by $90^\circ$. However, the probabilities of occurrence of some diagrams are related to one another through simple transformations. Since all four cases are similar, we provide a detailed discussion only for the upper half-plane $\G=\Z\times\mathbb{N}^*$ with a horizontal open boundary. For more generality, we discuss the case of an $n$-site probability with $n{-}1$ heights equal to 1. As explained in Section~\ref{sec3}, the probabilities associated with predecessor diagrams only depend on the Green function $G^{\textrm{op}}$ and its derivative $G'^{\,\textrm{op}}$. Since we shall not refer to other types of Green functions in this section, we shall drop the superscript ``$\textrm{op}$'' for both functions, i.e. write $G,G'$ for $G^{\text{op}},G'^{\,\text{op}}$.

Let us consider multisite probabilities $\P_{a,1,\cdots ,1}(i_1,\ldots,i_n)$ on the upper half-plane $\G=\Z\times\mathbb{N}^*$ with open boundary conditions, where the height $a$ is at $i_1$. The reference sites $i_k$ have coordinates $(x_k,y_k)$ and are assumed to be far away from one another and from the boundary. For each $k$, we denote by $\V_k=\{i_k,{\rm N}_k,{\rm E}_k,{\rm S}_k,{\rm W}_k\}$ the close neighborhood of $i_k$, containing $i_k$ and its four nearest neighbors. The actual calculation of $\P_{a,1,\ldots ,1}(i_1,\ldots,i_n)$ requires that the graph $\G$ be modified by removing certain edges around the reference sites. The removal of edges around $i_k$ is implemented by a matrix $U_k$ so that the Laplacian on the modified graph $\xbar{\G}$ can be written as $\xbar{\mathbf{\Delta}} = \mathbf{\Delta} - U^{\rm t}U = \mathbf{\Delta} - \sum_{k=1}^n U^{\rm t}_k U^{}_k$ (see Appendix~\ref{modgra}; the matrix $U$ here is obtained simply by piling up the rectangular matrices $U_k$). Note that each of the $U_k$'s is defined up to a sign.

\subsubsection*{Height-one probabilities}
We first consider the joint probabilities $\P_{1,\ldots ,1}(i_1,\ldots,i_n)$ of heights 1. In this case, each $U_k$ is a rectangular $3 \times \infty$ matrix, of the form given in \eqref{removal1}: it is identically zero for column labels outside of $\V_k$, the column with entries equal to 1 corresponds to $i_k$, while the three columns with one entry equal to $-1$ are labeled by three neighbors of $i_k$. As recalled above, which three neighbors of $i_k$ are chosen is irrelevant. We can write the multisite probability symmetrically, as a $3n \times 3n$ determinant,
\begin{equation}
\P_{1,\ldots ,1}(i_1,\ldots,i_n) = \det (\mathbb{I} - U G U^{\rm t} ),
\label{P11_sym}
\end{equation}
with $G = \left(\Delta_{\scriptsize{\G}}\right)^{-1}$ is the Green matrix on the (unmodified) upper half-plane with open boundary condition. We want to show that $\P_{1,\ldots,1}(i_1,\ldots,i_n)$ is an even function of each of the $y_k$ variables. 

Because of the structure of $U$, the determinant involves only the Green matrix entries contained in the blocks $G_{\V_\ell,\V_\ell'}$, for $1 \le \ell,\ell' \le n$. The transformation $y_k \mapsto -y_k$ affects those entries of $G$ labeled by sites in $\V_k$. If one denotes the neighbors of $i_k$ as $(x_k+a,y_k+b)$ for some $a,b=0,\pm 1$, then \eqref{Gop}, or \eqref{images}, implies the following transformations:
\begin{align}
& G_{(x_k+a,y_k+b),(x_k+c,y_k+d)}\Big|_{y_k\mapsto -y_k} = G_{(x_k+a,-y_k+b),(x_k+c,-y_k+d)} = G_{(x_k+a,y_k-b),(x_k+c,y_k-d)},\label{Gop_sym1}\\
\noalign{\medskip}
& G_{(x_k+a,y_k+b),v}\Big|_{y_k\mapsto -y_k} = - \,G_{(x_k+a,y_k-b),v}, \quad 
G_{u,(x_k+c,y_k+d),v}\Big|_{y_k\mapsto -y_k} = - \,G_{u,(x_k+c,y_k-d)}, 
\label{Gop_sym2}
\end{align}
if $u$ and $v$ are far from $i_k$. Up to a sign, replacing $y_k$ with its opposite amounts to exchanging the northern neighbor N$_k$ of $i_k$ with its southern neighbor S$_k$. This is equivalent to conjugating $G$ with a matrix $\Sigma_k$, equal to the infinite permutation matrix $\sigma_k$ permuting N$_k$ and S$_k$ times a matrix that is minus the identity on the $5 \times 5$ block labeled by sites of $\V_k$, and the identity elsewhere. We write 
\begin{equation}
\Sigma_k = {\mathbb I}_{\G \setminus \V_k} \oplus (-\sigma_k)\Big|_{\V_k}, \qquad \sigma_k\Big|_{\V_k} = \bordermatrix{& i_k & {\rm N}_k & {\rm E}_k & {\rm S}_k & {\rm W}_k\cr & 1 & 0 & 0 & 0 & 0\cr & 0 & 0 & 0 & 1 & 0\cr & 0 & 0 & 1 & 0 & 0\cr & 0 & 1 & 0 & 0 & 0\cr & 0 & 0 & 0 & 0 & 1}.
\label{Sigma_def}
\end{equation}
This leads to the following transformation
\begin{equation}
\P_{1,\ldots,1}(i_1,\ldots,i_n)\Big|_{y_k \mapsto -y_k} = \det (\mathbb{I} - U \Sigma_k \, G \,\Sigma^{\rm t}_k U^{\rm t} ).
\end{equation}

\noindent
It is not difficult to see that the factor $-\sigma_k$ can be absorbed into a redefinition of $U_k$ into $-U_k \sigma_k$ (since $U_k$ is identically zero for column indices outside of $\V_k$). As observed above, the sign is irrelevant, so we can say that $U_k$ is simply redefined into $\widehat{U}_k \equiv  U_k \sigma_k$, which means that a different choice is made for the removal of edges around $i_k$. Since this probability does not depend on this choice (see Section~\ref{sec2}), we obtain the result as claimed,
\begin{equation}
\P_{1,\ldots,1}(i_1,\ldots,i_n)\Big|_{y_k \mapsto -y_k}=\P_{1,\ldots,1}(i_1,\ldots,i_n).
\end{equation}
Let us note that the identity is true to all orders in the variables $y_k$, and not only at the order relevant for the scaling limit.

\subsubsection*{Predecessor diagrams with leaves}
Next we look at joint probabilities $\P_{a,1,\ldots,1}(i_1,\ldots,i_n)$ for which site $i_1$ has height $a \ge 2$. It requires to compute various fractions of spanning trees where $i_1$ has at least one predecessor among its neighbors while all other reference sites have none. We shall denote by $X_D^{\G}$ the fraction of spanning trees on $\G$ corresponding to a given predecessor diagram $D$ around $i_1$. As shown in the text in Section~\ref{sec5}, it can be done in two steps. The first step considers a modified lattice $\widetilde{\G}$ in order to enforce a height 1 at the sites $i_k$, for $k \ge 2$. This first modification is controlled by the same rectangular matrices $U_{k \ge 2}$, discussed in the previous case. We then obtain 
\begin{equation}
X_D^{\G} = X_D^{\scriptsize\widetilde{\G}}\times\frac{\det\Delta_{\scriptsize\widetilde{\G}}}{\det\Delta_{\G}} = X_D^{\scriptsize\widetilde{\G}}\times\P^{\G}_{1,\ldots,1}(i_2,\ldots,i_n).
\label{XDG}
\end{equation}
In the second step, the fraction $X_D^{\scriptsize\widetilde{\G}}$ is computed using the grove theorem \eqref{grovethm}, together with the insertion of a zipper anchored at $i_1$ as well as a further modification of $\widetilde{\G}$ to $\xbar{\G}$, by which two edges incident to $i_1$ are removed to form an annular-one graph. It involves a perturbation matrix $U_1$, of the form \eqref{firstU}, with nonzero entries corresponding to sites of $\V_1$ (see also below). Then  $X_D^{\scriptsize\widetilde{\G}}$ can be expressed in terms of $\xbar{G},\xbar{G}'$, evaluated at the nodes lying around the ``hole'' of the annulus, namely at $\V_1$. From the relation $\xbar{\mathbf{\Delta}}(z) = \mathbf{\Delta}(z) - \sum_{k=1}^n U_k^{\rm t} U^{}_k$, namely we include the zipper on both $\xbar{\G}$ and ${\G}$, the entries of $\xbar{G},\xbar{G}'$ on sites of $\V_1$ may be written in terms of entries of $G$ and $G'$. Indeed, the Woodbury formula enables one to relate $\xbar{\Gr}(z)$ to $\Gr(z)$. An expansion in powers of $z{-}1$ then yields
\begin{align}
&\xbar{G} = G + G U^{\rm t} A^{-1} U G, \qquad \hbox{with\ } A = {\mathbb I} - U G U^{\rm t}, \\
\noalign{\medskip}
&\xbar{G}' = G' + G' U^{\rm t} A^{-1} U G + G U^{\rm t} A^{-1} U G' U^{\rm t} A^{-1} U G + G U^{\rm t} A^{-1} U G'\,, \label{Gtildep}
\end{align}
where $U$ is the matrix obtained by piling up all $U_k$'s, for $1 \le k \le n$.

Let us now examine the effect on $X_D^{\G}$ of changing the sign of $y_k$. We first discuss the easier case $k \ge 2$. We have seen in the previous subsection that the transformation $y_k \mapsto -y_k$ has the effect of replacing the Green matrix $G$ with its conjugate $\Sigma_k G \Sigma_k$. The conjugation may itself be absorbed in the redefinition of $U_k$ into $\widehat{U}_k = U_k\sigma_k$ and leaves the factor $\P^{\G}_{1,\ldots,1}(i_2,\ldots,i_n)$ invariant. It is not difficult to see that the transformation has the same effect on $\xbar G$: it replaces $U_k$ with $\widehat{U}_k$ {\it and} conjugates $\xbar G$ with $\Sigma_k$. 

In fact, exactly the same conclusion applies to $\xbar{G}'$ provided the site $i_k$ is far from the zipper (this restriction is what makes the case $k=1$ different). It follows from the relation \eqref{images}, which shows that $G'$ obeys the same relations \eqref{Gop_sym1} and \eqref{Gop_sym2} as $G$, implying that it too gets conjugated with $\Sigma_k$ when one transforms $y_k$ to $-y_k$. The equation \eqref{Gtildep} readily shows that the transformations of $\xbar G$ and $\xbar G'$ are strictly identical.

We know that $X_D^{\scriptsize\widetilde{\G}}$ may be written in terms of the entries $\xbar G_{u,v}$ and $\xbar G'_{u,v}$ for $u,v \in \V_1$. Because $\Sigma_k$ is trivial on sites far from $\V_k$, the conjugation with $\Sigma_k$ has no effect at all on sites of $\V_1$,
\be
\xbar G_{u,v}\Big|_{U_k} \longrightarrow \big(\Sigma_k \xbar G \Sigma_k \big)_{u,v} = \xbar G_{u,v}\Big|_{\widehat{U}_k}, \qquad u,v \in \V_1,
\ee
and similarly for $\xbar G'_{u,v}$. The notation on the left-hand side (resp. right-hand side) of the equation indicates that the removed edges around $i_k$ in the modified graph $\xbar{\G}$ are encoded in $U_k$ (resp. $\widehat{U}_k$). It follows that the net effect of the transformation $y_k \mapsto -y_k$ is to replace $U_k$ with $\widehat{U}_k$, and is therefore irrelevant. Hence, we conclude that $X_D^{\scriptsize\widetilde{\G}}$ and $\P_{a,1,\ldots,1}(i_1,\ldots,i_n)$ are even functions of $y_k$ for any $k \ge 2$.

The effect of the transformation $y_1 \mapsto -y_1$ on a fraction $X_D^{\scriptsize\widetilde{\G}}$ is similar to some extent, but differs by the presence of the zipper in the neighborhood of $i_1$. The transformation of $\xbar G$ can be computed along the same lines as above, but that of $\xbar G'$ is tricky to write in general, because the precise location of the zipper depends on the edge cuts necessary to make of $\widetilde{\G}$ an annular-one graph, which themselves depend on the predecessor diagram one considers. So we shall illustrate the computation in a specific example, namely the case of $\P_2^{\textrm{op}}(i_1)$ discussed in Section~\ref{sec6.1}. There we claimed that the transformation $y_1 \mapsto -y_1$ exchanges the two fractions of spanning trees $X_1^{\rm S}(i_1)$ and $X_1^{\rm N}(i_1)$, for which the reference site $i_1$ has its southern (resp. northern) neighbor as its unique predecessor. As one does not enforce any height 1, there is no first modification to consider, so that $\widetilde \G = \G$.

\begin{figure}[t]
\centering
\begin{tikzpicture}[scale=0.75,font=\small]

\begin{scope}
\draw (-4,-3)--(-4,3)--(4,3)--(4,-3);
\draw[step=0.5cm,help lines,semithick,dotted] (-4,-3) grid (4,3);
\draw[thick] (-4,-3)--(-4,3)--(4,3)--(4,-3);
\fill[pattern=north east lines, pattern color=blue] (-4.0125,-3.5) rectangle (4.0125,-3.1);
\draw[fill=blue!15!white] (0,0)--(0.5,0)--(0.5,0.5)--(0,0.5)--(-0.5,0.5)--(-0.5,-0.5)--(0,-0.5)--(0,0);
\filldraw[black] (0,0) circle (0.075cm);
\draw (0.3,-0.3) node {5};
\filldraw (0.5,0) circle (0.075cm);
\draw (0.8,-0.3) node {1};
\filldraw (0,0.5) circle (0.075cm);
\draw (-0.25,0.9) node {2};
\filldraw (-0.5,0) circle (0.075cm);
\draw (-0.9,0) node {3};
\filldraw (0,-0.5) circle (0.075cm);
\draw (0,-0.9) node {4};
\filldraw (3,3) circle (0.075cm) node [above right] {6};
\draw (2.35,-2.5) node {\small $\mathbf{i_1\!=\!5\!=\!(x_1,y_1)}$};
\filldraw[red] (0.25,0.25) circle (0.075cm);
\draw[very thick,red] (0.25,0.25)--(4,0.25);
\draw[thick,->,red] (1.25,0)--(1.25,0.5);
\end{scope}

\begin{scope}[xshift=10cm]
\draw (-4,-3)--(-4,3)--(4,3)--(4,-3);
\draw[step=0.5cm,help lines,semithick,dotted] (-4,-3) grid (4,3);
\draw[thick] (-4,-3) -- (-4,3) -- (4,3) -- (4,-3);
\fill[pattern=north east lines, pattern color=blue] (-4.0125,-3.5) rectangle (4.0125,-3.1);
\draw[fill=blue!15!white] (-0.5,0)--(-0.5,0.5)--(0,0.5)--(0,0)--(0.5,0)--(0.5,-0.5)--(-0.5,-0.5)--(-0.5,0);
\filldraw[black] (0,0) circle (0.075cm);
\draw (-0.225,-0.225) node {5};
\filldraw (0.5,0) circle (0.075cm);
\draw (0.8,0.2) node {1};
\filldraw (0,0.5) circle (0.075cm);
\draw (-0.25,0.9) node {2};
\filldraw (-0.5,0) circle (0.075cm);
\draw (-0.9,0) node {3};
\filldraw (0,-0.5) circle (0.075cm);
\draw (0,-0.9) node {4};
\filldraw (3,3) circle (0.075cm) node [above right] {6};
\draw (2.35,-2.5) node {\small $\mathbf{i_1\!=\!5\!=\!(x_1,y_1)}$};
\filldraw[red] (0.25,-0.25) circle (0.075cm);
\draw[very thick,red] (0.25,-0.25)--(4,-0.25);
\draw[thick,->,red] (1.25,0)--(1.25,-0.5);
\end{scope}

\end{tikzpicture}
\caption{Schematic representation of the graphs $\xbar{\G}$ modified according to $U_1$ (left) and to $\widehat{U}_1=U_1\sigma_1$  (right), as used for the computation of $X_1^{\rm S}(i_1)$ and $X_1^{\rm N}(i_1)$, together with their nodes and zipper.}
\label{new_zip}
\end{figure}
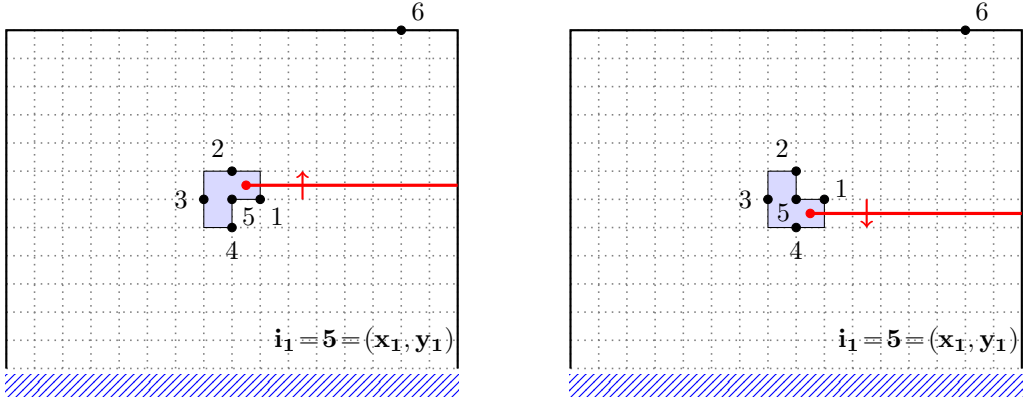

To compute $X_1^{\rm S}(i_1)$, we define the annular-one graph $\xbar{\G}$ by removing the edges connecting $i_1$ to its northern and western neighbors, meaning that we choose the matrix $U_1$ as
\begin{equation}
U_1=\bordermatrix{
& & i_1 & {\rm N}_1 & {\rm E}_1 & {\rm S}_1 & {\rm W}_1 & \cr
& \cdots & 1 & -1 & 0 & 0 & 0 & \cdots \cr
& \cdots & 1 & 0 & 0 & 0 & -1 & \cdots}, 
\end{equation}
The vertices of $\V_1=\{i_1,{\rm N}_1,{\rm E}_1,{\rm S}_1,{\rm W}_1\}$ are nodes, with the sink $s$ taken as the sixth node as usual. To facilitate the calculations, we choose, as represented on the left panel of Fig.~\ref{new_zip}, a semi-infinite zipper going horizontally to the right, with a nontrivial parallel transport $z\in\C^*$ on the oriented edges $((x_1{+}k{+}1,y_1),$ $(x_1{+}k{+}1,y_1{+}1))$, for $k\ge 0$.

From Section~\ref{sec4}, the fraction $X_1^{\rm S}(i_1)$ is equal to the following ratio:
\begin{equation}
X_1^{\rm S}(i_1) = 3 \frac{\xbar{Z}[4|12356]}{Z} = 3\frac{\xbar{Z}}{Z} \Big[(\xbar{G}_{4,4}-\xbar{G}_{4,5}) - (\xbar{G}_{3,4}-\xbar{G}_{3,5}-\xbar{G}'_{3,4}+\xbar{G}'_{3,5}-\xbar{G}'_{4,5})\Big].
\label{sym_ex1}
\end{equation}

The effect on $\xbar G$ of the change $y_1 \mapsto -y_1$ is exactly the same as the one computed above: $\xbar G$ gets conjugated with $\Sigma_1$ and at the same time, the matrix $U_1$ is replaced with $\widehat{U}_1\equiv U_1 \sigma_1$, where $\sigma_1$ acts on $\V_1$ by exchanging N$_1$ and S$_1$. However, since $\xbar G$ is evaluated at sites in $\V_1$, the conjugation is nontrivial, yielding
\be
\xbar G_{u,v}\Big|_{U_1} \longrightarrow \big(\Sigma_1 \xbar G \Sigma_1 \big)_{u,v} = \xbar G_{\sigma_1(u),\sigma_1(v)}\Big|_{\widehat{U}_1}, \qquad u,v \in \V_1,
\ee
Similarly we find using Eq.~\eqref{wood2} that
\begin{equation*}
\frac{\xbar{Z}}{Z}\Big|_{U_1} \longrightarrow \; \frac{\xbar{Z}}{Z}\Big|_{\widehat{U}_1}.
\end{equation*}

We compute the change of $\xbar G'$ by using its expression in terms of $\xbar G$ itself,
\begin{equation}
\xbar{G}'_{u,v} = \sum_{k=0}^{\infty}\Big(\xbar{G}_{u,(x_1+k+1,y_1+1)}\,\xbar{G}_{(x_1+k+1,y_1),v}-\xbar{G}_{u,(x_1+k+1,y_1)}\,\xbar{G}_{(x_1+k+1,y_1+1),v}\Big).
\end{equation}
For $u$ and $v$ in $\V_1$, it yields the following transformation
\begin{align}
\xbar{G}'_{u,v}\Big|_{U_1} \longrightarrow \; & \sum_{k=0}^{\infty}\Big(\xbar{G}_{\sigma_1(u),(x_1+k+1,y_1-1)}\,\xbar{G}_{(x_1+k+1,y_1),\sigma_1(v)}\nonumber \\
&\hspace{1cm}- \xbar{G}_{\sigma_1(u),(x_1+k+1,y_1)}\,\xbar{G}_{(x_1+k+1,y_1-1),\sigma_1(v)}\Big)\Big|_{\widehat{U}_1} \nonumber\\
& = \; \xbar{G}'^{\text{new}}_{\sigma_1(u),\sigma_1(v)}\Big|_{\widehat{U}_1},
\label{sym_ex4}
\end{align}
where in addition to the action of $\sigma_1$ as seen before, the derivative $\xbar{G}'^{\text{new}}$ is defined with respect to a new zipper, located one lattice spacing below the original one, and with a reversed orientation, now pointing downward, see the right panel of Fig.~\ref{new_zip}. Note that the hollow face has also been reversed, since the edge cuts are now prescribed by $\widehat{U}_1$; the edges connecting $i_1$ to its western and southern neighbors are accordingly removed.

Altogether, we obtain
\begin{equation}
X_1^{\rm S}(i_1)\Big|_{y_1\mapsto -y_1} = 3\frac{\xbar{Z}}{Z} \Big[(\xbar{G}_{2,2}-\xbar{G}_{2,5}) - (\xbar{G}_{3,2}-\xbar{G}_{3,5}-\xbar{G}'^{\text{new}}_{3,2}+\xbar{G}'^{\text{new}}_{3,5}-\xbar{G}'^{\text{new}}_{2,5})\Big]\Big|_{\widehat{U}_1},
\end{equation}
where the bar refers to the lattice $\xbar{\G}$ modified according to $\widehat{U}_1=U_1\sigma_1$. It is straightforward to see that the combinatorial expression appearing on the right-hand side is that of $3\,\xbar{Z}[2|13456]/Z$, and that it is equal to the fraction of spanning trees on the original graph $\G$ such that $i_1$ has its northern neighbor N$_1$ as its unique predecessor. We therefore find
\begin{equation}
X_1^{\rm S}(i_1)\Big|_{y_1\mapsto -y_1} = X_1^{\rm N}(i_1).
\end{equation}
Using a similar argument, one can show that
\begin{equation}
X_1^{\rm E}(i_1)\Big|_{y_1{\mapsto}{-}y_1} = X_1^{\rm E}(i_1), \quad X_1^{\rm W}(i_1)\Big|_{y_1{\mapsto}{-}y_1} = X_1^{\rm W}(i_1);
\end{equation}
the two being actually equal by the obvious left-right symmetry. From these relations, we obtain that the height-two probability
\begin{equation}
\P_2(i_1) = \P_1(i_1) + \frac{1}{3} \left\{X_1^{\rm N} (i_1) + X_1^{\rm E}(i_1) + X_1^{\rm S}(i_1) + X_1^{\rm W}(i_1)\right\}
\end{equation}
is an even function of the variable $y_1$.

\begin{figure}[t]
\centering
\begin{tikzpicture}[scale=1.25,font=\large]
\tikzstyle arrowstyle=[scale=1]
\tikzstyle directed=[postaction={decorate,decoration={markings,mark=at position .7 with {\arrow[arrowstyle]{stealth}}}}]

\begin{scope}[xshift=0cm]
\filldraw[black] (-1.5,0) circle (0.075cm) node[above left] {$i_1$};
\filldraw[white] (-1,0) circle (0.075cm);
\draw[thick] (-1,0) circle (0.075cm);
\filldraw[black] (-1.5,0.5) circle (0.075cm);
\filldraw[black] (-2,0) circle (0.075cm);
\filldraw[black] (-1.5,-0.5) circle (0.075cm);
\draw[directed,ultra thick] (-1.5,0.5)--(-1.5,0);
\draw[directed,ultra thick] (-2,0)--(-1.5,0);
\draw[directed,ultra thick] (-1.5,-0.5)to[out=230,in=230](-2,0);
\fill[pattern=north east lines, pattern color=blue] (-2.95,-1.5) rectangle (-0.05,-2);
%\draw[thick,<->] (-0.5,-2)--node[right] {$y_1$}(-0.5,0);
\node at (-2.625,-1.25) {\Large $D$};
\end{scope}

\begin{scope}[xshift=6cm]
\filldraw[black] (-1.5,0) circle (0.075cm) node[below right] {$i_1$};
\filldraw[black] (-1.5,-0.5) circle (0.075cm);
\filldraw[black] (-1.5,0.5) circle (0.075cm);
\filldraw[black] (-2,0) circle (0.075cm);
\filldraw[white] (-1,0) circle (0.075cm);
\draw[thick] (-1,0) circle (0.075cm);
\draw[directed,ultra thick] (-1.5,-0.5)--(-1.5,0);
\draw[directed,ultra thick] (-2,0)--(-1.5,0);
\draw[directed,ultra thick] (-1.5,0.5)to[out=130,in=130](-2,0);
\fill[pattern=north east lines, pattern color=blue] (-2.95,-1.5) rectangle (-0.05,-2);
%\draw[thick,<->] (-0.5,-2)--node[right] {$y_1$}(-0.5,0);
\node at (-2.625,-1.25) {\Large $D^*$};
\end{scope}

\end{tikzpicture}
\caption{Predecessor diagram $D$ on the upper half-plane contributing to $X_2$ and its mirror diagram $D^*$. The eastern neighbor of $i_1$, drawn as an open circle, is \emph{not} a predecessor of $i_1$.}
\label{mirror_diag}
\end{figure}
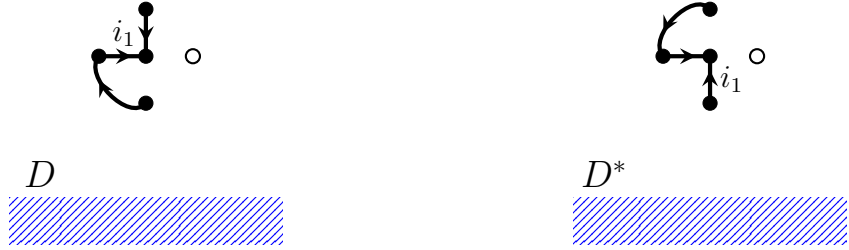

More generally, let us consider a given predecessor diagram $D$ for $i_1$ on $\G$ and its associated probability $X_D$. We define $D^*$ as the mirror predecessor diagram of $D$, such that the roles of N$_1$ and S$_1$ are swapped in $D^*$ with respect to $D$ (see for instance Fig.~\ref{mirror_diag}). We conjecture the following relation, which we have verified for all diagrams contributing to one- and two-site probabilities on the upper half-plane with open boundary condition:
\begin{equation}
X_D(i_1)\Big|_{y_1\mapsto -y_1}=X_{D^*}(i_1).
\label{sym_conj}
\end{equation}
Moreover, we make the following observation: a predecessor diagram $D$ and its mirror image $D^*$ contribute equally to the same fraction $X_p(i_1)$ of spanning trees with $p$ predecessors of $i_1$ among its neighbors. Consequently, if Eq.~\eqref{sym_conj} holds, multisite probabilities $\P_{a,1,\ldots,1}$ on the UHP with an open boundary are even functions of $y_1$ (and hence of each $y_k$ as shown above).

\subsubsection*{Closed boundary conditions and the diagonal upper half-plane}
Predecessor diagrams on the upper half-plane $\G=\Z\times\mathbb{N}^*$ with a closed boundary are also related to one another through a simple transformation under which the Green function \eqref{Gcl} is invariant:
\begin{equation}
G^{\textrm{cl}}_{(a,b),(c,d)}=G^{\textrm{cl}}_{(a,b),(c,1-d)}=G^{\textrm{cl}}_{(a,1-b),(c,1-d)}.
\end{equation}
Using similar arguments to those for the open boundary, we argue that a diagram $D$ and its mirror image $D^*$ are related through $y_1 \mapsto 1-y_1$, for a reference site $i_1$ located at $(x_1,y_1)$. It follows that joint probabilities of a single height $h_{i_1}=a\ge 1$ and many unit heights $h_{i_k}=1$ for $2\le k\le n$ are even functions of the variables $r_k=y_k{-}1/2$.

On the diagonal upper half-plane (DUHP) $\G=\left\{(x,y)\in\Z^2|y>x\right\}$, Green functions for open \eqref{Gop_diag} and closed \eqref{Gcl_diag} boundary conditions possess the following symmetries:
\begin{align}
G^{\textrm{op}}_{(a,b),(c,d)}& = - \, G^{\textrm{op}}_{(a,b),(d,c)} = G^{\textrm{op}}_{(b,a),(d,c)},\\
G^{\textrm{cl}}_{(a,b),(c,d)}& = G^{\textrm{cl}}_{(a,b),(d-1,c+1)} = G^{\textrm{cl}}_{(b-1,a+1),(d-1,c+1)},
\end{align}
so probabilities $\P_{a,1\ldots,1}(i_1,\ldots,i_n)$ on the DUHP are even functions of the variables
\begin{equation}
r_k=\frac{y_k-x_k}{\sqrt{2}}\quad\textrm{for open b.c. and }r_k=\frac{y_k-x_k-1}{\sqrt{2}}\quad\textrm{for closed b.c.}
\end{equation}

%%%%%%%%%%%%%%%%%%%%%%%%%%%%%%%%%%%%%%%%%%%%%%%%%%%%%%%%%%%%%%%%%%%%%%%%%%%%%%%%%%%%%%%%%%%%%%%%%%%%%%%%

%\bibliographystyle{acm}
%\bibliography{ASM}

\begin{thebibliography}{10}

\bibitem{BTW87}
{\sc Bak, P., Tang, C., and Wiesenfeld, K.}
\newblock {Self-organized criticality: An explanation of the 1/f noise}.
\newblock {\em Phys. Rev. Lett. 59}, 4 (1987), 381--384.

\bibitem{BBGJ07}
{\sc Bouttier, J., Bowick, M., Guitter, E., and Jeng, M.}
\newblock {Vacancy localization in the square dimer model}.
\newblock {\em Phys. Rev. E 76}, 4 (2007), 041140.

\bibitem{BIP93}
{\sc Brankov, J.~G., Ivashkevich, E.~V., and Priezzhev, V.~B.}
\newblock {Boundary effects in a two-dimensional Abelian sandpile}.
\newblock {\em J. Phys. I France 3}, 8 (1993), 1729--1740.

\bibitem{BPPR14}
{\sc Brankov, J.~G., Poghosyan, V.~S., Priezzhev, V.~B., and Ruelle, P.}
\newblock {Transfer matrix for spanning trees, webs and colored forests}.
\newblock {\em J. Stat. Mech.\/} (2014), P09031.

\bibitem{CS12}
{\sc Caracciolo, S., and Sportiello, A.}
\newblock {Exact integration of height probabilities in the Abelian Sandpile
  model}.
\newblock {\em J. Stat. Mech.\/} (2012), P09013.

\bibitem{Car84}
{\sc Cardy, J.~L.}
\newblock {Conformal invariance and surface critical behavior}.
\newblock {\em Nucl. Phys. B 240}, 4 (1984), 514--532.

\bibitem{CR13}
{\sc Creutzig, T., and Ridout, D.}
\newblock {Logarithmic conformal field theory: beyond an introduction}.
\newblock {\em J. Phys. A 46}, 49 (2013), 494006.

\bibitem{Dha90}
{\sc Dhar, D.}
\newblock {Self-organized critical state of sandpile automaton models}.
\newblock {\em Phys. Rev. Lett. 64}, 14 (1990), 1613--1616.

\bibitem{Dha06}
{\sc Dhar, D.}
\newblock {Theoretical studies of self-organized criticality}.
\newblock {\em Physica A 369}, 1 (2006), 29--70.

\bibitem{Dur09}
{\sc Dürre, M.}
\newblock {Conformal covariance of the Abelian sandpile height one field}.
\newblock {\em Stochastic Process. Appl. 119}, 9 (2009), 2725--2743.

\bibitem{Flo03}
{\sc Flohr, M. A.~I.}
\newblock {Bits and pieces in logarithmic field theory}.
\newblock {\em Int. J. Mod. Phys. A 18}, 25 (2003), 4497--4591.

\bibitem{For93}
{\sc Forman, R.}
\newblock {Determinants of Laplacians on graphs}.
\newblock {\em Topology 32}, 1 (1993), 35--46.

\bibitem{GK96}
{\sc Gaberdiel, M.~R., and Kausch, H.~G.}
\newblock {Indecomposable fusion products}.
\newblock {\em Nucl. Phys. B 477}, 1 (1996), 293--318.

\bibitem{GK99}
{\sc Gaberdiel, M.~R., and Kausch, H.~G.}
\newblock {A local logarithmic conformal field theory}.
\newblock {\em Nucl. Phys. B 538}, 3 (1999), 631--658.

\bibitem{GR06}
{\sc Gaberdiel, M.~R., and Runkel, I.}
\newblock {The logarithmic triplet theory with boundary}.
\newblock {\em J. Phys. A 39}, 47 (2006), 14745--14780.

\bibitem{GJRSV13}
{\sc Gainutdinov, A., Jacobsen, J., Read, N., Saleur, H., and Vasseur, R.}
\newblock {Logarithmic conformal field theory: a lattice approach}.
\newblock {\em J. Phys. A 46}, 49 (2013), 494012.

\bibitem{GRR13}
{\sc Gainutdinov, A., Ridout, D., and Runkel, I. (eds.)}
\newblock {Logarithmic conformal field theory}.
\newblock {\em J. Phys. A 46}, 49 (2013).
\newblock Special issue.

\bibitem{GPP09}
{\sc Grigorev, S.~Y., Poghosyan, V.~S., and Priezzhev, V.~B.}
\newblock {Three-leg correlations in the two-component spanning tree on the
  upper half-plane}.
\newblock {\em J. Stat. Mech.\/} (2009), P09008.

\bibitem{Gur93}
{\sc Gurarie, V.}
\newblock {Logarithmic operators in conformal field theory}.
\newblock {\em Nucl. Phys. B 410}, 3 (1993), 535--549.

\bibitem{Iva94}
{\sc Ivashkevich, E.~V.}
\newblock {Boundary height correlations in a two-dimensional Abelian sandpile}.
\newblock {\em J. Phys. A 27}, 11 (1994), 3643.

\bibitem{Jen05a}
{\sc Jeng, M.}
\newblock {Conformal field theory correlations in the Abelian sandpile model}.
\newblock {\em Phys. Rev. E 71}, 1 (2005), 016140.

\bibitem{Jen05b}
{\sc Jeng, M.}
\newblock {Four height variables, boundary correlations, and dissipative
  defects in the Abelian sandpile model}.
\newblock {\em Phys. Rev. E 71}, 3 (2005), 036153.

\bibitem{JPR06}
{\sc Jeng, M., Piroux, G., and Ruelle, P.}
\newblock {Height variables in the Abelian sandpile model: scaling fields and
  correlations}.
\newblock {\em J. Stat. Mech.\/} (2006), P10015.

\bibitem{Ken11}
{\sc Kenyon, R.}
\newblock {Spanning forests and the vector bundle Laplacian}.
\newblock {\em Ann. Probab. 39}, 5 (2011), 1983--2017.

\bibitem{KW15}
{\sc Kenyon, R.~W., and Wilson, D.~B.}
\newblock {Spanning trees of graphs on surfaces and the intensity of
  loop-erased random walk on planar graphs}.
\newblock {\em J. Amer. Math. Soc. 28}, 4 (2015), 985--1030.

\bibitem{KR09}
{\sc Kyt{\"o}l{\"a}, K., and Ridout, D.}
\newblock {On staggered indecomposable Virasoro modules}.
\newblock {\em J. Math. Phys. 50}, 12 (2009), 123503.

\bibitem{MR01}
{\sc Mahieu, S., and Ruelle, P.}
\newblock {Scaling fields in the two-dimensional abelian sandpile model}.
\newblock {\em Phys. Rev. E 64}, 6 (2001), 066130.

\bibitem{MD91}
{\sc Majumdar, S.~N., and Dhar, D.}
\newblock {Height correlations in the Abelian sandpile model}.
\newblock {\em J. Phys. A 24}, 7 (1991), L357.

\bibitem{MD92}
{\sc Majumdar, S.~N., and Dhar, D.}
\newblock {Equivalence between the Abelian sandpile model and the $q\to 0$
  limit of the Potts model}.
\newblock {\em Physica A 185}, 1 (1992), 129--145.

\bibitem{MRR15}
{\sc Morin-Duchesne, A., Rasmussen, J., and Ruelle, P.}
\newblock {Dimer representations of the Temperley-Lieb algebra}.
\newblock {\em Nucl. Phys. B 890}, Supplement C (2015), 363--387.

\bibitem{MRR16}
{\sc Morin-Duchesne, A., Rasmussen, J., and Ruelle, P.}
\newblock {Integrability and conformal data of the dimer model}.
\newblock {\em J. Phys. A 49}, 17 (2016), 174002.

\bibitem{PR07}
{\sc Pearce, P.~A., and Rasmussen, J.}
\newblock {Solvable critical dense polymers}.
\newblock {\em J. Stat. Mech.\/} (2007), P02015.

\bibitem{PRZ06}
{\sc Pearce, P.~A., Rasmussen, J., and Zuber, J.-B.}
\newblock {Logarithmic minimal models}.
\newblock {\em J. Stat. Mech.\/} (2006), P11017.

\bibitem{PV17}
{\sc Pearce, P.~A., and Vittorini-Orgeas, A.}
\newblock {Yang–Baxter solution of dimers as a free-fermion six-vertex
  model}.
\newblock {\em J. Phys. A 50}, 43 (2017), 434001.

\bibitem{PR04}
{\sc Piroux, G., and Ruelle, P.}
\newblock {Pre-logarithmic and logarithmic fields in a sandpile model}.
\newblock {\em J. Stat. Mech.\/} (2004), P10005.

\bibitem{PR05a}
{\sc Piroux, G., and Ruelle, P.}
\newblock {Boundary height fields in the Abelian sandpile model}.
\newblock {\em J. Phys. A 38}, 7 (2005), 1451.

\bibitem{PR05b}
{\sc Piroux, G., and Ruelle, P.}
\newblock {Logarithmic scaling for height variables in the Abelian sandpile
  model}.
\newblock {\em Phys. Lett. B 607}, 1 (2005), 188--196.

\bibitem{PGPR08}
{\sc Poghosyan, V.~S., Grigorev, S.~Y., Priezzhev, V.~B., and Ruelle, P.}
\newblock {Pair correlations in sandpile model: A check of conformal field
  theory}.
\newblock {\em Phys. Lett. B 659}, 3 (2008), 768--772.

\bibitem{PGPR10}
{\sc Poghosyan, V.~S., Grigorev, S.~Y., Priezzhev, V.~B., and Ruelle, P.}
\newblock {Logarithmic two-point correlators in the Abelian sandpile model}.
\newblock {\em J. Stat. Mech.\/} (2010), P07025.

\bibitem{PP11}
{\sc Poghosyan, V.~S., and Priezzhev, V.~B.}
\newblock {The Problem of Predecessors on Spanning Trees}.
\newblock {\em Acta Polytech. 51}, 2 (2011), 59--62.

\bibitem{PPR08}
{\sc Poghosyan, V.~S., Priezzhev, V.~B., and Ruelle, P.}
\newblock {Jamming probabilities for a vacancy in the dimer model}.
\newblock {\em Phys. Rev. E 77}, 4 (2008), 041130.

\bibitem{PPR11}
{\sc Poghosyan, V.~S., Priezzhev, V.~B., and Ruelle, P.}
\newblock {Return probability for the loop-erased random walk and mean height
  in the Abelian sandpile model: a proof}.
\newblock {\em J. Stat. Mech.\/} (2011), P10004.

\bibitem{Pri94}
{\sc Priezzhev, V.~B.}
\newblock {Structure of two-dimensional sandpile. I. Height probabilities}.
\newblock {\em J. Stat. Phys. 74}, 5 (1994), 955--979.

\bibitem{RS92}
{\sc Rozansky, L., and Saleur, H.}
\newblock {Quantum field theory for the multi-variable Alexander-Conway
  polynomial}.
\newblock {\em Nucl. Phys. B 376}, 3 (1992), 461--509.

\bibitem{Rue02}
{\sc Ruelle, P.}
\newblock {A c=-2 boundary changing operator for the Abelian sandpile model}.
\newblock {\em Phys. Lett. B 539}, 1 (2002), 172--177.

\bibitem{Rue13}
{\sc Ruelle, P.}
\newblock {Logarithmic conformal invariance in the Abelian sandpile model}.
\newblock {\em J. Phys. A 46}, 49 (2013), 494014.

\bibitem{Spi76}
{\sc Spitzer, F.}
\newblock {\em {Principles of Random Walk}}.
\newblock Springer, 1976.

\end{thebibliography}

\end{document}